\documentclass[12pt]{article}
\usepackage{amsfonts}
\usepackage{latexsym}
\usepackage{amsmath}
\usepackage{amssymb}
\usepackage{amssymb}
\usepackage{relsize}

\newcommand{\newsection}{    
\setcounter{equation}{0}\section}
\def\appendix#1{\addtocounter{section}{1}\setcounter{equation}{0}
\renewcommand{\thesection}{\Alph{section}}
\section*{Appendix \thesection\protect\indent \parbox[t]{11.15cm}{#1}}
\addcontentsline{toc}{section}{Appendix \thesection\ \ \ #1}}

\newcommand{\be}{\begin{eqnarray}}
\newcommand{\ee}{\end{eqnarray}}
\newcommand{\bea}{\begin{eqnarray}}
\newcommand{\eea}{\end{eqnarray}}
\newcommand{\ba}{\begin{array}}
\newcommand{\ea}{\end{array}}

\newenvironment{frcseries}{\fontfamily{frc}\selectfont}{}

\def\cL{{\cal L}}

\def\hn{{\hat{\nu}}}

\def\bbe{{\bf{e}}}
\font\mybb=msbm10 at 11pt

\def\bb#1{\hbox{\mybb#1}}

\def\bR {\bb{R}}

\def\bC {\bb{C}}
\def\bI {\bb{I}}

\def\hn {{\hat{\nabla}}}
\def\hT {{\hat{\Theta}}}
\def\cD {{\cal{D}}}
\def\cG {{\cal{G}}}
\def\cbG {{\bar{{\cal{G}}}}}
\def\nnu{{\mathlarger{\mathlarger{\mathlarger{\nu}}}}}
\def\uu{{\mathlarger{\mathlarger{\mathlarger{u}}}}}
\begin{document}
\begin{titlepage}
\begin{center}
\vspace*{-1.0cm}
\hfill DMUS--MP--16/08 \\
\hfill LTH 1090 \\


\vskip 1cm

\vspace{2.0cm}
{\Large \bf Dynamical symmetry enhancement near  ${\cal N}=2$, $D=4$  gauged supergravity horizons} \\[.2cm]

\vskip 2cm
 {\large  J.~Gutowski$^1$, T. Mohaupt$^2$ and  G. Papadopoulos$^3$}
\vspace{0.5cm}
\\
$^1$ Department of Mathematics \\
University of Surrey \\
Guildford, GU2 7XH, UK \\

\vspace{0.5cm}

${}^2$ Department of Mathematical Sciences\\
University of Liverpool\\
Peach Street, Liverpool L69 7ZL, UK\\

\vspace{0.5cm}
${}^3$ Department of Mathematics\\
King's College London\\
Strand\\
London WC2R 2LS, UK\\

\vspace{0.5cm}

\end{center}

\vskip 1cm
\begin{abstract}

\end{abstract}
 We show that all smooth Killing horizons with compact horizon sections of 4-dimensional gauged ${\cal N}=2$ supergravity coupled to any number of vector  multiplets
preserve $2 c_1({\cal K})+4 \ell$ supersymmetries, where ${\cal K}$ is a pull-back of the Hodge bundle of the special K\"ahler manifold on the horizon spatial section. We also demonstrate that all such horizons with $c_1({\cal K})=0$  exhibit an $\mathfrak{sl}(2, \bR)$ symmetry and preserve either 4 or 8 supersymmetries. If the orbits
of the $\mathfrak{sl}(2, \bR)$ symmetry are 2-dimensional, the horizons are warped products of $\mathrm{AdS}_2$ with the horizon spatial section.
Otherwise, the horizon section admits an isometry which preserves all the fields. The proof of these results is centered on the use of
index theorem in  conjunction with an appropriate generalization of the Lichnerowicz theorem for horizons that preserve at least one supersymmetry.
 In all $c_1({\cal K})=0$ cases, we specify the local geometry of spatial horizon sections and demonstrate that
 the solutions are determined by first order non-linear ordinary differential equations on some of the fields.

\end{titlepage}


 \setcounter{section}{0}

\newsection{Introduction}

It has been known for some time that there is (super)symmetry enhancement near extreme black hole and brane horizons \cite{carter, gibbons1, gibbons}. This observation has been made
on a case by case basis and it has been instrumental in the formulation of AdS/CFT correspondence \cite{maldacena}
.

In the last three years it has been realized that (super)symmetry enhancement is a generic phenomenon  for all smooth supergravity  Killing
horizons with compact spatial sections that preserve at least one supersymmetry. The essential features of this (super)symmetry enhancement  mechanism have been described in \cite{iibindex} in the form of the ``{\it horizon conjecture}''  following earlier related work in \cite{5index, 11index}.  The horizon conjecture has two parts. One part involves a formula for the number of supersymmetries preserved by such horizons. In the second part, this is used to show that some of the horizons with non-trivial fluxes admit an
 $\mathfrak{sl}(2, \bR)$ symmetry subalgebra.
So far, the horizon conjecture has been proven  for all 10- and 11-dimensional supergravities \cite{heterotic, 11index, iibindex, iiaindex, iiamindex} and  minimal 5-dimensional gauged supergravity \cite{5index}.

In this paper, we shall demonstrate the validity of the horizon conjecture  \cite{iibindex} for all 4-dimensional gauged ${\cal N}=2$ supergravities coupled
to any number of abelian vector multiplets, see for example
 \cite{ferrara4}. The supersymmetric black hole solutions  of such theories,  and hence their near horizon geometries,  have been extensively investigated in the context of  entropy counting
and  attractor mechanism, starting from \cite{ferrara1, strominger1, ferrara2, ferrara3}.

The assumptions which are made for the proof of the horizon conjecture are as follows:

\begin{itemize}
\item The near horizon geometry as well as the rest of the fields are smooth,

\item the near horizon spatial section is compact without boundary,

\item the matrix of gauge couplings $\mathrm{Im}\, {\cal N}$  is negative definite and hence invertible\footnote{In turn this implies that the scalar manifold admits a (positive definite)  K\"ahler metric.},

\item the scalar potential $V$ is negative semi-definite, $V\leq 0$.

\end{itemize}
The first two assumptions may be replaced by the  requirement that the data are such that the Hopf maximum principle applies \cite{maxp}, and that a certain surface term integral over the horizon spatial section vanishes.

A consequence of the proof of the conjecture is that all  Killing horizons that satisfy these assumptions:

\begin{itemize}

\item[(i)] preserve
\bea
N=2 c_1({\cal K})+ 4 \ell~,
\label{nf}
\eea
supersymmetries, where $N\leq 8$, $\ell=1,2$ and ${\cal K}$ is the pull-back of the Hodge
bundle of the special K\"ahler geometry on the  spatial horizon section ${\cal S}$,

\item[(ii)] and those with  $\ell\not=0$, or equivalently $c_1({\cal K})=0$, admit
an $\mathfrak{sl}(2,\bR)$ symmetry\footnote{In 11-dimensional and type II horizons the presence of
$\mathfrak{sl}(2,\bR)$ requires that the horizons must have non-trivial fluxes. This is not necessary here as this assumption is implied by our restrictions on the couplings of ${\cal N}=2$ gauged supergravity.}.

\end{itemize}
Note that if $c_1({\cal K})=0$, which as we shall show is the case for all the horizons with $\ell\not=0$, the number of
supersymmetries preserved are either 4 or 8. This Chern class corresponds
to the index of a certain Dirac operator defined on ${\cal{S}}$.

We further proceed to investigate the geometry of the horizons with $c_1({\cal K})=0$.  There are two cases to consider depending on whether the orbits of
$\mathfrak{sl}(2,\bR)$ are 2- or 3-dimensional. In the former case, the horizons are warped products of $\mathrm{AdS}_2$ with the horizon spatial section
${\cal S}$, $\mathrm{AdS}_2\times_w {\cal S}$. Furthermore, if the warp factor is trivial,
 ${\cal S}$ is a sphere $S^2$, a torus $T^2$ or a (quotient of) hyperbolic space $H^2$ equipped with the Einstein metric depending on the sign  of the right-hand-side term in (\ref{rdeltavx}) and the rest of the fields either vanish
 or they are constant.  If the warp factor is non-trivial,  ${\cal S}$ admits an isometry which leaves the rest of the fields invariant. We give
 the local form of the metric on ${\cal S}$ and show that it depends on the scalars of the
 gauge multiplet. Moreover, we show that all the remaining fields
 are specified by first order ordinary differential equations. In particular, the scalars flow
 on the horizon.

If $\mathfrak{sl}(2,\bR)$ has a 3-dimensional orbit on the spacetime, then  ${\cal S}$ admits an isometry which leaves all the remaining fields invariant. There are several cases that one can consider. In all cases, we give the local form of the spacetime metric and demonstrate that the remaining
fields are determined by first order ordinary differential equations. In most cases,
the scalars flow on the horizon. Furthermore as the scalars depend on at most one coordinate, the first Chern class of ${\cal K}$ vanishes and so all such
horizons preserve either 4 or 8 supersymmetries.

We also  present an application of the horizon conjecture. In particular, we show that it is a consequence of the horizon conjecture  that all horizons with fluxes and $N_-\not=0$, see \cite{iibindex} and (\ref{npm}), for which the spatial horizon section is a marginally trapped surface contain
untrapped surfaces both just inside and outside the horizon.  This is a characteristic behavior of  extreme black hole horizons. As a result
such supersymmetric horizons  meet the necessary conditions of \cite{lucietti1}, see also \cite{booth2, mars},
 to be extended to  full extreme black hole solutions.

The proof of the horizons conjecture utilizes in a essential way that near a smooth Killing horizon
one can adapt a null gaussian coordinate system.  Then the Killing spinor equations (KSEs) of
${\cal N}=2$ supergravity are integrated along the lightcone directions to express the Killing spinors
in terms of spinors that depend only on the coordinates of ${\cal S}$.  The remaining equations
involve the reduction of the gravitino and gaugini KSEs on ${\cal S}$ as well as a large number
of integrability conditions. The latter are shown to be implied by the reduced gravitino and gaugini KSEs on ${\cal S}$ as well as the field equations. Unlike similar calculations for $D=11$ and type II supergravities, the assumption that the horizons admit one supersymmetry is used in an essential way.
Then the number of solutions of the reduced gravitino and gaugini KSEs on ${\cal S}$ are counted
by first making use of Lichnerowicz type theorems to turn the problem into
one of counting zero modes of Dirac-like operators on ${\cal S}$, and then
using the index theorem \cite{atiyah1}. After
taking into account that the KSEs of the ${\cal N}=2$ theory are linear over the complex numbers, the formula for the number of supersymmetries $N$ is produced
(\ref{nf}), where the number of supersymmetries $N$ is counted over the reals.

The proof of the second part of the horizon conjecture proceeds after first observing that if $c_1({\cal K})=0$ then  one can always construct pairs of Killing spinors over the spacetime which in turn
 give rise to three linearly independent vector bilinears.  Then the commutators of these vector fields are calculated and it is found that they satisfy a  $\mathfrak{sl}(2,\bR)$ algebra. The geometry
of these horizons is also  investigated.  For this, appropriate coordinates are adapted on the
horizon, and local expressions for the metric and other fields are obtained in all cases.

The paper is organized as follows. In section 2, after a brief description of gauged ${\cal N}=2$
supergravity,  we describe the near horizon geometries
and   evaluate the field equations of the theory on the near horizon fields. In section 3, we
solve the KSEs of ${\cal N}=2$ supergravity  along the lightcone directions of near horizon geometries
and state the remaining independent KSEs. In section 4, we establish that near horizon geometries
either preserve 4 or 8 supersymmetries.  In section 5, we slow that the near horizon geometries
exhibit an $\mathfrak{sl}(2,\bR)$ symmetry. In section 6, we describe the local geometries of
all near horizon geometries of  ${\cal N}=2$  gauged supergravity. In section 7, we present an
application of the horizon conjecture on trapped surfaces. In appendix A, we give our conventions.
In appendix B, we summarize the properties of special K\"ahler geometry which are essential in
all our derivations. In appendix C, we determine the independent KSEs of the near horizon
backgrounds.  In appendix D, we present the derivation of Lichnerowicz type theorems essential
for counting the supersymmetries.  In appendix E, we examine some symmetry properties of the
near horizon fields. In appendix F, we derive the near horizon data of a class of solutions
found previously in \cite{hristov}. In appendix G, we present the details of the derivation
of the local expressions for geometries of all near horizon configurations, and in appendix H we verify some of the field equations.

\newsection{Near-Horizon Geometry and Field Equations}

\subsection{${\mathcal N}=2$ gauged supergravity with vector multiplets}

The bosonic field content of the gravitational multiplet of ${\mathcal N}=2$ supergravity is a metric and a $U(1)$ field. The theory can also couple to $k$ vector abelian multiplets
in which case contains $k$ additional  $U(1)$ fields and
$2k$ real scalars. In the coupled theory, all the fields interact and the $U(1)$ field of the gravitational multiplet
mixes with the rest. The scalars take values on a sigma model manifold which exhibits a special K\"ahler structure. The two (real)
gravitini of the theory  can be described together as  a Dirac $\mathfrak{so}(3,1)$ spinor 1-form. The gaugini can also be described  as Dirac spinors. The supersymmetry parameter is then a Dirac spinor which is taken in what follows to be commuting.

The action of  ${\mathcal N}=2$, 4-dimensional, $U(1)$ gauged supergravity  with
no gauging of special K\"ahler isometries \cite{ferrara4} in the  conventions of  \cite{klemm1} is given by
\bea
e^{-1} {\cal{L}} &=& {1 \over 2} R +{1 \over 4}
\big( {\rm Im} {\cal{N}} \big)_{IJ} F^I_{\mu \nu}
F^{J \mu \nu} -{1 \over 8} \big( {\rm Re} {\cal{N}} \big)_{IJ}
e^{-1} \epsilon^{\mu \nu \rho \sigma} F^I_{\mu \nu}
F^J_{\rho \sigma}
\cr&&~~~~~~- g_{\alpha \bar{\beta}} \nabla_\mu z^\alpha
\nabla^\mu z^{\bar{\beta}} -V~,
\eea
where $R$ is the Ricci scalar of spacetime, $F^I=dA^I$ are the field strengths of $U(1)$ fields and so  $I=1, \dots, k+1$, $z$ are
$k$ complex scalars, and $V$ is the scalar potential, for a review see also \cite{vp}. We have suppressed all terms in the action that depend on the fermions. The scalar manifold $M$ exhibits special K\"ahler geometry with metric $g_{\alpha\bar\beta}$; see appendix B
for the definition and a summary of some key properties. The rest of the couplings include the
gauge couplings matrix ${\rm Im}{\cal{N}}$ and the theta angles  ${\rm Re} {\cal{N}} $ which can depend
on the scalars. These couplings are also determined in terms of the special K\"ahler geometry. Furthermore, the scalar
potential is given by
\bea
V = 4 g^2 \bigg(U^{IJ} -3  {\bar{X}}^I X^J \bigg) \xi_I \xi_J= -2 g^2\bigg( \big({\rm Im} {\cal{N}}\big)^{-1 \ IJ}+ 8 {\bar{X}}^I X^J \bigg) \xi_I \xi_J~,
\eea
where  $g$ is a non-zero constant, and the constants $\xi_I$ are
obtained from the $U(1)$ Fayet-Iliopoulos terms.  Moreover $X^I$, $I=1,\dots, k+1$, depend only the scalar fields $z, \bar z$ and are  defined
in the context of special K\"ahler geometry, see appendix B.
To establish the second identity we have used the expression for $U^{IJ}$ in appendix B.

As we have already mentioned in the introduction,  apart from the smoothness of the near horizon data, we shall make two assumptions on the couplings
of the theory.  These are that the matrix of gauge couplings ${\rm Im} {\cal{N}}$ is  negative definite, and that  $V\leq 0$.
A consequence of our two assumptions is that $\xi_I X^I$
never vanishes,
\bea
\xi_I X^I\not=0~.
\eea
 This is because
if $\xi_I X^I=0$ at any point, then at such a point
$V= -2g^2 ({\rm{Im}} {\cal{N}})^{-1 IJ} \xi_I \xi_J >0$, in contradiction
to our assumption that $V \leq 0$.

The Einstein, gauge and scalar field equations of the theory are

\bea
\label{eineq}
R_{\mu \nu} =-2 {\rm Im} ({\cal{N}})_{IJ} (F^+)^I{}_{\rho \mu}
(F^-)^{J\rho}{}_\nu +2 g_{\alpha \bar{\beta}} \nabla_{(\mu} z^\alpha
\nabla_{\nu)}{\bar{z}}^{\bar{\beta}} + g_{\mu \nu} V~,
\eea

\bea
\label{geq}
-2 \nabla_\mu \bigg( {\rm Im} ({\cal{N}})_{IJ} (F^-)^{J \mu \nu} \bigg)
+i (\nabla_\mu {\cal{N}}_{IJ}) {\tilde{F}}^{J \mu \nu}=0~,
\eea

\bea
\label{scaleq}
\nabla_\mu \nabla^\mu z^\alpha &+&{1 \over 4i} (F^+)^I{}_{\mu \nu}
(F^+)^{J \mu \nu} g^{\alpha {\bar{\gamma}}} {\partial \over \partial {\bar{z}}^{\bar{\gamma}}} {\cal{N}}_{IJ}
\cr
&-&{1 \over 4i} (F^-)^I{}_{\mu \nu} (F^-)^{J \mu \nu} g^{\alpha {\bar{\gamma}}} {\partial \over \partial {\bar{z}}^{\bar{\gamma}}} {{\bar{\cal{N}}}}_{IJ}
- g^{\alpha {\bar{\gamma}}} {\partial \over \partial {\bar{z}}^{\bar{\gamma}}} V =0~,
\eea
respectively.
It should be noted that ({\ref{eineq}}) and ({\ref{scaleq}}) correct
typographical errors found in \cite{klemm1}.

\subsection{Horizon Fields and field equations}

The black hole horizons that we shall be investigating are extremal Killing horizons with regular spatial horizon sections ${\cal S}$.
For such horizons, one can adapt a Gaussian Null
 coordinate system \cite{isen, gnull} such that the spacetime metric $ds^2$ and 2-form field strengths $F^I$ can be written as
\begin{eqnarray}
ds^2 &=& 2 \bbe^+ \bbe^- + \delta_{ij} \bbe^i \bbe^j~,~~~
\cr
F^I&=&\Phi^I \, \bbe^+\wedge \bbe^- + r \bbe^+\wedge d_h \Phi^I+{1\over2} Q^I \epsilon_{ij} \, \bbe^i\wedge \bbe^j~,
\label{horgeom}
\end{eqnarray}
where $u,r$ are the lightcone coordinates and $y^I$, $I=1, 2$, are the remaining coordinates of the spacetime, $d_h\Phi^I=d\Phi^I-h \Phi^I$, and the spatial horizon section ${\cal S}$ is given by $u=r=0$ with induced metric and volume form
\be
ds_{\cal{S}}^2 = \delta_{ij} \bbe^i \bbe^j~,~~~d\mathrm{vol}({\cal S})
={1\over2} \epsilon_{ij} \bbe^i\wedge \bbe^j~,
\ee
respectively. Furthermore,  we have used the frame
\bea
\label{nhbasis}
\bbe^+ &=& du~,~~~
\bbe^- = dr + r h - {1 \over 2} r^2 \Delta du~,~~~
\bbe^i = e^i{}_J dy^J~,~~~i,j=1,2~.
\eea
 The components of fields $h, \Delta, \Phi^I, Q^I$ and $\bbe^i$  depend only on  $y^I$. The black hole stationary
Killing vector field is identified with $\partial_u$ and becomes null on the hypersurface $r=0$. The 1-form gauge potential associated to $F^I$ is
\begin{eqnarray}
A^I = - r \Phi^I du + B^I , \qquad  d B^I = Q^I d\mathrm{vol}({\cal S})~.
\end{eqnarray}

Our smoothness assumption asserts that $\Delta, \Phi^I$, $Q^I$ are globally defined smooth scalars,  and $h$ is a globally defined smooth 1-form
on the horizon section ${\cal S}$, respectively. In addition,  the induced metric on ${\cal S}$, $ds_{\cal{S}}^2$,
is   smooth, and ${\cal S}$ is  compact, connected without boundary. We denote the Levi-Civita connection of ${\cal{S}}$ by $\hn$.

In what follows, it is convenient to define
\bea
(F^\pm)^I{}_{\mu \nu} = {1 \over 2}(F^I \pm {\tilde{F}}^I)_{\mu \nu}~,
\qquad {\tilde{F}}^I{}_{\mu \nu} = -{i \over 2} \epsilon_{\mu \nu}{}^{\rho \sigma} F^I_{\rho \sigma}~.
\eea
We note that the components of $(F^\pm)^I$ are given by
\bea
(F^\pm)^I_{+-} &=& {1 \over 2} (\Phi^I \mp i Q^I)~,~~~
(F^\pm)^I_{-j} = 0~,
\nonumber \\
(F^\pm)^I_{+i} &=& {r \over 2} \bigg( d_h\Phi_i \pm i \epsilon_i{}^j d_h\Phi_j\bigg)~,~~~
(F^\pm)^I_{ij} = \pm {i \over 2} (\Phi^I \mp i Q^I) \epsilon_{ij}~.
\eea

Before proceeding with the analysis of the supersymmetry, we decompose  the field equations of the bosonic fields along the lightcone and ${\cal{S}}$ directions. In particular,  $\nu=-$ component of field equations of the $U(1)$ gauge fields ({\ref{geq}}) is
\bea
\label{geq1}
\hn^j \bigg({\rm Im}({\cal{N}}_{IJ}) d_h\Phi^J_j \bigg)
-{\rm Im}({\cal{N}}_{IJ}) h^j d_h\Phi^J_j
+{1 \over 2} {\rm Im}({\cal{N}}_{IJ}) (dh)_{ij} \epsilon^{ij}
Q^J
\nonumber \\ + \bigg(\hn_j {\rm Re}({\cal{N}}_{IJ})\bigg)
\epsilon^{jk} d_h\Phi^J_k =0~,
\eea
and the $\nu=j$ component of ({\ref{geq}}) is equivalent to
\bea
\label{geq2}
{\rm Im}({\cal{N}}_{IJ}) d_h\Phi^J_j
= - d_h\big({\rm Im}({\cal{N}}_{IJ}) Q^J\big)_k \epsilon^{k}{}_{j}
- \bigg(\hn_k {\rm Re}({\cal{N}}_{IJ})\bigg) \Phi^J \epsilon^{k}{}_{j}~.
\eea

The scalar field equation\footnote{We shall use $\partial_\alpha = {\partial \over \partial z^\alpha}$, and $\partial_{\bar{\alpha}}= {\partial \over \partial {\bar{z}}^{\bar{\alpha}}}$ to denote differentiation w.r.t. the scalars
$z^\alpha$, ${\bar{z}}^{\bar{\alpha}}$.} ({\ref{scaleq}}) can be expressed as

\bea
\label{scaleq1}
&&\hn_i \hn^i z^\alpha - h^i \hn_i z^\alpha
+ g^{\alpha {\bar{\gamma}}} \partial_\lambda  g_{\sigma {\bar{\gamma}}}
\hn_i z^\lambda \hn^i z^\sigma -g^{\alpha {\bar{\gamma}}} \partial_{\bar{\gamma}} V
+{1 \over 2}\bigg(Q^I \Phi^J + Q^J \Phi^I\bigg) g^{\alpha {\bar{\gamma}}} \partial_{\bar{\gamma}}
{\rm Re}({\cal{N}}_{IJ})
\cr
&&~~~~
+{1 \over 2} \bigg(Q^I Q^J - \Phi^I \Phi^J\bigg)
g^{\alpha {\bar{\gamma}}} \partial_{\bar{\gamma}}
{\rm Im}({\cal{N}}_{IJ})
 =0~,
\eea
where the K\"ahler connection of the scalar manifold involving partial derivatives of $g_{\alpha\bar\beta}$ has been given explicitly.

The $+-$ component of the Einstein equations ({\ref{eineq}})
is
\bea
\label{ein1}
{1 \over 2} \hn^i h_i - \Delta -{1 \over 2} h^2
-{1 \over 2} {\rm Im}({\cal{N}}_{IJ})\bigg(\Phi^I \Phi^J
+Q^I Q^J\bigg) -V=0~,
\eea
while  $++$ component of the Einstein equations is
\bea
\label{ein2}
 \hn^i \hn_i \Delta - 3 h^i \hn_i \Delta
- \Delta \hn^i h_i + 2\Delta h^2 +{1 \over 2} (dh)_{ij}
(dh)^{ij}
+ 2{\rm Im}({\cal{N}}_{IJ}) \delta^{ij} d_h\Phi^I_i
d_h\Phi^J_j =0.
\eea
Next the $+i$ component of the Einstein equations is
\bea
\label{ein3}
{1 \over 2} \hn^j (dh)_{ij}
-(dh)_{ij} h^j - \hn_i \Delta + \Delta h_i
-{\rm Im}({\cal{N}}_{IJ}) \bigg(\Phi^I d_h\Phi^I_i - Q^I \epsilon_i{}^j d_h\Phi^I_j \bigg)
=0~,
\eea
and finally $ij$ component of the Einstein equations is
\bea
\label{ein4}
&&{1 \over 2} {\hat{R}}\delta_{ij} + \hn_{(i} h_{j)}
-{1 \over 2} h_i h_j
+{1 \over 2} {\rm Im}({\cal{N}}_{IJ})\bigg(\Phi^I \Phi^J
+Q^I Q^J\bigg) \delta_{ij}
\cr
&&~~~~~~~~~~~~~
-2 g_{\alpha {\bar{\beta}}} \hn_{(i} z^\alpha
\hn_{j)} {\bar{z}}^{{\bar{\beta}}} - V \delta_{ij}=0~,
\nonumber \\
\eea
where $\hat R$ is the Ricci scalar of the spatial horizon section ${\cal S}$.

Not all of these field equations are independent. In particular,  ({\ref{geq1}}) is obtained by taking the divergence
of ({\ref{geq2}}). ({\ref{ein2}}) is obtained from taking the divergence of
({\ref{ein3}}), together with ({\ref{geq2}}) and ({\ref{geq1}}).
 Equation ({\ref{ein3}}) is obtained by taking the divergence of
the traceless part of ({\ref{ein4}}), together with
({\ref{ein1}}), ({\ref{scaleq1}}) and ({\ref{geq2}}).
So the independent  bosonic field equations are
({\ref{geq2}}), ({\ref{scaleq1}}), ({\ref{ein1}})
and ({\ref{ein4}}).

\newsection{Supersymmetric Near-Horizon Geometries}

\subsection{Killing spinor equations}

The KSEs of supergravity theories are the vanishing conditions of the supersymmetry variations of the fermionic fields of these theories evaluated at the
locus where all the fermionic fields vanish. The fermionic fields of  4-dimensional  ${\cal N}=2$,  gauged supergravity coupled to $U(1)$ multiplets
are the gravitino and the gaugini.  In particular, the gravitino KSE is
\bea
\label{grav}
\nabla_\mu \epsilon +\bigg({i \over 2}A_\mu \Gamma_5
+ig \xi_I (A^I)_\mu +g \Gamma_\mu
\xi_I \bigg( {\rm Im} X^I+i \Gamma_5 {\rm Re} X^I\bigg)
\nonumber \\
+{i \over 4} \Gamma^{\rho \sigma} \bigg(
{\rm Im} \big( (F^-)^I_{\rho \sigma} X^J \big)
-i \Gamma_5 {\rm Re} \big( (F^-)^I_{\rho \sigma} X^J \big)\bigg)
{\rm Im} {\cal{N}}_{IJ} \Gamma_\mu \bigg) \epsilon =0~,
\eea
and the  gaugini KSEs are
\bea
\label{gaugino}
{i \over 2}  {\rm Im} {\cal{N}}_{IJ}
\Gamma^{\rho \sigma} \bigg({\rm Im} \big((F^-)^J_{\rho \sigma}
{\cal{D}}_{\bar{\beta}} {\bar{X}}^I g^{\alpha {\bar{\beta}}}\big)
-i \Gamma_5 {\rm Re} \big((F^-)^J_{\rho \sigma}
{\cal{D}}_{\bar{\beta}} {\bar{X}}^I g^{\alpha {\bar{\beta}}}\big)\bigg)
\epsilon
\nonumber \\
+ \Gamma^\mu \nabla_\mu \bigg({\rm Re} z^\alpha -i \Gamma_5 {\rm Im} z^\alpha \bigg) \epsilon
+2g  \xi_I \bigg( {\rm Im} \big({\cal{D}}_{\bar{\beta}}
{\bar{X}}^I g^{\alpha {\bar{\beta}}} \big) -i \Gamma_5
{\rm Re} \big({\cal{D}}_{\bar{\beta}}
{\bar{X}}^I g^{\alpha {\bar{\beta}}} \big)\bigg) \epsilon =0~,
\eea
where $\epsilon$ is the supersymmetry parameter that is taken to be  Dirac commuting spinor,
\bea
\nabla_\mu \epsilon=
\partial_\mu\epsilon +{1 \over 4} \Omega_{\mu, \rho \sigma}\Gamma^{\rho \sigma}  \epsilon~,~~~
A_\mu = -{i \over 2}\bigg(\partial_\alpha K \nabla_\mu z^\alpha
-\partial_{\bar{\alpha}} K \nabla_\mu z^{\bar{\alpha}} \bigg)~,
\eea
and $\Omega$ is the frame connection of the spacetime metric.
The gravitino KSE is a parallel transport equation for the spinor $\epsilon$, while the gaugini KSEs do not involve derivatives of $\epsilon$ and so are algebraic.
Our spinor conventions including those for the gamma matrices $\Gamma^\mu$ as well as  the realization of $\mathrm{Cliff}(3,1)$ used are specified in Appendix A. Observe that the KSEs is linear over the complex numbers. So the supersymmetric configurations always admit an even number of supersymmetries as counted over the real numbers. The classification
of supersymmetric solutions of gauged ${\cal N}=2$ supergravity coupled to any number of vector multiplets has been investigated in \cite{klemm1, klemm2, kzor}.

\subsection{Integrability along the lightcone  and independent KSEs}

For the near horizon geometries that we are investigating, the KSEs of the 4-dimensional supergravity theory can be explicitly integrated
along the lightcone directions. This determines the dependence of the Killing spinors in terms of the $u,r$ coordinates. Then we substitute back the  resulting expressions
for the Killing spinors into the KSEs to find  remaining conditions on the Killing spinors. The remaining conditions include those that one
expects by the naive restriction of both  the gravitino and gaugini KSEs  on the spatial horizon section ${\cal S}$ as well as large number of integrability conditions.

To determine all the conditions on the Killing spinors, we first solve the $\mu=-$ component of the gravitino KSE ({\ref{grav}}) to find
\bea
\epsilon_+ = \phi_+~,~~~ \qquad \epsilon_-
= \phi_- + r \Gamma_- \Theta_+ \phi_+~,
\eea
where $\partial_r \phi_\pm=0$, $\Gamma_\pm\epsilon_\pm=\Gamma_\pm\phi_\pm=0$, and we have defined
\bea
\Theta_\pm &=& {1 \over 4} h_i \Gamma^i -g \xi_I \big({\rm Im}X^I +i \Gamma_5 {\rm Re} X^I \big)
\nonumber \\
&\mp &{i \over 2} \bigg({\rm Im}((\Phi^I+i Q^I)X^J)+i \Gamma_5
{\rm Re}((\Phi^I+i Q^I)X^J) \bigg){\rm Im} {\cal{N}}_{IJ}~.
\eea

Next,we solve the $\mu=+$ component of the gravitino KSE ({\ref{grav}}) to find that
\bea
\phi_+ = \eta_+ + u \Gamma_+ \Theta_- \eta_-,
\qquad \phi_- = \eta_-~,
\label{phipm}
\eea
where $\partial_r \eta_\pm = \partial_u \eta_\pm =0$, $\Gamma_\pm\eta_\pm=0$,  and so $\eta_\pm$ depend only on the coordinates of ${\cal S}$.
Thus after solving the gravitino KSE along the lightcone directions the Killing spinor can be written as
\bea
\epsilon=\epsilon_++ \epsilon_-~,~~~\epsilon_+=\eta_+ + u \Gamma_+ \Theta_- \eta_-~,~~~\epsilon_-= \eta_- + r \Gamma_- \Theta_+\big(\eta_+ + u \Gamma_+ \Theta_- \eta_-\big)~.
\label{kspinors}
\eea
Substituting $\epsilon$ back into all the KSEs, one obtains a large number of conditions (\ref{alg1})-(\ref{alg7}) described in appendix C.

Not all these conditions are independent.  Using in an essential way that the horizons preserve at least one supersymmetry\footnote{Such an
assumption is not necessary for the proof of a similar result in 10- and 11-dimensional supergravities \cite{11index, iibindex, iiaindex, iiamindex} but it has been used before
in the context of minimal 5-dimensional supergravity \cite{5index}.}, and in particular the  relations
between the fields (\ref{kappabos}), (\ref{kappader}), (\ref{alg5b}), (\ref{alg5c}),  (\ref{alg1a}),  (\ref{alg1b}),  (\ref{alg3a}) and (\ref{alg4a})  that are implied by such an assumption, and after utilizing  the field equations, one finds that the remaining independent conditions  implied by the gravitino KSE  on the Killing spinors are
\bea
\nabla^{(\pm)}_i \eta_\pm=0~,
\label{ptx}
\eea
where
\bea
\nabla^{(\pm)}_i\equiv \hn_i+{i \over 2} A_i \Gamma_5
+ig \xi_I B^I_i - \Gamma_i \hat\Theta_\mp \mp{1 \over 4}  h_i~,
\label{lichd1x}
\eea
and
\bea
\Theta_\mp={1\over4} \Gamma^i h_i+\hat\Theta_\mp~,~~~ A_i = -{i \over 2}\big(\partial_\alpha K \partial_i z^\alpha
-\partial_{\bar{\alpha}} K \partial_i z^{\bar{\alpha}} \big)~.
\eea
Similarly, the gaugini KSEs ({\ref{gaugino}}) give
\bea
{\cal A}^\alpha_{(\pm)}\eta_\pm=0~,
\label{lichd2x}
\eea
where
\bea
\label{algx}
{\cal A}^\alpha_{(\pm)}=\mp i  {\rm Im} {\cal{N}}_{IJ}
 \bigg[{\rm Im} \big((\Phi^J+iQ^J){\cal{D}}_{\bar{\beta}} {\bar{X}}^I g^{\alpha {\bar{\beta}}}\big)
 -i \Gamma_5 {\rm Re} \big((\Phi^J+iQ^J){\cal{D}}_{\bar{\beta}} {\bar{X}}^I g^{\alpha {\bar{\beta}}}\big) \bigg]
 \nonumber \\
 +\Gamma^i \hn_i \bigg[{\rm Re}z^\alpha -i \Gamma_5 {\rm Im}z^\alpha \bigg] +2g  \xi_I \bigg[
 {\rm Im} \big({\cal{D}}_{\bar{\beta}} {\bar{X}}^I g^{\alpha {\bar{\beta}}}\big)-i \Gamma_5 {\rm Re} \big({\cal{D}}_{\bar{\beta}} {\bar{X}}^I g^{\alpha {\bar{\beta}}}\big)\bigg]~.
 \eea
The KSEs (\ref{ptx}) and (\ref{algx})  can be thought of as the naive restriction
of the gravitino and gaugini KSEs on the spatial horizon section ${\cal S}$.

Furthermore, one also establishes from the analysis of the integrability conditions that if $\eta_-$ satisfies the above KSEs, then
\bea
\eta_+=\Gamma_+\Theta_-\eta_-~,
\label{epen}
\eea
also is a Killing spinor. To see whether $\eta_+\not=0$, one has to show that $\mathrm {Ker}\, \Theta_-=\{0\}$ which is demonstrated below.

\subsection{ $\mathrm {Ker}\, \Theta_-=\{0\}$ }

To show this, we shall use contradiction.  Suppose that there is
exists $\eta_- \neq 0$ such that $\Theta_- \eta_-=0$.
It follows that
\bea
\Theta_+ \eta_- &=& \big(\Theta_+ - \Theta_-\big) \eta_-
 \nonumber \\
 &=& {\rm Im}{\cal{N}}_{IJ} \bigg(\Gamma_5 {\rm Re} \big((\Phi^I+i Q^I) X^J\big)
 -i {\rm Im} \big((\Phi^I+i Q^I) X^J\big) \bigg) \eta_-~.
 \eea
 It then follows from ({\ref{ptx}}) that
 \bea
 \label{nrm1}
 \hn_i \parallel \eta_- \parallel^2 = - h_i \parallel \eta_- \parallel^2~,
 \eea
and so  $dh=0$ as $\eta_-$ is a parallel spinor and so is nowhere vanishing. The integrability condition ({\ref{alg2}}) further implies that
 $\Delta=0$. On taking the divergence of ({\ref{nrm1}}), one then
 obtains
 \bea
 \label{lapl1}
 \hn^i \hn_i \parallel \eta_- \parallel^2 =
 \bigg(-{\rm Im} {\cal{N}}_{IJ} \big(\Phi^I \Phi^J + Q^I Q^J\big)
 -2V \bigg) \parallel \eta_- \parallel^2~,
 \eea
 As we have assumed that ${\rm Im} {\cal{N}}_{IJ}$ is negative definite, and also $V \leq 0$, an application of the maximum principle reveals
 that
\bea
\Phi^I=Q^I=0~, \qquad V=0~,
\eea
and also $\parallel \eta_- \parallel^2=\mathrm{const}$. Substituting the
constant norm condition into ({\ref{nrm1}}), we obtain $h=0$.

Substituting all of these conditions back into the condition
$\Theta_- \eta_-=0$, one obtains
\bea
\xi_I \bigg( {\rm Im} X^I +i \Gamma_5 {\rm Re} X^I\bigg) \eta_-=0~,
\eea
which implies $\xi_I X^I=0$ which contradicts our assumptions on the couplings. Thus, we  establish that $\mathrm {Ker}\, \Theta_-=\{0\}$.

One consequence of the above result is that for all horizons $\phi_+\not=0$.  To see this, since our backgrounds are supersymmetric either $\eta_+$ or $\eta_-$ must not vanish.
If $\eta_+\not=0$, then $\phi_+\not=0$.  On the other hand if $\eta_-\not=0$, then also $\phi_+\not=0$ as can be seen from
(\ref{phipm}) and $\mathrm {Ker}\, \Theta_-=\{0\}$. In particular, this means
that all supersymmetric near-horizon geometries must admit
a non-zero spinor $\phi_+$ satisfying ({\ref{pt1}}), ({\ref{alg1}}),
({\ref{alg3}}), ({\ref{alg4}}), ({\ref{alg5}}) and ({\ref{alg7}}).

\newsection{Counting the supersymmetries of horizons}

In this section, we shall demonstrate the first consequence of the horizons conjecture which is the
 counting of supersymmetries of ${\cal N}=2$ supergravity horizons as  stated in the introduction. For this,
we shall establish two Lichnerowicz type theorems and then we shall use index theory to count the number of supersymmetries preserved by the near horizon
geometries.

\subsection{Lichnerowicz type Theorem for $\phi_+$}

The Killing spinor equations on $\eta_+$ have been reduced to the naive restriction
of the gravitino and gaugini KSEs on ${\cal S}$ (\ref{ptx}) and (\ref{algx}), respectively.
Let us define the horizon Dirac operators
\bea
\label{lichd3x}
\cD^{(\pm)} \equiv \Gamma^i {\hat{\nabla}}^{(\pm)}_i=\Gamma^i \hat\nabla_i+{i\over 2} \Gamma^i A_i \Gamma_5+ i g\Gamma^i  \xi_I B^I_i-2 \hat\Theta_\mp \mp {1\over4} \Gamma^i h_i \ .
\eea
Here we shall establish  the Lichnerowicz type theorem
\bea
{\hat{\nabla}}^{(+)}_i \phi_+ =0 \qquad {\rm and} \qquad {\cal A}_{(+)}^\alpha \phi_+ =0
\qquad \Longleftrightarrow \qquad  \cD^{(+)} \phi_+=0~.
\eea
The proof of the Lichnerowicz type theorem for $\eta_+$ spinor is similar.
It is clear that if $\phi_+$ is Killing, then it is a zero mode of the $\cD^{(+)}$ and so
one direction is straightforward. To prove the converse, we shall assume that the near horizon
geometries preserve one supersymmetry\footnote{This assumption is not necessary
 for 11-dimensional and type II horizons in \cite{11index, iibindex, iiaindex, iiamindex} but this assumption has been used
 before for the 5-dimensional horizons  \cite{5index}.}and that the maximum principle applies.

The assumption of the existence of one supersymmetry requires some explanation. We have shown in appendix C that the fields of the near horizon geometries  that preserve one supersymmetry satisfy certain at most first order differential conditions which depend on the choice of the Killing spinor via a function $\kappa$.
These conditions are necessary to establish the Lichnerowicz type theorems. However although
the  conditions  (\ref{kappabos}), (\ref{kappader}), (\ref{alg5b}), (\ref{alg5c}), (\ref{alg1a}), (\ref{alg1b}), (\ref{alg7a}), (\ref{alg3a}) and (\ref{alg4a})  are used,  $\kappa$ is not required to be related to the spinor under
investigation in the Lichnerowicz type theorem. In other words, we use the at most first order differential conditions on the fields that are derived from the requirement of one supersymmetry  but the
Lichnerowicz type theorems are valid for every zero mode of the horizon Dirac operators irrespectively
on whether this zero mode  is associated to the Killing spinor used to establish the differential relations.

To proceed
one can show utilizing ({\ref{alg5b}}) that the gaugini algebraic condition can be rewritten as
\bea
{\cal A}^\alpha_{(+)}\phi_+=0~,
\eea
where now
\bea
\label{lichd2xx}
{\cal A}_{(+)}^\alpha &=&\Gamma^i \hn_i {\rm Re} z^\alpha +i \Gamma_5 \Gamma^i \hn_i {\rm Im} z^\alpha
+2 g(1-\kappa \Gamma_5) \bigg(\xi_I {\rm Im} \big({\cal{D}}_{\bar{\beta}} {\bar{X}}^I g^{\alpha {\bar{\beta}}}\big)
\cr
&&~~~~~~~~
-i \Gamma_5 \xi_I {\rm Re} \big({\cal{D}}_{\bar{\beta}} {\bar{X}}^I g^{\alpha {\bar{\beta}}}\big) \bigg)~.
\eea
Next assume that $\phi_+$ is a zero mode of the horizon Dirac operator, $\cD^{(+)} \phi_+=0$, then after some computation which is described in appendix D, one can show that
\bea
\hn_i \hn^i \parallel \phi_+ \parallel^2 -h^i \hn_i \parallel \phi_+ \parallel^2
&=& 2 \langle {\hat{\nabla}}^{(+)i}  \phi_+, {\hat{\nabla}}^{(+)}_i \phi_+ \rangle
\nonumber \\
&+&\langle {\cal A}_{(+)}^\beta \phi_+, \big( {\rm Re}(g_{\alpha \bar{\beta}}) +i \Gamma_5 {\rm Im} (g_{\alpha \bar{\beta}}) \big) {\cal A}_{(+)}^\alpha \phi_+ \rangle \ .
\eea
The right-hand-side of this expression is a sum of  positive definite terms.
The maximum principle then implies that $\eta_+$ is a Killing spinor and that
\bea
\partial_i\parallel\phi_+\parallel =0~.
\eea

We conclude by stating the  Lichnerowicz type theorem
for $\eta_+$ spinors. In particular, we have that
\bea
{\hat{\nabla}}^{(+)}_i \eta_+ =0 \qquad {\rm and} \qquad {\cal A}_{(+)}^\alpha \eta_+ =0~,
\qquad \Longleftrightarrow \qquad  \cD^{(+)} \eta_+=0~,
\label{plichn}
\eea
and
\bea
\parallel \eta_+\parallel=\mathrm{const}~,
\label{etaconst}
\eea
where ${\hat{\nabla}}^{(+)}$, ${\cal A}_{(+)}^\alpha$ and $ \cD^{(+)}$ are defined by
({\ref{lichd1x}}), ({\ref{lichd2xx}}) and ({\ref{lichd3x}}), respectively.

\subsection{Lichnerowicz type Theorem for $\eta_-$ spinors}

There is an analogous Lichnerowicz type theorem  for $\eta_-$ spinors. In particular,
one can show that
\bea
{\hat{\nabla}}^{(-)}_i \eta_- =0 \qquad {\rm and} \qquad {\cal A}_{(-)}^\alpha \eta_- =0
\qquad \Longleftrightarrow \qquad  \cD^{(-)} \eta_-=0~,
\label{nlichn}
\eea
where the operator  $\hat{\nabla}^{(-)}$ is defined in (\ref{lichd1x}), ${\cal A}_{(-)}$ is given
in (\ref{algx}) and upon using ({\ref{alg5b}}) can be expressed as
\bea
\label{lichd5x}
{\cal A}_{(-)}^\alpha &\equiv& \Gamma^i \hn_i {\rm Re} z^\alpha +i \Gamma_5 \Gamma^i \hn_i {\rm Im} z^\alpha
\cr &&~~~~~
+2g(1+\kappa \Gamma_5) \bigg( \xi_I {\rm Im} \big({\cal{D}}_{\bar{\beta}} {\bar{X}}^I g^{\alpha {\bar{\beta}}}\big)
-i  \Gamma_5 {\rm Re} \big({\cal{D}}_{\bar{\beta}} {\bar{X}}^I g^{\alpha {\bar{\beta}}}\big) \bigg)~,
\eea
and the horizon Dirac operator is
\bea
\label{lichd6x}
\cD^{(-)} \equiv \Gamma^i {\hat{\nabla}}^{(-)}_i \ .
\eea
We have again assumed that the near horizon
geometries preserve one supersymmetry and we shall use this in a way that has been explained for
$\phi_+$ spinors in the previous section.
It is clear that if $\eta_-$ is a Killing spinor, then it is also a zero mode of the
horizon Dirac operator. To establish the converse, take $\eta_-$ to be a zero mode of the
horizon Dirac operator $\cD^{(-)}$, $\cD^{(-)}\eta_-=0$, and after some computation that is described
in appendix D, one can establish the identity
\bea
&&\hn_i \hn^i \parallel \eta_- \parallel^2 + \hn^i \bigg( h_i  \parallel \eta_- \parallel^2 \bigg)
= 2 \langle {\hat{\nabla}}^{(-)i}  \eta_-, {\hat{\nabla}}^{(-)}_i \eta_- \rangle
\cr&&~~~~~~~~~~~~
+\langle {\cal A}_{(-)}^\beta \eta_-, \big( {\rm Re}(g_{\alpha \bar{\beta}}) +i \Gamma_5 {\rm Im} (g_{\alpha \bar{\beta}}) \big) {\cal A}_{(-)}^\alpha \eta_- \rangle \ .
\eea
The right-hand-side of this expression is a sum of  positive definite terms.
On integrating both sides of this expression over ${\cal{S}}$, which is taken to be compact without boundary, the contribution
from the  left-hand-side vanishes. So the integral of the right-hand-side vanishes as well.  As it is the sum of positive terms, this implies that $\eta_-$ is Killing spinor as required.

\subsection{Counting Supersymmetries}
The number of supersymmetries of near horizon geometries is $N=N_++N_-$ where $N_\pm$ is
the number of linearly independent $\eta_\pm$ Killing spinors. On the other hand,
the two  Lichnerowicz type theorems (\ref{plichn}) and (\ref{nlichn}) we have established for both the $\eta_\pm$  spinor imply that
\bea
N_\pm=\mathrm {dim~Ker}~\cD^{(\pm)}~.
\label{npm}
\eea
Moreover  one can easily show that
\bea
\Gamma_- (\cD^{(+)})^\dagger = \cD^{(-)} \Gamma_-~,
\eea
which implies that
\bea
{\rm dim} \big({\rm Ker} (\cD^{(+) \dagger}) \big) = {\rm dim} \big({\rm Ker} (\cD^{(-)}) \big)~.
\eea
On the other hand \cite{atiyah1}
\bea
\mathrm{Index~} \big(\cD^{(+)}\big)\equiv \mathrm {dim~Ker}~\cD^{(+)}- \mathrm {dim~Ker}~(\cD^{(+})^\dagger=N_+-N_-~.
\eea
Therefore, the number of supersymmetries preserved by the near horizon geometries
can be expressed as
\bea
N
= {\rm index}\big(\cD^{(+)}\big) + 2 N_-~.
\eea

It remains to calculate  ${\rm index}\,(\cD^{(+)})$.  For this observe that from (\ref{lichd3x}) and using the conventions in appendix A that one can write
\bea
\cD^{(+)}&=&\Gamma^i \hat\nabla_i+{i\over 2} \Gamma^i A_i \Gamma_5+ i g\Gamma^i  \xi_I B^I_i-2 \hat\Theta_- - {1\over4} \Gamma^i h_i
\cr&=&\sigma^i\otimes\sigma_3\hat\nabla_i-{i\over 2} \sigma^i \sigma_3\otimes \sigma_3 A_i + i g   \sigma^i\otimes \sigma_3 \xi_I B^I_i
-2 \hat\Theta_- - {1\over4} \sigma^i \otimes \sigma_3 h_i
\cr
&=&\sigma^i\hat\nabla_i-{i\over 2} \sigma^i \sigma_3 A_i + i g   \sigma^i \xi_I B^I_i
-2 \hat\Theta_- - {1\over4} \sigma^i  h_i
~.
\eea
where in the last equality we have used  $\Gamma_+\eta_+=0$, or equivalently $\bI_2\otimes \sigma_3\eta_+=\eta_+$, and identified $\sigma^i\otimes 1=\sigma^i$
and $\sigma^i\sigma_3\otimes 1=\sigma^i\sigma_3$.
Using the chirality operator $\sigma_3$ on ${\cal S}$ the above operator further decomposes into two other operators
as
\bea
\cD^{(+)}=\cD^{(+)}_+\oplus\cD^{(+)}_-~,
\eea
where
 \bea
 \cD^{(+)}_\pm=\sigma^i\hat\nabla_i\mp{i\over 2} \sigma^i  A_i + i g   \sigma^i \xi_I B^I_i
-2 \hat\Theta_- - {1\over4} \sigma^i  h_i~.
\eea
To continue observe that
\bea
&&\cD^{(+)}_+:~~\Gamma(S_+\otimes {\cal K}^{{1\over2}}\otimes {\cal L})\rightarrow \Gamma(S_-\otimes {\cal K}^{{1\over2}}\otimes {\cal L})~,
\cr
&&\cD^{(+)}_-:~~\Gamma(S_-\otimes \bar{{\cal K}}^{{1\over2}}\otimes {\cal L})\rightarrow \Gamma(S_+\otimes \bar{{\cal K}}^{{1\over2}}\otimes {\cal L})~,
\eea
where $S_\pm$ are the bundles of chiral/antichiral spinors on ${\cal S}$, respectively,  ${\cal K}$ is the pull-back of the Hodge bundle on ${\cal S}$,
${\cal L}$ is the line bundle with connection $\xi_I B^I$ and $\Gamma(E)$ denotes the smooth sections of the vector bundle $E$.

The index of $\cD^{(+)}$ can be calculated as follows.
\bea
{\rm index}\big(\cD^{(+)}\big)&=&{\rm index}\big(\cD^{(+)}_+\big)+{\rm index}\big(\cD^{(+)}_-\big)={\rm index}\big(\cD^{(+)}_+\big)-{\rm index}\big((\cD^{(+)}_-)^\dagger\big)
\cr
&=&
\big({1\over2} c_1({\cal K})+c_1 ({\cal L})\big)-\big({1\over2} c_1(\bar{{\cal K}})+c_1({\cal L})\big)=c_1({\cal K})~,
\eea
where we have used that $\cD^{(+)}_+$ and $(\cD^{(+)}_-)^\dagger$ have the same principal symbol as that of twisted Dirac operators with the bundles
${\cal K}^{{1\over2}}\otimes {\cal L}$ and $\bar{{\cal K}}^{{1\over2}}\otimes {\cal L}$, respectively, and so the same index.

Therefore, we have found that
\bea
N=2 c_1({\cal K})+ 2 N_-=2 c_1({\cal K})+ 4\ell~,
\eea
as $N_-$ is an even number because the $\cD^{(-)}$ is linear over the complex numbers.
The additional factor of $2$ in front of $c_1({\cal K})$ appears because the index is computed over the complex numbers while our
counting of supersymmetries is over the real numbers.

In many cases of interest $c_1({\cal K})$ vanishes. In particular, we shall see that if $N_-\not=0$, or equivalently $\ell\not=0$, then $c_1({\cal K})=0$.  This is because
the pull-back of the Hodge bundle on ${\cal S}$ in all these cases is trivial.  This will be proven after a detailed analysis of the
geometries of the horizons in section 6. Conversely, if $c_1({\cal K})=0$
then $N=4 \ell$, so all supersymmetric solutions with $c_1({\cal K})=0$
must have $\ell \neq 0$.

\newsection{$\mathfrak{sl}(2, \bR)$ symmetry}

We shall demonstrate   that
all supersymmetric horizons with $N_-\not=0$, or equivalently $c_1({\cal K})=0$ in (\ref{nf}),  of ${\cal N}=2$ gauged supergravity exhibit an $\mathfrak{sl}(2, \bR)$ symmetry which is the second part of the horizons conjecture as  stated in the introduction. To prove this, first observe that if $\epsilon_1$ and $\epsilon_2$ are Killing spinors, then
the 1-form bilinear
\bea
K(\epsilon_1, \epsilon_2)=\mathrm{Re}\langle (\Gamma_+-\Gamma_-)\epsilon_1, \Gamma_\mu \epsilon_2\rangle\, e^\mu
\eea
is associated with a Killing vector which in addition leaves all other fields invariant, see \cite{klemm1, klemm2, kzor}
and also appendix E.  The former property is a consequence of the gravitino KSE.
Suppose now that $N_-\not=0$. We have also shown that if $\eta_-$ is Killing spinor, then $\eta_+=\Gamma_+\Theta_-\eta_-$ is also a Killing spinor (\ref{epen}). Using these, we can
construct  two linearly independent Killing spinors over the whole spacetime associated with the pairs $(\eta_-, 0)$ and $(\eta_-, \eta_+)$ which
after a rearrangement can be written as
\bea
\epsilon_1=\eta_-+ u \eta_++ r u \Gamma_-\Theta_+\eta_+~,~~~\epsilon_2= \eta_++ r \Gamma_- \Theta_+\eta_+~;~~~\eta_+=\Gamma_+\Theta_-\eta_-~.
\eea
These give rise to three 1-form bi-linears as
\begin{eqnarray}
 K_1=\mathrm{Re}\,\langle(\Gamma_+-\Gamma_-)\epsilon_1, \Gamma_\mu \epsilon_2\rangle \, e^\mu&=& (2r\, \mathrm{Re}\,\langle\Gamma_+\eta_-, \Theta_+\eta_+\rangle+ 4  u r^2  \parallel\Theta_+\eta_+\parallel^2) \,{\bf{e}}^+
 \nonumber \\
 &-&2u \parallel\eta_+\parallel^2\, {\bf{e}}^-  + W_i {\bf{e}}^i~,
 \nonumber \\
 K_2=\mathrm{Re}\,\langle(\Gamma_+-\Gamma_-)\epsilon_2, \Gamma_\mu \epsilon_2\rangle \, e^\mu &=&4 r^2  \parallel\Theta_+\eta_+\parallel^2 {\bf{e}}^+-2 \parallel\eta_+\parallel^2 {\bf{e}}^-~,
\nonumber \\
K_3=\mathrm{Re}\,\langle(\Gamma_+-\Gamma_-)\epsilon_1, \Gamma_\mu \epsilon_1\rangle \, e^\mu&=&(2\parallel\eta_-\parallel^2+4r u \mathrm{Re}\,\langle\Gamma_+\eta_-, \Theta_+\eta_+\rangle
 + 4 r^2 u^2  \parallel\Theta_+\eta_+\parallel^2) {\bf{e}}^+
\nonumber \\
 &-&2u^2 \parallel\eta_+\parallel^2 {\bf{e}}^- +2u W_i {\bf{e}}^i~,
 \nonumber \\
 \label{b1forms}
 \end{eqnarray}
where to simplify the expressions for $K_1, K_2$ and $K_3$ somewhat we have used the fact that $\parallel\eta_+\parallel$ is constant (\ref{etaconst}), and have set
\begin{eqnarray}
\label{extraiso}
W_i = \mathrm{Re}\, \langle \Gamma_+ \eta_- , \Gamma_i \eta_+ \rangle~ \ ,
\end{eqnarray}
and also  have used
\bea
\mathrm{Re}\langle\eta_+, \Gamma_i \Theta_+\eta_+\rangle=\mathrm{Re}\langle\eta_+, \Gamma_i \Theta_-\eta_+\rangle=0~,
\eea
which follows from a direct computation utilizing   the expressions for $\Theta_\pm$.

Furthermore, the requirement that all the above three 1-forms give rise to Killing vector fields implies the conditions, see also appendix E,
\begin{eqnarray}
&&\hat\nabla_{(i} W_{j)}=0~,~~~\hat{\cal L}_Wh=0 ~,~~~\hat{\cal L}_W\Delta=0~,~~~4\parallel \Theta_+\eta_+\parallel^2= \Delta \parallel\eta_+\parallel^2~,
\cr
&&-2 \parallel\eta_+\parallel^2-h_i W^i+2 \mathrm{Re}\,\langle\Gamma_+\eta_-, \Theta_+\eta_+\rangle=0~,~~~i_W (dh)+2 d \mathrm{Re}\,\langle\Gamma_+\eta_-, \Theta_+\eta_+\rangle=0~,
\cr
&& 2 \mathrm{Re}\, \langle\Gamma_+\eta_-, \Theta_+\eta_+\rangle-\Delta \parallel\eta_-\parallel^2=0~,~~~
W+ \parallel\eta_-\parallel^2 h+d \parallel\eta_-\parallel^2=0~.
\label{concon}
\end{eqnarray}
Using the above expressions, observe that $K_1, K_2$ and $K_3$ can be simplified further and also one can show  that
\begin{eqnarray}
{\cal L}_W\parallel\eta_-\parallel^2=0~.
\label{inphim}
\end{eqnarray}
In addition to the Killing vectors associated with $K_1$, $K_2$ and $K_3$, the geometry of spacetime
is further restricted  by the KSEs and field equations of the theory. An exhaustive description of the geometry of
the horizons will be given in the next section.

To demonstrate that the horizons exhibit an  $\mathfrak{sl}(2,\mathbb{R})$ symmetry,   we use the various identities derived above
 (\ref{concon}) to write the vector fields associated to the 1-forms $K_1, K_2$ and $K_3$ (\ref{b1forms}) as
\begin{eqnarray}
K_1&=&-2u \parallel\eta_+\parallel^2 \partial_u+ 2r \parallel\eta_+\parallel^2 \partial_r+ W^i \hat \partial_i~,
\cr
K_2&=&-2 \parallel\eta_+\parallel^2 \partial_u~,
\cr
K_3&=&-2u^2 \parallel\eta_+\parallel^2 \partial_u +(2 \parallel\eta_-\parallel^2+ 4ru \parallel\eta_+\parallel^2)\partial_r+ 2u W^i \hat\partial_i~,
\end{eqnarray}
where we have used the same symbol for the 1-forms and the associated vector fields. A direct computation then reveals using (\ref{inphim}) that
\begin{eqnarray}
[K_1,K_2]=2 \parallel\eta_+\parallel^2 K_2~,~~~[K_2, K_3]=-4 \parallel\eta_+\parallel^2 K_1  ~,~~~[K_3,K_1]=2 \parallel\eta_+\parallel^2 K_3~. \ \ \ \
\end{eqnarray}
Therefore all such  horizons with non-trivial fluxes admit an $\mathfrak{sl}(2,\mathbb{R})$ symmetry subalgebra.

The orbits of the $\mathfrak{sl}(2,\mathbb{R})$ symmetry are either two or three dimensional depending on whether $W$ vanishes or not. In the
former case, the spacetime is a warped product of $\mathrm{AdS}_2$ with ${\cal S}$.

\newsection{Geometry of the Near-Horizon Solutions}

In this section, we shall summarize the
local forms of all near-horizon geometries
of ${\cal N}=2$ gauged supergravity with $c_1({\cal K})=0$, which implies that $N_- \neq 0$. In fact, as a consequence of the following analysis, it can also
be easily seen that the converse  holds, i.e. $N_- \neq 0$ implies
that $c_1({\cal K})=0$. This is because if $N_- \neq 0$ then
 the scalars locally depend on at most one coordinate  or they are constant.  As a result the first Chern class of the pull-back of the
Hodge bundle on ${\cal S}$ vanishes. Hence $c_1({\cal K})=0$ if, and only if, $N_- \neq 0$.
All such near-horizon geometries preserve  either 4 or 8 supersymmetries.
As $c_1({\cal K})=0$ implies that $N_- = N_+$, the global argument
given previously implies that the KSEs (\ref{ptx}) and (\ref{lichd2x})
admit the same number of $\eta_-$ and $\eta_+$ spinor solutions.

The function
\bea
 \kappa=\parallel \phi_+\parallel^{-2} \langle\phi_+, \Gamma_5\phi_+\rangle~,
 \eea
plays a particularly important role in the analysis,
because the  metric and other fields depend on it, see also appendices C and G.  Observe that $|\kappa|\leq 1$ as a consequence of the Cauchy-Schwarz inequality and
 $\kappa=\pm1$ iff $\phi_+$ is an eigenspinor of $\Gamma_5$ with eigenvalue $\pm1$.

 To examine the geometry of near horizon backgrounds, we are mostly concerned with solving the conditions (\ref{kappabos}), (\ref{kappader}), (\ref{alg5b}), (\ref{alg5c}),  (\ref{alg1a}),  (\ref{alg1b}),  (\ref{alg3a}) and (\ref{alg4a})  on the fields which arise from the KSEs on $\phi_+$ and for this we also make use of some of the Einstein equations (\ref{ein1})-(\ref{ein4})  according to need.
Note that
  the independent field equations are   (\ref{geq2}) for the vector fields, (\ref{scaleq1}) for the scalars,  and the Einstein equations (\ref{ein1}) and (\ref{ein4}). Moreover, we have verified that for all supersymmetric  near horizon backgrounds (\ref{geq2}) is automatically satisfied in appendix H. The scalar field equations
 (\ref{scaleq1}) have not at any point been used in the analysis of any
of the KSE. Furthermore, it has been shown in \cite{klemm1} that
the scalar field equations are implied by supersymmetry and the remaining
field equations.

In many of the cases we consider, it turns out that the scalar
fields $z^\alpha$ have a non-trivial dependence on the co-ordinate
$\kappa$. Such near-horizon solutions have also been considered in
the context of the entropy function formalism \cite{Astefanesei}, in which a $SL(2,\bR) \times U(1)$ symmetry was assumed, together with spherical topology of the horizon spatial
cross-section ${\cal{S}}$. 

The solution of the (\ref{kappabos}), (\ref{kappader}), (\ref{alg5b}), (\ref{alg5c}),  (\ref{alg1a}),  (\ref{alg1b}),  (\ref{alg3a}) and (\ref{alg4a}) equations  and field equations is arranged so that all the other fields are determined in terms of the scalar fields
 of the vector multiplets and $\kappa$.  These   in turn obey  non-linear
first order differential  equations.  In what follows, we shall not give details
of the proof. Instead, we shall simply state the results with some minimal explanation. A more detailed
derivation can be found in appendix G.

\subsection{Warped AdS$_2$ horizons; $W \equiv 0$}

It can be shown using the maximum principle that for all these backgrounds
\bea
\Delta>0~,~~~\xi_I \Phi^I=0~.
\eea
Furthermore, the associated vector field to
\bea
\tau= \star_{\cal{S}} h~,
\eea
leaves the field $z^\alpha, \Delta, h, \Phi^I, Q^I$  invariant. As a result, there are two
cases to consider depending on whether or not $h=0$.

\subsubsection{Solutions with $W=h \equiv 0$}

The conditions from supersymmetry and the field equations imply that the fields $z^\alpha$, $\kappa$,  $\Delta$, $\Phi^I$
and $Q^I$ are all constant, with $\Delta>0$.
The spacetime metric is then given by
\bea
ds^2 = 2 du (dr - {1 \over 2} r^2 \Delta du) + ds^2_{\cal{S}}~,
\eea
where the Ricci scalar of ${\cal{S}}$ is given by
\bea
{\hat{R}}= 2 \Delta +4V~,
\label{rdeltavx}
\eea
where
\bea
\Delta = -{1 \over 2} {\rm Im}{\cal{N}}_{IJ} \big(\Phi^I \Phi^J + Q^I Q^J\big) -V \ .
\eea
So the spacetime is $AdS_2 \times {\cal{S}}$ where ${\cal{S}}$ is $T^2$, $S^2$ or $H^2$
according to whether $2 \Delta+4V>0$, $2 \Delta+4V=0$ or $2 \Delta+4V<0$, respectively.

The constant fields $z^\alpha$, $\kappa$,  $\Delta$, $\Phi^I$
and $Q^I$ are not arbitrary.  In particular, as $h=0$, (\ref{kappabos}) implies that $\kappa^2=1$. Also, (\ref{alg5b}) must be imposed, which relates the electric and magnetic parts of the
$U(1)$ fields in terms of the scalars.

\subsubsection{Solutions with $W \equiv 0$ and $h \not= 0$.}
Such solutions are warped products $\mathrm{AdS}_2\times_w {\cal S}$.
Adapting
suitable coordinates along $h$ and $\tau$ and after a further coordinate transformation, one finds that
the near-horizon data  is then given by
\bea
&&\Delta = 4 \nu^2 e^x, \qquad h=dx, \qquad
\cr
&&ds^2_{\cal{S}}={1 \over 16g^2 |\xi_I X^I|^2 (1-\kappa^2)} dx^2 + 16g^2 L^2 (1-\kappa^2) e^{-x} d \psi^2~,
\label{solwzero}
\eea
together with
\bea
\Phi^I+iQ^I = -4i \nu {|\xi_J X^J| \over \xi_T {\bar{X}}^T} e^{{x \over 2}} {\bar{X}}^I -2ig \kappa \xi_J {\rm Im}{\cal{N}}^{-1IJ}~,
\label{wzeropq}
\eea
where $\nu, L$ are a real constants.
The scalars $z^\alpha$ and $\kappa$ depend only on $x$ and satisfy
\bea
{dz^\alpha \over dx} = {1 \over 2 \xi_J {\bar{X}}^J} \xi_I {\cal{D}}_{\bar{\beta}} {\bar{X}}^I
g^{\alpha {\bar{\beta}}}, \qquad {d \kappa \over dx} = \kappa -{\nu \over 2g |\xi_I X^I|} e^{x \over 2}~.
\label{wzerozk}
\eea
On setting $r = e^{-x} \rho$, the spacetime metric is
\bea
ds^2 = 2 e^{-x} du \big(d \rho -2 \nu^2 \rho^2 du)+{1 \over 16g^2 |\xi_I X^I|^2 (1-\kappa^2)} dx^2 + 16g^2 L^2 (1-\kappa^2) e^{-x} d \psi^2~,
\eea
 which is a warped product $\mathrm{AdS}_2\times_w {\cal S}$ with warp factor $e^{-x}$.

 In this case, we have solved all the (\ref{kappabos}), (\ref{kappader}), (\ref{alg5b}), (\ref{alg5c}),  (\ref{alg1a}),  (\ref{alg1b}),  (\ref{alg3a}) and (\ref{alg4a}) equations.

\subsection{Solutions with $W \not \equiv 0$}

The spacetime metric as well as all the other fields are invariant under the action
of $W\not=0$. $W$ also leaves invariant the metric on ${\cal S}$ as well as
the other near horizon data $h$,
$z^\alpha$, $\Phi^I$, $Q^I$, and $\Delta$.
Furthermore, the Lie derivatives of $\kappa$, and $\parallel \eta_- \parallel^2$ with respect to $W$ also vanish. We present the proof of these in Appendix G. There are several cases to consider
and we summarize the local form of the fields below.

\subsubsection{Solutions with $W \not \equiv 0$ and $\kappa=\mathrm{const}$}

For all these solutions $h\not=0$, $dh=0$,  and
\bea
\label{simpcharges1x}
\Phi^I+iQ^I =-2ig \kappa \bigg( \xi_J {\rm Im}{\cal{N}}^{-1IJ}
+4 \xi_J X^J {\bar{X}}^I \bigg)~,
\eea
and
\bea
\Delta = 16 g^2 \kappa^2 |\xi_I X^I|^2~,
\label{deltaxiX}
\eea
see appendix G. Furthermore because of (\ref{topt}), ${\hat{R}}=(1+\kappa^2)  (1-\kappa^2)^{-1} \hn^i h_i$,  the Euler number of ${\cal S}$ vanishes and so ${\cal S}$
is a topological 2-torus.
There are two different subcases to consider, corresponding as
to whether $\parallel \eta_- \parallel^2$ is constant
or not.
\vskip 0.3cm
\underline{{\bf $\parallel \eta_- \parallel^2$ constant}}
\vskip 0.2cm
If $\parallel \eta_- \parallel^2$ is constant, then one finds that
({\ref{auxd1}}) implies that
\bea
\parallel \eta_- \parallel^2 h +W=0~,~~~\hat R=0~,~~~|\xi_I X^I|^2=\mathrm{const.}
\eea
Thus  $h$ is covariantly constant and ${\cal S}$ is a torus. Then   one can introduce local co-ordinates
$x$, $y$ on ${\cal{S}}$ such that
\bea
h=dx, \qquad \star_{\cal{S}} h =dy~,
\eea
so that the $z^\alpha$, $\Phi^I$ and $Q^I$ depend only on $y$. In these co-ordinates, the metric is
\bea
ds^2=2 du \bigg(dr +r dx -8g^2 \kappa^2 |\xi_I X^I|^2 r^2 du \bigg)
+{1 \over 16 g^2 (1-\kappa^2) |\xi_I X^I|^2}\bigg(dx^2+dy^2\bigg)~,
\eea
and the scalars $z^\alpha$ satisfy
\bea
\label{stopb}
{d z^\alpha \over dy} =  {i \over 2 \xi_J {\bar{X}}^J} \xi_I \cD_{\bar{\beta}} {\bar{X}}^I g^{\alpha {\bar{\beta}}}~.
\eea
The $\Phi^I$ and $Q^I$ are given by ({\ref{simpcharges1x}}) and $\Delta$ is constant given by (\ref{deltaxiX}); the scalars also must satisfy
\bea
g^{\alpha {\bar{\beta}}}
\xi_I \cD_\alpha X^I \xi_J \cD_{\bar{\beta}} {\bar{X}}^J
= |\xi_I X^I|^2, \quad {\rm and} \quad  {\rm Im}{\cal{N}}^{-1 IJ} \xi_I \xi_J
= -4 |\xi_I X^I|^2~.
\eea

\newpage

\vskip 0.3cm
\underline{{\bf $\parallel \eta_- \parallel^2$ non-constant}}
\vskip 0.2cm

For this class of solutions $i_W h$ is a negative constant.
So we set
\bea
i_Wh=-\mu^2~,
\eea
and introduce coordinates $x,\psi$ on ${\cal S}$ as
\bea
W = \mu^2 {\partial \over \partial \psi},  \qquad \mu^2 x= \parallel \eta_- \parallel^2~.
\eea
Then after some extensive analysis which utilizes the maximum principle and is presented
in appendix G, one can  show that
\bea
\Delta=\kappa=0~, \qquad \Phi^I=Q^I=0~,\qquad h = -d \psi~,
\eea
\bea
ds^2_{\cal{S}} = {1 \over x} \bigg( \big(x d \psi-dx\big)^2 +{1 \over 16 g^2 |\xi_I X^I|^2 x -1}
dx^2 \bigg)~,
\eea
and
\bea
{dz^\alpha \over dx} = -{i \over 2x} \bigg({1 \over \sqrt{16g^2|\xi_L X^L|^2-1}}-i \bigg)
{1 \over \xi_J {\bar{X}}^J} \xi_I \cD_{\bar{\beta}} {\bar{X}}^I g^{\alpha \bar{\beta}}~.
\label{wconstz}
\eea
It follows that the spacetime metric is given by
\bea
ds^2 = 2du(dr-rd \psi) + {1 \over x} \bigg( \big(x d \psi-dx\big)^2 +{1 \over 16 g^2 |\xi_I X^I|^2 x -1}
dx^2 \bigg)~,
\label{wconstm}
\eea
which concludes the analysis.

\subsubsection{Solutions with $W \not \equiv 0$ and $\kappa \neq \mathrm{const}.$}

The local coordinates in this case are chosen to be
\bea
\kappa~,~~~\phi~;~~~W={\partial\over\partial\psi}~.
\eea
The relation between $\phi$ and $\psi$ can be found in appendix G.
Furthermore, one sets
\bea
A+iB =\kappa \xi_I {\bar{X}}^I \cG~,~~~\cG = -{2ig \over 1-iY}~;~~~Y\not=0, -i~,
\eea
and after some extensive analysis which has been presented in appendix G, one finds that
\bea
\label{argsolx}
{\bar{Y} \over Y} = {\kappa+ic \over \kappa-ic}~,
\eea
where $c$ is a real constant, and
\bea
\label{compderv3x}
{d \cG \over d \kappa}
={1 \over 2\kappa (\kappa+ic)} \bigg({\kappa \cG +ig(\kappa+ic) \over {1 \over 2} \cG +ig} \bigg)
\bigg({ig(\kappa-ic) \cG \over \kappa \cG+ig (\kappa+ic)} (1-{i \over g}\cG)
\nonumber \\
-{1 \over |\xi_L X^L|^2} \cG \xi_I \cD_\alpha X^I \xi_J \cD_{\bar{\beta}} {\bar{X}}^J g^{\alpha \bar{\beta}} \bigg)~.
\eea
There are two cases to investigate depending on whether $\kappa \cG +2ig(\kappa+ic)$ vanishes
or not.

\vskip 0.3cm
\underline{
$\kappa \cG +2ig(\kappa+ic)\not=0$}
\vskip 0.2cm

In this case, after some computation which is explained in appendix G, one finds that
\bea
\label{nhd1x}
\Delta = {16g^2 \kappa^2 |\xi_I X^I|^2 \over |1-iY|^2}~,
\eea
\bea
\label{nhd2x}
{d z^\alpha \over d \kappa} = {1 \over 2 \kappa \xi_J {\bar{X}}^J}(1+iY^{-1}) \xi_I \cD_{\bar{\beta}} {\bar{X}}^I g^{\alpha \bar{\beta}}~,
\eea
\bea
\label{nhd4x}
h=\kappa^{-1} \bigg(1-{c \over (\kappa+ic)Y}\bigg)d \kappa
-(1-\kappa^2)d \phi~,
\eea
and
\bea
\label{nhd5x}
\Phi^I+iQ^I = -{8ig \kappa \over 1+i {\bar{Y}}}
\xi_J X^J {\bar{X}}^I
-2ig \kappa {\rm Im}{\cal{N}}^{-1 IJ} \xi_J~,
\eea
where $\psi={p \over 16g^2}\phi$ and $p$ is an integration constant which appears at an intermediate step.
Moreover, the spacetime metric is
\bea
\label{nhd3x}
ds^2 &=&2 du \bigg(dr + r \big[\kappa^{-1} \bigg(1-{c \over (\kappa+ic)Y}\bigg)d \kappa
-(1-\kappa^2)d \phi\big] -r^2 {8g^2 \kappa^2 |\xi_I X^I|^2 \over |1-iY|^2}du\bigg)
\cr &&~~~~+  \Delta^{-1} \bigg({1 \over |Y|^2 (1-\kappa^2)}
d \kappa^2 +(\kappa^2+c^2)(1-\kappa^2)d \phi^2 \bigg)~.
\eea
From these data after solving the first order
non-linear differential equations, one can construct  explicit solutions for each of the theories.

\vskip 0.3cm
\underline{
$\kappa \cG +2ig(\kappa+ic)=0$}
\vskip 0.2cm

This special case
corresponds to taking $c \neq 0$, with
\bea
Y ={c \over \kappa+ic}~.
\eea
Furthermore, $\parallel \eta_- \parallel^2=\mathrm{const}$, $|\xi_I X^I|^2=\mathrm{const}$, and
({\ref{compderv3}}) implies that
\bea
\xi_I \cD_\alpha X^I \xi_J \cD_{\bar{\beta}} {\bar{X}}^J g^{\alpha \bar{\beta}} = |\xi_I X^I|^2~.
\eea
The remainder of the near-horizon data is given by
({\ref{nhd1x}})-({\ref{nhd5x}}) for this choice of $Y$ with $c \neq 0$.

In all the four cases above, we have solved all the (\ref{kappabos}), (\ref{kappader}), (\ref{alg5b}), (\ref{alg5c}),  (\ref{alg1a}),  (\ref{alg1b}),  (\ref{alg3a}) and (\ref{alg4a}) equations.

\newsection{Degenerate Marginally Trapped Surfaces}

The definition of what is a  black hole spacetime is a long standing problem in general relativity, see \cite{booth1}
for a review. In particular it is desirable to have a quasi-local definition of what is a black hole horizon. An investigation
of this question for extreme black holes has revealed that the degenerate  Killing horizons that occur in extreme black holes
exhibit a marginally trapped surface which after a suitable deformation becomes untrapped both inside and outside the horizon
\cite{booth2, mars, lucietti1, lucietti2}. From the perspective of the Killing horizons, one then turn these conditions
into criteria for a near horizon geometry to extend to a full black hole spacetime.  In particular, these conditions can be stated as
 follows \cite{lucietti1,lucietti2}. Given
the 1-form $h$ on ${\cal{S}}$, there is a unique positive function
$\Gamma$, and a divergence-free 1-form $h'$ such that
\bea
\label{decomp1}
h= \Gamma^{-1} h' - d \log \Gamma~.
\eea
For ${\cal S}$ to be a marginally trapped surface, it is required that
\bea
\int_{\cal S} \Gamma \gamma^{(1)}>0~,
\eea
where $\gamma^{(1)}$ is a function associated with the deformation of the metric of ${\cal S}$.
Then the condition to have untrapped surfaces both inside and outside the horizon is that
the integral
\bea
\int_{\cal S}  \gamma^{(1)} (F'-(h')^2)<0~,
\label{necess1}
\eea
where $F'=-\Gamma^2 \Delta$.

For the supersymmetric horizons of ${\cal N}=2$ supergravity we are considering , as well as
the horizons of 11-dimensional and type II supergravities with fluxes \cite{11index, iibindex, heterotic, iiaindex, iiamindex}, which satisfy the criteria of the
second part of the horizon conjecture and have a marginally trapped surface
using  ({\ref{auxd1}}), one finds that
\bea
\label{pmatch}
h'=-W, \qquad \Gamma = \parallel \eta_- \parallel^2~,
\eea
which in turn gives
\bea
\label{pos2}
\int_{\cal{S}} \Gamma \gamma^{(1)} = \int_{\cal{S}} \parallel \eta_- \parallel^2 \gamma^{(1)} >0~.
\eea
Moreover,
\bea
F'= - \Gamma^2 \Delta = - \parallel \eta_- \parallel^4 \Delta~.
\eea
For all such supersymmetric near-horizon solutions, the conditions
({\ref{concon}}) imply that
\bea
W^2=- \parallel \eta_- \parallel^2 h^i W_i
\eea
and
\bea
\Delta \parallel \eta_- \parallel^2 - h^i W_i = 2 \parallel \eta_+ \parallel^2 \ ,
\eea
and hence
\bea
F'-(h')^2 = - 2 \parallel \eta_- \parallel^2 \parallel \eta_+ \parallel^2 \ .
\eea
So one obtains
\bea
\int_{\cal{S}}  \gamma^{(1)} \big(F'-(h')^2 \big)
=-2 \parallel \eta_+ \parallel^2 \int_{\cal{S}} \parallel \eta_- \parallel^2 \gamma^{(1)} <0
\eea
as a consequence of ({\ref{pos2}}), where we have made use of
the condition $\parallel \eta_+ \parallel^2 = \mathrm{const}.$

Hence, ({\ref{necess1}}) holds automatically for all supersymmetric
near horizon geometries with fluxes and $N_-\not=0$ satisfying ({\ref{pos2}}). Therefore assuming the validity of  the horizon conjecture,
 we have shown the following: {\it All supersymmetric
horizons with fluxes and $N_-\not=0$  for which the spatial horizon section is a marginally trapped surface contain
untrapped surfaces both just inside and outside the horizon.}

\newsection{Concluding remarks}

We have confirmed the validity of the {\it horizon conjecture} for all near  horizon geometries of ${\cal N}=2$, $D=4$, gauged supergravity coupled
to any number of vector multiplets under some mild restrictions on the couplings. As a result, we have provided a formula which counts the number of superymmetries of all such backgrounds (\ref{nf})
as well as demonstrated that those with $N_-\not=0$, or equivalently $c_1({\cal K})=0$, in (\ref{nf})  exhibit a $\mathfrak{sl}(2,\bR)$ symmetry. We have also provided an exhaustive local
description of supersymmetric  near horizon geometries.

The horizon conjecture has been confirmed for a large number of theories.  It demonstrates that the emergence of conformal symmetry near the horizon
of supersymmetric black holes
is a consequence of the fluxes of supergravity theories and the smoothness of the horizons.  Therefore it is a generic property of these theories and
it does not depend on the details of the black hole solution under consideration.

Apart from this,  we have  demonstrated another application of the horizon conjecture.  In particular, we have shown that the horizon conjecture implies that
 all those horizons for which the horizon section is a marginally trapped surface have untrapped surfaces both just inside and outside the horizon.
 As a result, it is possible that they may be extended to full extreme black hole solutions. As it is likely that the horizon conjecture holds for all supergravity theories, perhaps
 under some mild restrictions on the couplings, the above result holds for all such supersymmetric near horizon geometries. As the first obstruction to extend
 the near horizon geometries to full black hole solutions can be removed, it indicates that many of the supersymmetric horizons could be extended to full black hole solutions.
 However not all criteria for this are known and so the question of which of the near horizon geometries are extendable and which are not remains an open question.

 Other aspects of our results are the plethora of new Lichnerowicz type  theorems that have been demonstrated, and  the extensive applications that
 the maximum principle has in the context of horizons. The former results can be adapted to the theory of Clifford bundles and so they can used for applications to geometry.  The latter indicate that the maximum principle has a close relationship with supersymmetry.  Perhaps this is not too surprising as supersymmetry imposes restrictions
  on the couplings of various theories which are essential for the validity of the various maximum principle formulae. However the precise relation is
  not apparent and it would be of interest to investigate it in the future.

\vskip 0.5cm
\noindent{\bf Acknowledgements} \vskip 0.1cm
\noindent  JG is supported by the STFC grant, ST/1004874/1. 
TM is partially supported by the STFC consolidated grant ST/L000431/1. GP is partially supported by the  STFC consolidated grant ST/J002798/1.

\vskip 0.5cm

\vskip 0.5cm
\noindent{\bf Data Management} \vskip 0.1cm

\noindent No additional research data beyond the data presented and cited in this work are
needed to validate the research  findings in this work.

\vskip 0.5cm

\newpage

 \setcounter{section}{0}

 \appendix{Conventions}

 \setcounter{subsection}{0}

\subsection{Spin Connection and Curvature}

The non-vanishing components of the spin connection of the near horizon geometry (\ref{horgeom}) in
the frame basis ({\ref{nhbasis}}) are
\begin{eqnarray}
&&\Omega_{-,+i} = -{1 \over 2} h_i~,~~~
\Omega_{+,+-} = -r \Delta, \quad \Omega_{+,+i} ={1 \over 2} r^2(  \Delta h_i - \partial_i \Delta),
\cr
&&\Omega_{+,-i} = -{1 \over 2} h_i, \quad \Omega_{+,ij} = -{1 \over 2} r dh_{ij}~,~~~
\Omega_{i,+-} = {1 \over 2} h_i, \quad \Omega_{i,+j} = -{1 \over 2} r dh_{ij},
\cr
&&\Omega_{i,jk}= \hat\Omega_{i,jk}~,
\end{eqnarray}
where $\hat\Omega$ denotes the spin-connection of the spatial horizon cross section  ${{\cal{S}}}$ in with basis ${\bf{e}}^i$.
If $f$ is any function of spacetime, then frame derivatives are expressed in terms of co-ordinate derivatives  as
\begin{eqnarray}
\partial_+ f &=& \partial_u f +{1 \over 2} r^2 \Delta \partial_r f~,~~
\partial_- f = \partial_r f~,~~
\partial_i f = {\tilde{\partial}}_i f -r \partial_r f h_i \ .
\end{eqnarray}
The non-vanishing components of the Ricci tensor is the
 basis ({\ref{nhbasis}}) are
\bea
R_{+-} &=& {1 \over 2} \hn^i h_i - \Delta -{1 \over 2} h^2~,~~~
R_{ij} = {\hat{R}}_{ij} + \hn_{(i} h_{j)} -{1 \over 2} h_i h_j~,
\nonumber \\
R_{++} &=& r^2 \big( {1 \over 2} \hn^2 \Delta -{3 \over 2} h^i \hn_i \Delta -{1 \over 2} \Delta \hn^i h_i + \Delta h^2
+{1 \over 4} (dh)_{ij} (dh)^{ij} \big)~,
\nonumber \\
R_{+i} &=& r \big( {1 \over 2} \hn^j (dh)_{ij} - (dh)_{ij} h^j - \hn_i \Delta + \Delta h_i \big) \ ,
\eea
where ${\hat{R}}$ is the Ricci tensor of the horizon section ${\cal S}$ in the $\bbe^i$ frame.

\subsection{Spinor Conventions}

We first present a matrix representation of $\mathrm{Cliff}(3,1)$ adapted to the basis ({\ref{nhbasis}}).
The module of Dirac spinors has been identified with $\bC^4$ and we have set
\bea
\Gamma_i &=& \sigma_i\otimes \sigma_3= \begin{pmatrix}  \sigma^i \ \ \ \ \  0 \cr   \ \ 0 \ \ -\sigma^i \end{pmatrix}, \ i=1,2 \ ;~~~  \Gamma_0=i\bI_2\otimes \sigma_2~,~~~\Gamma_3=\bI_2\otimes\sigma_1~;  \cr
\Gamma_+ &=&\begin{pmatrix} \ \ 0  \ \ \ \sqrt{2}\, \bI_2 \cr  0 \ \ \ \ \ 0 \end{pmatrix}~, \qquad
\Gamma_- = \begin{pmatrix}  \ \ 0 \ \ \ \ \ 0 \cr \sqrt{2}\, \bI_2 \ \ \ \ 0 \end{pmatrix}~,
\eea
where $\sigma^i$, are the Hermitian Pauli matrices $\sigma^i \sigma^j = \delta^{ij} \bI_2 + i \epsilon^{ijk} \sigma^k$.
Note that
\bea
\Gamma_{+-} = \begin{pmatrix} \bI_2 \ \ \ \ \ 0 \cr \ \ \ 0 \ \ \ -\bI_2 \end{pmatrix}~,
\eea
and we define
\bea
\Gamma_5 = i \Gamma_{+-12}=-\sigma_3\otimes\sigma_3~.
\eea
It will be convenient to decompose the spinors into positive and negative chiralities
with respect to the lightcone directions as
\bea
\epsilon = \epsilon_+ + \epsilon_-~,
\eea
where
\bea
\Gamma_{+-} \epsilon_\pm = \pm \epsilon_\pm, \qquad {\rm or \ equivalently} \qquad \Gamma_\pm \epsilon_\pm =0~.
\eea
With these conventions, note that
\bea
\Gamma_{ij} \epsilon_\pm = \mp i  \epsilon_{ij} \Gamma_5 \epsilon_\pm \ .
\eea

The  inner product,  $\langle \cdot , \cdot \rangle$, we use is that for which spacelike gamma matrices
are Hermitian while time-like ones are anti-Hermitian. When restricted on $\mathrm{Spin}(2)$ is also
$\mathrm{Spin}(2)$-invariant. In particular,  note that $(\Gamma_{ij})^\dagger = - \Gamma_{ij}$.

 \appendix{Special K\"ahler geometry}

\subsection{Definition}
The matter couplings of the ${\cal N}=2$, $d=4$ supergravity are described by special K\"ahler geometry data. For this, we shall give a brief
summary of special K\"ahler geometry. For a review of the various approaches to special K\"ahler geometry,  see \cite{craps} and references within.

Let $M$ be a Hodge K\"ahler manifold\footnote{A K\"ahler manifold $M$ is Hodge,
if the cohomology class represented by the  K\"ahler form is the Chern class of a line bundle ${\cal K}$ on $M$. We have also denoted with ${\cal K}$  the pull back of the Hodge
 bundle over ${\cal S}$. Which line bundle ${\cal K}$ refers to is clear from the context.}, ${\cal K}$ be the Hodge complex line bundle over $M$ and $E$ be a flat $Sp(2(k+1), \bR)$ vector bundle $E$ with typical fibre $\bC^{2(k+1)}$ and compatible (symplectic) fibre inner product $\langle\cdot, \cdot\rangle$.

  Next, consider $E\otimes {\cal K}$ and introduce the connection on the sections $\nnu$ of $E\otimes {\cal K}$
\bea
{\cal D}_{\bar\alpha} \nnu&=&D_{\bar\alpha}\nnu-{1\over2} \partial_{\bar\alpha} K \nnu~,~~~~
\cr
{\cal D}_{\alpha} \nnu&=&D_{\alpha}\nnu+{1\over2} \partial_{\alpha} K \nnu~,~~~~
\eea
where  $D$ is the flat connection of $E$, $\partial_\alpha=\partial/\partial z^\alpha$ and $z^\alpha$ are homomorphic coordinates
of the K\"ahler manifold. Observe that the curvature of ${\cal D}$ is proportional to the K\"ahler form of $M$.

{\bf Definition:} $M$ is a  special K\"ahler manifold  provided that  $E\otimes {\cal K}$ admits a section $\nnu$ such that it satisfied the following conditions
\bea
{\cal D}_{\bar\alpha} \nnu&=&0~,~~~~
\langle \nnu, \bar\nnu\rangle =i~,
\cr
\langle {\cal D}_\alpha \nnu, \nnu\rangle&=&0~,~~~~
\langle {\cal D}_\alpha \nnu, {\cal D}_\beta\nnu\rangle =0~.
\label{skg1}
\eea
\rightline{$\triangle$}

To investigate the consequences of the above definition, first perform a $GL(2(k+1), \bR)$ transformation to bring the symplectic
inner product $\langle\cdot, \cdot\rangle$ into canonical form\footnote{Of course one then can use a local gauge $Sp(2(k+1), \bR)$ transformation to set $D_\alpha=\partial_\alpha$ and $D_{\bar\alpha}=\partial_{\bar\alpha}$ as $D$ is flat. But this is not necessary in what follows.}.
Then the above conditions can be re-written as
\bea
{\cal D}_{\bar\alpha} X^I={\cal D}_{\bar\alpha} F_I&=&0~,~~~~
\cr
X^I \bar F_I-F_I \bar X^I&=&i~,
\cr
{\cal D}_\alpha F_I \,X^I- F_I\, {\cal D}_\alpha X^I&=&0~,
\cr
{\cal D}_\alpha F_I{\cal D}_\beta X^I-{\cal D}_\beta F_I {\cal D}_\alpha X^I&=&0~,
\label{skg2}
\eea
where the section $\nnu$ has been written in the canonical form as
\bea
\nnu=\begin{matrix}\left(\begin{array}{c}X^I\\ F_I\end{array}\right)\end{matrix}~.
\eea
Observe that the first condition in (\ref{skg2}) is a covariant holomorphicity condition while the last condition in (\ref{skg2}) is implied by the third condition.

Taking the covariant derivative of the second  condition in (\ref{skg2}), we find that
\bea
{\cal D}_\alpha X^I \bar F_I- {\cal D}_\alpha F_I \bar X^I=0~.
\label{conx1}
\eea
Next taking that ${\cal D}_{\bar\beta}$ covariant derivative of the above expression we find that
\bea
 g_{\alpha\bar\beta}\equiv \partial_\alpha\partial_{\bar\beta}K = i[ {\cal D}_\alpha X^I {\cal D}_{\bar\beta}\bar F_I- {\cal D}_\alpha F_I {\cal D}_{\bar\beta}\bar X^I]~.
\label{conx2}
\eea

The gauge couplings ${\cal N}$ are then defined as
\bea
F_I={\cal N}_{IJ} X^I~,~~~{\cal D}_{\bar\alpha} \bar F_I= {\cal N}_{IJ} {\cal D}_{\bar\alpha} \bar X^J~.
\label{gcouple}
\eea
The  conditions of special K\"ahler geometry together with the requirement that $M$ is a K\"ahler manifold imply that ${\cal N}$ is a symmetric matrix.
In terms of the gauge couplings, the second and third equations in (\ref{skg2}),
and (\ref{conx2}) can be written as
\bea
{\rm Im} {\cal{N}}_{IJ} X^I \bar X^J=-{1\over2}~,
\label{conx3}
\eea
\bea
{\rm Im} {\cal{N}}_{IJ} X^I {\cal D}_\alpha X^J=0~,
\label{conx4}
\eea
\bea
g_{\alpha\bar\beta}=-2{\rm Im} {\cal{N}}_{IJ}{\cal D}_\alpha X^I {\cal D}_{\bar\beta}\bar X^J~,
\label{conx5}
\eea
respectively. As the K\"ahler metric must be positive definite, it is required that ${\rm Im} {\cal{N}}$ is negative definite.
 The fourth equation in (\ref{skg2}) and (\ref{conx1}) are automatically implied as ${\cal N}$ is a symmetric matrix.

 Furthermore from the definition of ${\cal N}$, one can establish the identity
 \bea
U^{IJ} \equiv g^{\alpha \bar{\beta}}  {\cal{D}}_\alpha X^I
{\cal{D}}_{\bar{\beta}} {\bar{X}}^J = -{1 \over 2} \big({\rm Im} {\cal{N}}\big)^{-1 \ IJ} -  {\bar{X}}^I X^J~.
\label{UUU}
\eea
This identity is required in the definition of the scalar potential of the supergravity theory.

\subsection{Prepotential}

A special class of  solutions for the  conditions of special K\"ahler geometry (\ref{skg2})  can be  expressed in terms of a holomorphic prepotential as follows. It is well-known that the solutions of a covariant holomorphicity  condition
on sections of a vector bundle with respect to a connection which has  (1,1) curvature can be expressed in terms of the holomorphic sections of the associated holomorphic bundle. In this case, write
\bea
\nnu=e^{\frac{K}{2}}  \uu~,
\eea
and observe that
\bea
{\cal D}_\alpha \nnu= e^{\frac{K}{2}} {\cal D}_\alpha \uu~,~~~{\cal D}_{\bar\alpha} \nnu= e^{\frac{K}{2}} {\cal D}_{\bar\alpha} \uu~,
\eea
with
\bea
{\cal D}_\alpha \uu=D_\alpha \uu+\partial_\alpha K \uu~,~~~{\cal D}_{\bar\alpha} \uu=D_{\bar\alpha} \uu~.
\eea
It is clear from this that in the gauge ${\cal D}_{\bar\alpha}=\partial_{\bar\alpha}$, the covariant holomorphicity condition on $\nnu$
can be solved by setting
\bea
\nnu= e^\frac{K}{2}\begin{matrix}\left(\begin{array}{c}Z^I\\ \frac{\partial}{\partial Z^I} F\end{array}\right)\end{matrix}~,~~~\uu=\begin{matrix}\left(\begin{array}{c}Z^I\\ \frac{\partial}{\partial Z^I} F\end{array}\right)\end{matrix}~,
\eea
where $\uu$ is a holomorphic section, ie function only  of $z$,  and $F(Z)$ is the prepotential which is taken to be a homogeneous function of degree two in $Z$.
The use of the homogeneity  condition will become apparent later.

Let us now investigate the remaining conditions of the special K\"ahler geometry (\ref{skg1}) or (\ref{skg2}) in terms of $\uu$. The second condition
in (\ref{skg2}) can now be rewritten as
\bea
 e^{-K}=-i(Z^I \bar\partial_I\bar F-\bar Z^I \partial_I F)=-2 Z^I \bar Z^J \mathrm {Im} (\partial_I \partial_J F)~,
\eea
where we have used the homogeneity of the prepotential.
The remaining two conditions in (\ref{skg2}) are identically satisfied as a consequence of the homogeneity of $F$.  While (\ref{conx1}) and (\ref{conx2}) can now be written as
\bea
\mathrm{Im}(\partial_I \partial_J F ) {\cal D}_\alpha Z^I \bar Z^J=0~,~~~g_{\alpha\bar\beta}=- \partial_\alpha \partial_{\bar\beta} \log[Z^I \bar Z^J \mathrm {Im} (\partial_I \partial_J F)]~.
\eea

Furthermore, the identities involving the gauge couplings in terms of $\uu$ can now be written as follows.  First the definition of the gauge couplings becomes
\bea
\partial_I F={\cal N}_{IJ} Z^I~,~~~ \bar\partial_I\bar\partial_J\bar F{\cal D}_{\bar\alpha} \bar Z^J= {\cal N}_{IJ} {\cal D}_{\bar\alpha} \bar Z^J~.
\label{gcouplez}
\eea
Then the remaining identities can be expressed as
\bea
e^{-K}=-2 {\rm Im} {\cal{N}}_{IJ} Z^I \bar Z^J~,
\label{kpot}
\eea
\bea
{\rm Im} {\cal{N}}_{IJ} Z^I {\cal D}_\alpha Z^J=0~,
\label{conx4z}
\eea
\bea
g_{\alpha\bar\beta}=-2 e^K{\rm Im} {\cal{N}}_{IJ}{\cal D}_\alpha Z^I {\cal D}_{\bar\beta}\bar Z^J~.
\label{conx5z}
\eea
Furthermore, $U^{IJ}$ in (\ref{UUU}) can be easily written in terms of $Z$.
This concludes the description of the geometry.

\appendix{Independent KSEs}
 \setcounter{subsection}{0}

\subsection{KSEs and integrability conditions on ${\cal S}$}

Substituting the Killing spinor $\epsilon$ (\ref{kspinors})  back into all the KSEs, one obtains from the gravitino KSE along the lightcone directions  the integrability conditions
\bea
\label{alg1}
\bigg({1 \over 2} \Delta +{i \over 8}dh_{ij} \epsilon^{ij} \Gamma_5 -ig \xi_I \Phi^I
- \Gamma_+ \Theta_- \Gamma_- \Theta_+ \bigg) \phi_+ =0~,
\eea
and
\bea
\label{alg2}
&&\bigg(\Gamma_- \Theta_+ \Gamma_+ \Theta_-
-{1 \over 2} \Delta -{i \over 8} dh_{ij}\epsilon^{ij}
\Gamma_5 -ig \xi_I \Phi^I
\cr && ~~~~~~~
-i  {\rm Im} {\cal{N}}_{IJ} \Gamma^i {\rm Im} \big( \big(d_h\Phi^I_i
-i \epsilon_i{}^j d_h\Phi^I_j \big) X^J \big)
\bigg) \eta_-=0~,
\eea
and
\bea
\label{alg3}
&&\bigg({1 \over 4} \Gamma^i \big(\Delta h_i - \hn_i \Delta\big)
+{i \over 8} dh_{ij} \epsilon^{ij} \Gamma_5 \Theta_+ -ig \xi_I \Phi^I \Theta_+
\cr &&~~~~~~~~~~
+ i  {\rm Im} {\cal{N}}_{IJ} \Gamma^i {\rm Im} \big( \big(d_h\Phi^I_i
-i \epsilon_i{}^j d_h\Phi^I_j \big) X^J \big) \Theta_+
\bigg) \phi_+ =0~,
\eea
where  $\phi_+$ is defined in  (\ref{phipm}).

We remark that the conditions ({\ref{alg2}}) and ({\ref{alg3}}) are
obtained by making use of the following identity:
\bea
\label{auxid1}
{\rm Im}{\cal{N}}_{IJ} \Gamma^i \bigg({\rm Im}
\big((\hn_i \Phi^I - h_i \Phi^I-i \epsilon_i{}^j
(\hn_j \Phi^I - h_j \Phi^I))X^J\big)
\nonumber \\
\pm i \Gamma_5 {\rm Re}
\big((\hn_i \Phi^I - h_i \Phi^I-i \epsilon_i{}^j
(\hn_j \Phi^I - h_j \Phi^I))X^J\big)\bigg) \xi_\pm =0.
\eea

Furthermore, substituting $\epsilon$ given in (\ref{kspinors}) into the $\mu=i$ component of the gravitino KSE ({\ref{grav}}) gives
two parallel transport equations
\bea
\label{pt1}
\hn_i \phi_+ + \bigg({i \over 2} A_i \Gamma_5
+ig \xi_I B^I_i - \Gamma_i \Theta_- -{i \over 4} \epsilon_{ij} h^j
\Gamma_5 \bigg) \phi_+ =0~,
\eea
and
\bea
\label{pt2}
\hn_i \eta_- + \bigg({i \over 2} A_i \Gamma_5
+ig \xi_I B^I_i +{1 \over 2}h_i - \Gamma_i \Theta_+ +{i \over 4} \epsilon_{ij} h^j
\Gamma_5 \bigg) \eta_- =0~, \ \
\eea
together with an algebraic integrability  condition
\bea
\label{alg4}
\bigg(-\hn_i \Theta_- +{1 \over 2} \hn_{(i}h_{j)}\Gamma^j
-2g \xi_I \hn_i \big({\rm Im} X^I +i \Gamma_5 {\rm Re}X^I\big)
-{i \over 4} \epsilon_{ij} h^j \Theta_+ \Gamma_5
\nonumber \\
-{i \over 2} A_i \big(
\Gamma_5 \Theta_+ + \Theta_+ \Gamma_5\big)
-2g \Theta_+ \Gamma_i
\xi_I  \big({\rm Im} X^I +i \Gamma_5 {\rm Re}X^I\big)
-{3 \over 4} h_i \Theta_+ +{1 \over 4} \Gamma_i h_j \Gamma^j \Theta_+
\nonumber \\
+{i \over 2}  {\rm Im} {\cal{N}}_{IJ}\bigg[
{\rm Im} \big((d_h\Phi^I_i-i \epsilon_i{}^j
d_h\Phi^I_j)X^J\big)
\nonumber \\
+i \Gamma_5 {\rm Re} \big((d_h\Phi^I_i-i \epsilon_i{}^j
d_h\Phi^I_j)X^J\big) \bigg] \bigg) \phi_+=0~.
\eea

Next we consider the gaugini KSEs ({\ref{gaugino}}). Substituting the spinor $\epsilon$ (\ref{kspinors}) again, we obtain
\bea
\label{alg5}
\bigg(-i  {\rm Im} {\cal{N}}_{IJ}
 \bigg[{\rm Im} \big((\Phi^J+iQ^J){\cal{D}}_{\bar{\beta}} {\bar{X}}^I g^{\alpha {\bar{\beta}}}\big)
 -i \Gamma_5 {\rm Re} \big((\Phi^J+iQ^J){\cal{D}}_{\bar{\beta}} {\bar{X}}^I g^{\alpha {\bar{\beta}}}\big) \bigg]
 \nonumber \\
 +\Gamma^i \hn_i \bigg[{\rm Re}z^\alpha -i \Gamma_5 {\rm Im}z^\alpha \bigg] +2g  \xi_I \bigg[
 {\rm Im} \big({\cal{D}}_{\bar{\beta}} {\bar{X}}^I g^{\alpha {\bar{\beta}}}\big)-i \Gamma_5 {\rm Re} \big({\cal{D}}_{\bar{\beta}} {\bar{X}}^I g^{\alpha {\bar{\beta}}}\big)\bigg] \bigg) \phi_+=0~,
 \eea
and
\bea
\label{alg6}
\bigg(i  {\rm Im} {\cal{N}}_{IJ}
 \bigg[{\rm Im} \big((\Phi^J+iQ^J){\cal{D}}_{\bar{\beta}} {\bar{X}}^I g^{\alpha {\bar{\beta}}}\big)
 -i \Gamma_5 {\rm Re} \big((\Phi^J+iQ^J){\cal{D}}_{\bar{\beta}} {\bar{X}}^I g^{\alpha {\bar{\beta}}}\big) \bigg]
 \nonumber \\
 +\Gamma^i \hn_i \bigg[{\rm Re}z^\alpha -i \Gamma_5 {\rm Im}z^\alpha \bigg] +2g  \xi_I \bigg[
 {\rm Im} \big({\cal{D}}_{\bar{\beta}} {\bar{X}}^I g^{\alpha {\bar{\beta}}}\big)-i \Gamma_5 {\rm Re} \big({\cal{D}}_{\bar{\beta}} {\bar{X}}^I g^{\alpha {\bar{\beta}}}\big)\bigg] \bigg) \eta_-=0~,
 \eea
 and
 \bea
 \label{alg7}
 \bigg(i  {\rm Im} {\cal{N}}_{IJ}
 \bigg[{\rm Im} \big((\Phi^J+iQ^J){\cal{D}}_{\bar{\beta}} {\bar{X}}^I g^{\alpha {\bar{\beta}}}\big)
 +i \Gamma_5 {\rm Re} \big((\Phi^J+iQ^J){\cal{D}}_{\bar{\beta}} {\bar{X}}^I g^{\alpha {\bar{\beta}}}\big) \bigg]\Theta_+
 \nonumber \\
  -\Gamma^i \hn_i \bigg[{\rm Re}z^\alpha +i \Gamma_5 {\rm Im}z^\alpha \bigg]\Theta_+ +2g  \xi_I \bigg[
 {\rm Im} \big({\cal{D}}_{\bar{\beta}} {\bar{X}}^I g^{\alpha {\bar{\beta}}}\big)+i \Gamma_5 {\rm Re} \big({\cal{D}}_{\bar{\beta}} {\bar{X}}^I g^{\alpha {\bar{\beta}}}\big)\bigg]\Theta_+
 \nonumber \\
 +i   {\rm Im} {\cal{N}}_{IJ} \Gamma^i
  {\rm Im} \bigg( \big(d_h\Phi^J_i
 -i \epsilon_i{}^j d_h\Phi^J_j\big){\cal{D}}_{\bar{\beta}} {\bar{X}}^I g^{\alpha {\bar{\beta}}} \bigg) \bigg) \phi_+=0~.
 \eea
The KSEs (\ref{pt1}), (\ref{pt2}), (\ref{alg5}) and (\ref{alg6}) on $\eta_\pm$ can be thought of as the naive reduction
of the gravitino and gaugini KSEs on the spatial horizon section ${\cal S}$.  The remaining conditions should be thought of
as integrability conditions.  Typically, the integrability conditions are not independent. Rather they are implied
by (\ref{pt1}), (\ref{pt2}), (\ref{alg5}) and (\ref{alg6}) on $\eta_\pm$ and the field equations.

\subsection{ Conditions on $\parallel \phi_+ \parallel$}

Having established that $\phi_+$ cannot vanish identically as a consequence of $\mathrm{Ker}\, \Theta_-=\{0\}$ and the assumption that the solutions are supersymmetric,
we consider further the conditions on $\phi_+$.
In particular, we shall establish, via a maximum principle
argument, that $\parallel \phi_+ \parallel^2$ does not depend
on the co-ordinates of ${\cal{S}}$.

To proceed, note that ({\ref{pt1}}) implies that
\bea
\label{normder1}
\hn_i \parallel \phi_+ \parallel^2 =
{1 \over 2} h_i \parallel \phi_+ \parallel^2
+ \langle \phi_+, -2g \Gamma_i \xi_I \big( {\rm Im} X^I +i \Gamma_5
{\rm Re} X^I \big) \phi_+ \rangle~,
\eea
and hence it follows that
\bea
\label{lapl1a}
\hn_i \hn^i \parallel \phi_+ \parallel^2 &=&
{1 \over 2} \hn^i h_i \parallel \phi_+ \parallel^2 +{1 \over 2} h^i \hn_i \parallel \phi_+ \parallel^2
\nonumber \\
&+& \langle \phi_+, -g h^i \Gamma_i \xi_I \big( {\rm Im}X^I
+i \Gamma_5 {\rm Re} X^I \big) \phi_+ \rangle
\nonumber \\
&+& \langle \phi_+ , -4g \xi_I {\rm Im}X^I \big({\hat{\Theta}}_-^\dagger + {\hat{\Theta}}_- \big) \phi_+ \rangle
\nonumber \\
&+& \langle \phi_+ , -4ig \Gamma_5 \xi_I {\rm Re}X^I
\big({\hat{\Theta}}_-^\dagger - {\hat{\Theta}}_- \big) \phi_+ \rangle
\nonumber \\
&+& \langle \phi_+ , -2g \Gamma^i \hn_i \big[ \xi_I \big({\rm Im}X^I +i \Gamma_5 {\rm Re} X^I \big) \big] \phi_+ \rangle
\nonumber \\
&+& {\rm Re}\bigg(\langle \phi_+, -2i\Gamma^i A_i {\hat{\Theta}}_+
\Gamma_5 \phi_+ \rangle \bigg)~,
\eea
where
\bea
{\hat{\Theta}}_\pm &=&-g \xi_I \big({\rm Im}X^I +i \Gamma_5 {\rm Re} X^I \big)
\nonumber \\
&\mp &{i \over 2} \bigg({\rm Im}((\Phi^I+i Q^I)X^J)+i \Gamma_5
{\rm Re}((\Phi^I+i Q^I)X^J) \bigg){\rm Im} {\cal{N}}_{IJ} \ .
\eea

Next, contract ({\ref{alg4}}) with $\Gamma^i$, to obtain
\bea
\label{auxtr1}
\bigg({1 \over 4} \hn^i h_i -{1 \over 8} dh_{ij} \Gamma^{ij}
-2g \xi_I \Gamma^i \hn_i \big({\rm Im}X^I +i \Gamma_5 {\rm Re} X^I \big)
-{1 \over 8} h_i h^i
\nonumber \\
 -\Gamma^i \hn_i {\hat{\Theta}}_-
-2g \Gamma^i {\hat{\Theta}}_+ \Gamma_i \xi_I
\big({\rm Im}X^I +i \Gamma_5 {\rm Re} X^I \big)
-i \Gamma^i A_i \Gamma_5 {\hat{\Theta}}_+ \bigg) \phi_+=0~.
\eea
This expression implies
\bea
\label{auxlap1}
\bigg({1 \over 2} \hn^i h_i -{1 \over 4} h_i h^i \bigg) \parallel \phi_+ \parallel^2
+ \langle \phi_+, -2g \xi_I \Gamma^i \hn_i \big({\rm Im}X^I +i \Gamma_5 {\rm Re} X^I \big) \phi_+ \rangle
\nonumber \\
+ {\rm Re} \bigg( \langle \phi_+, -4g \Gamma^i
{\hat{\Theta}}_+ \Gamma_i \xi_I \big({\rm Im}X^I +i \Gamma_5 {\rm Re} X^I \big) \phi_+ \rangle \bigg)
\nonumber \\
- {\rm Re} \bigg(\langle \phi_+, 2i \Gamma^i A_i {\hat{\Theta}}_+
\Gamma_5 \phi_+ \rangle \bigg)
=0~.
\eea
On substituting ({\ref{auxlap1}}) into ({\ref{lapl1a}}) to eliminate the
$\xi_I \Gamma^i \hn_i \big({\rm Im}X^I +i \Gamma_5 {\rm Re} X^I \big)$ term, and making use of ({\ref{normder1}}), we obtain
\bea
\hn^i \hn_i \parallel \phi_+ \parallel^2 -h^i \hn_i
\parallel \phi_+ \parallel^2 =0~.
\eea
On applying the maximum principle{\footnote{See e.g.
\cite{maxp}.}} we find that
\bea
\label{const1}
\hn_i \parallel \phi_+ \parallel^2 =0~,
\eea
and hence
\bea
\label{hident}
{1 \over 2} h_i \parallel \phi_+ \parallel^2
+ \langle \phi_+, -2g \Gamma_i \xi_I \big( {\rm Im} X^I +i \Gamma_5
{\rm Re} X^I \big) \phi_+ \rangle =0~,
\eea
or equivalently
\bea
{\rm Re} \bigg( \langle \phi_+, \Gamma_i \Theta_- \phi_+ \rangle
\bigg)=0~,
\eea
or, again, equivalently
\bea
\label{hgam}
h_i \Gamma^i \phi_+ = 4g \bigg(\xi_I {\rm Im}X^I +i \Gamma_5 \xi_I {\rm Re} X^I \bigg)
\bigg(1- {\langle \phi_+, \Gamma_5 \phi_+ \rangle \over \parallel \phi_+ \parallel^2} \Gamma_5 \bigg) \phi_+ \ .
\eea
These conditions imply that
\bea
\label{normhsq}
h^2 = 16g^2 |\xi_I X^I|^2
\bigg( 1 - {\langle \phi_+, \Gamma_5 \phi_+ \rangle^2 \over \langle \phi_+,\phi_+ \rangle^2} \bigg) \ .
\eea
As a consequence of the last equation we conclude that  $\langle \phi_+, \Gamma_5 \phi_+ \rangle \parallel \phi_+ \parallel^{-2}$ does not depend on the coordinate $u$.

\subsection{Independent KSEs on  $\phi_+$}

In this appendix, we shall first prove that, given the gravitino KSE ({\ref{pt1}}) on $\phi_+$   and
({\ref{hidentaux}}) defined below, the algebraic KSEs which arise
as integrability conditions ({\ref{alg1}}), ({\ref{alg3}}), ({\ref{alg4}}), ({\ref{alg5}}), ({\ref{alg7}})
can be reduced to conditions involving only the bosonic fields and a function $\kappa$ on ${\cal S}$.
Then we shall show that, given these bosonic conditions
together with the bosonic field equations, the KSEs
involving $\phi_+$ are equivalent to the naive restriction of the gravitino ({\ref{pt1}}) and
gaugini ({\ref{alg5}}) KSEs on $\phi_+$.

To continue, consider
\bea
\label{hidentaux}
h_i \Gamma^i \phi_+ = 4g \big(\xi_I {\rm Im} X^I +i \Gamma_5 \xi_I {\rm Re} X^I \big) \big(1-\kappa \Gamma_5 \big) \phi_+~,
\eea
an additional condition,
where $\kappa$ is a real function.   As   ({\ref{hidentaux}})
implies that
\bea
\label{kappaferm}
\kappa = {\langle \phi_+, \Gamma_5 \phi_+ \rangle \over \parallel \phi_+ \parallel^2}~,
\eea
(\ref{hidentaux})   is a rewriting of (\ref{hgam}) but without $\parallel\phi_+\parallel$ being constant.
So ({\ref{hidentaux}}) is equivalent to ({\ref{hident}}).
Furthermore, ({\ref{hidentaux}}) also implies that
\bea
\label{kappabos}
\kappa^2 = 1-{h^2 \over 16g^2 |\xi_I X^I|^2} \ .
\eea
Then ({\ref{pt1}}) implies that $\kappa$ satisfies
\bea
\label{kappader}
\hn_i \kappa = \kappa h_i - {\rm Im} \bigg({(A-iB) \over 2g \xi_I X^I}
\bigg) h_i + {\rm Re} \bigg({(A-iB) \over 2g \xi_I X^I}
\bigg) \epsilon_i{}^j h_j \ ,
\eea
where we define the scalars $A$ and $B$ via
\bea
\label{extrasc}
A &=& -g \kappa \xi_I {\rm Im}X^I +{1 \over 2} {\rm Im} {\cal{N}}_{IJ}
{\rm Re} \big( (\Phi^I+iQ^I) X^J \big)~,
\nonumber \\
B &=& -g \kappa \xi_I {\rm Re}X^I -{1 \over 2} {\rm Im} {\cal{N}}_{IJ}
{\rm Im} \big( (\Phi^I+iQ^I) X^J \big) \ .
\eea
In the analysis which will follow, we shall also make use
of the integrability condition of ({\ref{pt1}}),
which is
\bea
\Gamma^j \big(\hn_j \hn_i - \hn_i \hn_j) \phi_+
= {1 \over 2} \Gamma^j {\hat{R}}_{ij} \phi_+ \ ,
\eea
where the LHS is evaluated using ({\ref{pt1}}).
This condition is equivalent to
\bea
\label{intc1}
\bigg(-g_{\alpha \bar{\beta}} \hn_i z^{\bar{\beta}} \hn^i z^\alpha- i g_{\alpha \bar{\beta}} \hn_i z^{\bar{\beta}} \hn_j z^\alpha \epsilon^{ij}
 +{1 \over 4} h_i h^i + \Delta + {\rm{Im}} {\cal{N}}_{IJ}
 (\Phi^I \Phi^J + Q^I Q^J)
  \nonumber \\
  -{1 \over 4} dh_{ij} \Gamma^{ij}
+2g \Gamma_5 \xi_I Q^I
 -2 \Gamma^i \hn_i {\hat{\Theta}}_- -2 \Gamma^j {\hat{\Theta}}_-
 \Gamma_j {\hat{\Theta}}_-+2i
 \Gamma^i A_i\Gamma_5 {\hat{\Theta}}_-
 \bigg) \phi_+ =0~.
 \eea

Now we are ready to determine the conditions on the fields implied by the
remaining KSEs on $\phi_+$.   We begin with the condition ({\ref{alg5}}).
This condition
is equivalent to the following two conditions:

\bea
\label{alg5b}
\Phi^I+iQ^I &=&-2 {\rm Im} {\cal{N}}_{JN}X^J (\Phi^N+iQ^N) {\bar{X}}^I
\cr
&&~~~~~~~~~~~-2ig \kappa \bigg(\xi_J {\rm Im}{\cal{N}}^{-1  IJ} +2 \xi_J X^J {\bar{X}}^I \bigg)~,
\eea
and
\bea
\label{alg5c}
\hn_i {\rm Re} z^\alpha -\epsilon_i{}^j \hn_j {\rm Im} z^\alpha
&=& {1 \over 2} {\rm Re} \bigg({1 \over \xi_J {\bar{X}}^J} \xi_I {\cal{D}}_{\bar{\beta}}
{\bar{X}}^I g^{\alpha \bar{\beta}} \bigg) h_i
\cr
&&~~~~~~~~~
-{1 \over 2} {\rm Im} \bigg({1 \over \xi_J {\bar{X}}^J} \xi_I {\cal{D}}_{\bar{\beta}}
{\bar{X}}^I g^{\alpha \bar{\beta}} \bigg) \epsilon_i{}^j h_j~,
\eea
where ({\ref{hidentaux}}) has been used in order to obtain ({\ref{alg5c}}).

Next we shall consider ({\ref{alg1}}); this is equivalent to the following conditions:
\bea
\label{alg1a}
\Delta=4(A^2+B^2)~,
\eea
and
\bea
\label{alg1b}
{1 \over 8} dh_{ij} \epsilon^{ij}
+2g \kappa {\rm Im}{\cal{N}}_{IJ} {\rm Re} \big(\xi_N {\bar{X}}^N
(\Phi^I +i Q^I)X^J\big)=0 \ .
\eea
In particular, ({\ref{alg1a}}) implies that $\Delta \geq 0$.

Next, we consider ({\ref{alg7}}). We remark that with the definition
of the scalars $A$, $B$ in ({\ref{extrasc}}), together with
({\ref{hidentaux}}), one has
\bea
\label{thetnorm}
\Theta_+ \phi_+ = (A \Gamma_5+iB) \phi_+ \ .
\eea
This expression can be used, together with  ({\ref{alg5b}}), to simplify ({\ref{alg7}}) considerably. After some computation, we find that
({\ref{alg7}}) is equivalent to:
\bea
\label{alg7a}
\bigg(-{(A-iB) \over 2 \xi_J X^J} \xi_I {\cal{D}}_{\bar{\beta}} {\bar{X}}^I
g^{\alpha \bar{\beta}}+2(A+iB) {\rm Im}{\cal{N}}_{IJ} X^J {\cal{D}}_{\bar{\beta}} {\bar{X}}^I
g^{\alpha \bar{\beta}} \bigg) \bigg(h_i -i \epsilon_i{}^j h_j\bigg)
\nonumber \\
-(A+iB) \bigg(\hn_i z^\alpha -i \epsilon_i{}^j \hn_j z^\alpha\bigg)
+{\rm Im}{\cal{N}}_{IJ} {\cal{D}}_{\bar{\beta}} {\bar{X}}^I
g^{\alpha \bar{\beta}} \bigg( \hn_i \Phi^J -i \epsilon_i{}^j \hn_j \Phi^J \bigg)=0~.
\eea

Next we consider ({\ref{alg3}}). This algebraic KSE  is equivalent to
\bea
\label{alg3a}
&&\bigg({1 \over 4} \Delta +{1 \over \xi_J X^J} (A \xi_I {\rm Re} X^I
-B \xi_I {\rm Im} X^I) (A-iB)
-(A+iB) {\rm Im} {\cal{N}}_{IJ} \Phi^I X^J \bigg)  \bigg(h_i -i \epsilon_i{}^j h_j\bigg)
\cr &&
-{1 \over 4} \bigg(\hn_i \Delta - i \epsilon_i{}^j \hn_j \Delta \bigg)
+(A+iB) {\rm Im}{\cal{N}}_{IJ} X^J \bigg( \hn_i \Phi^I -i \epsilon_i{}^j \hn_j \Phi^I \bigg)=0~.
\eea

Finally, we consider the algebraic KSE ({\ref{alg4}}).
On making use of ({\ref{pt1}}), after some further involved computation,
one finds
\bea
\label{alg4a}
\hn_i(A+iB) -{1 \over 2} (A+iB)h_i
-i(A+iB) A_i
\nonumber \\
-{1 \over 2} {\rm Im}{\cal{N}}_{IJ} {\bar{X}}^J
\bigg( d_h\Phi^I_i +i \epsilon_i{}^j
 d_h\Phi^I_j\bigg)
+{\xi_J \Phi^J \over 8 \xi_I X^I} \bigg(h_i -i \epsilon_i{}^j h_j\bigg)=0~.
\eea

Having rewritten the algebraic conditions in this fashion,
we shall now reconsider the condition ({\ref{hidentaux}}).
This was obtained via a global analysis in the previous section.
However, no such analogous condition exists for $\eta_-$.
Hence, we wish to exchange the condition
({\ref{hidentaux}}) for another algebraic condition,
({\ref{alg5}}), for which there does exist an analogous
condition for $\eta_-$, which is ({\ref{alg6}}).
First, note that if one assumes ({\ref{hidentaux}}), together with
({\ref{alg5b}}) and ({\ref{alg5c}}), then one directly
obtains ({\ref{alg5}}). Conversely,
if one assumes ({\ref{alg5}}), together with
({\ref{alg5b}}) and ({\ref{alg5c}}), then one obtains
the condition
\bea
\label{alg5aux}
&&\bigg({\rm Im} \big(\xi_I {\cal{D}}_{\bar{\beta}} {\bar{X}}^I
g^{\alpha {\bar\beta}}\big)
-i \Gamma_5 {\rm Re} \big(\xi_I {\cal{D}}_{\bar{\beta}} {\bar{X}}^I
g^{\alpha {\bar\beta}}\big) \bigg)
\cr
&&~~~~~ \bigg(h_i \Gamma^i
-4g \big(\xi_J {\rm Im} X^J +i \Gamma_5 \xi_J {\rm Re} X^J \big)(1-\kappa \Gamma_5)
\bigg) \phi_+=0~.
\eea
Hence, either $\xi_I {\cal{D}}_\alpha X^I=0$, or ({\ref{alg5aux}}) implies ({\ref{hidentaux}}).

We remark that in the special case for which $\xi_I {\cal{D}}_\alpha X^I=0$
then the equations ({\ref{alg5c}}), ({\ref{alg5b}}) and
({\ref{UUU}}) imply that the scalars $z^\alpha$ are constant, and
also
\bea
\label{simpgauge}
\Phi^I+iQ^I =-2 {\rm Im} {\cal{N}}_{JN}X^J (\Phi^N+iQ^N) {\bar{X}}^I \ .
\eea
In this special case, it is then straightforward to show that one can obtain
the condition ({\ref{hidentaux}}) directly from the KSE
({\ref{pt1}}) and the bosonic conditions listed above.
To see this,  note that ({\ref{normder1}}) holds as a consequence
of ({\ref{pt1}}), and as the scalars are constant one finds
that ({\ref{lapl1a}}) can be simplified to give
\bea
\label{lapl1ab}
\hn_i \hn^i \parallel \phi_+ \parallel^2 &=&
{1 \over 2} \hn^i h_i \parallel \phi_+ \parallel^2 +{1 \over 2} h^i \hn_i \parallel \phi_+ \parallel^2
\nonumber \\
&+& \langle \phi_+, -g h^i \Gamma_i \xi_I \big( {\rm Im}X^I
+i \Gamma_5 {\rm Re} X^I \big) \phi_+ \rangle
\nonumber \\
&+& \langle \phi_+ , -4g \xi_I {\rm Im}X^I \big({\hat{\Theta}}_-^\dagger + {\hat{\Theta}}_- \big) \phi_+ \rangle
\nonumber \\
&+& \langle \phi_+ , -4ig \Gamma_5 \xi_I {\rm Re}X^I
\big({\hat{\Theta}}_-^\dagger - {\hat{\Theta}}_- \big) \phi_+ \rangle \ ,
\eea
which can then be further rewritten as
\bea
\label{lapl1abc}
&&\hn_i \hn^i \parallel \phi_+ \parallel^2 -h^i \hn_i
 \parallel \phi_+ \parallel^2
= \bigg(\Delta +{1 \over 4} h_i h^i -4g^2 |\xi_I X^I|^2
-|{\rm Im}{\cal{N}}_{IJ} (\Phi^I+iQ^I)X^J|^2
\cr
&&~~~~~~~~-4g {\langle \phi_+, \Gamma_5 \phi_+ \rangle \over
\parallel \phi_+ \parallel^2} {\rm Im} \big( \xi_L {\bar{X}}^L {\rm Im}{\cal{N}}_{IJ}
(\Phi^I+i Q^I)X^J \big) \bigg) \parallel \phi_+ \parallel^2 \ ,
\eea
where we have used ({\ref{ein1}}) to eliminate the divergence in $h$ term,
together with ({\ref{simpgauge}}).
However, on taking the inner product of ({\ref{intc1}}) with $\phi_+$
and expanding out the terms, one finds that the RHS of
({\ref{lapl1abc}}) vanishes as a consequence of ({\ref{pt1}}) and
the Einstein field equations. Hence, we have
\bea
\hn^i \hn_i \parallel \phi_+ \parallel^2 -h^i \hn_i
\parallel \phi_+ \parallel^2 =0~,
\eea
which, via an application of the maximum principle, we get $\parallel\phi_+\parallel=\mathrm{const}$ on ${\cal S}$.
Then (\ref{normder1}), which follows from (\ref{pt1}), implies
({\ref{hidentaux}}) as claimed.

\subsection{Independent  KSEs on  $\eta_-$}

In this section, we analyse the various KSEs involving $\eta_-$.
The conditions involving $\eta_-$
are the $u$-dependent parts of the conditions on $\phi_+$
together with ({\ref{pt2}}), ({\ref{alg2}}) and ({\ref{alg6}}).
We shall assume all of the conditions on the bosonic fields
({\ref{alg5b}}), ({\ref{alg5c}}),
({\ref{alg1a}}), ({\ref{alg1b}}), ({\ref{alg7a}}), ({\ref{alg3a}})
and ({\ref{alg4a}}), together with ({\ref{kappabos}}) and ({\ref{kappader}}); which we have previously obtained.

A consequence of our assumptions is that all the KSEs involving $\eta_-$ which come
from the u-dependent parts of $\phi_+$, apart from that of ({\ref{pt1}}), are automatically
satisfied. In the case of the $u$-dependent part of the
gaugino equation, ({\ref{alg5}}), we remark that
this is implied from ({\ref{pt1}}), by making use
of the Lichnerowicz theorem analysis as set out in
Section 4.1.

We shall show that the conditions on $\eta_-$ corresponding to
the $u$-dependent part of ({\ref{pt1}}),
as well as ({\ref{alg2}}), are implied by ({\ref{pt2}}) and
({\ref{alg6}}) together with the bosonic conditions. We
begin with the $u$-dependent part of ({\ref{pt1}}).

\subsubsection{The $u$-dependent part of ({\ref{pt1}})}

The $u$-dependent part of ({\ref{pt1}}) can be rewritten as
\bea
&&\bigg(\hn_i {\hat{\Theta}}_- +{1 \over 8} dh_{ij} \Gamma^j
+{1 \over 4} \big(\hn_{(i} h_{j)} -{1 \over 2} h_i h_j\big) \Gamma^j
-\big({i \over 4} h_i +{1 \over 4} \epsilon_{ij} h^j \Gamma_5 \big)
{\rm Im}{\cal{N}}_{IJ} \big( {\rm Im}((\Phi^I+iQ^I)X^J)
\cr &&~~~~~~~~
+i \Gamma_5 {\rm Re} ((\Phi^I+iQ^I)X^J) \big)
-i A_i
\Gamma_5 {\hat{\Theta}}_-
\cr &&~~~~~~~~~~
-2g {\hat{\Theta}}_- \Gamma_i \xi_I \big({\rm Im}X^I+i \Gamma_5 {\rm Re}X^I\big)
\bigg) \eta_-=0~.
\label{udaux1}
\eea
To proceed, note that the integrability condition of ({\ref{pt2}})
can be written as
\bea
\label{udaux2}
&&\bigg({1 \over 4} \Gamma^j \big(\hn_{(i} h_{j)}-{1 \over 2} h_i h_j\big)
+{1 \over 8}dh_{ij} \Gamma^j
-{1 \over 2} \Gamma^j g_{\alpha {\bar{\beta}}} \hn_{(i}z^\alpha \hn_{j)} z^{\bar{\beta}}
+\Gamma_i \big( {1 \over 8}{\rm Im}{\cal{N}}_{IJ}(\Phi^I \Phi^J+Q^I Q^J)
\cr&&~~~~~~
-{1 \over 4} V +{i \over 4} g_{\alpha {\bar{\beta}}} \epsilon^{mn}
\hn_m z^{\bar{\beta}} \hn_n z^\alpha -{1 \over 2}g \xi_I Q^I \Gamma_5
-{1 \over 2} \Gamma^j \hn_j {\hat{\Theta}}_+
\cr &&~~~~~~~~~
+{i \over 2}
A_j \Gamma^j {\hat{\Theta}}_+ \Gamma_5
-{1 \over 2} \Gamma^j {\hat{\Theta}}_+ \Gamma_j {\hat{\Theta}}_+ \big) \bigg) \eta_-=0~.
\eea
On computing the difference of  ({\ref{udaux1}}) from ({\ref{udaux2}}), one then obtains, after making use
of ({\ref{alg5b}})
\bea
\label{udaux3}
&&\bigg( {1 \over 2} \Gamma^j g_{\alpha {\bar{\beta}}} \hn_{(i} z^\alpha
\hn_{j)} z^{\bar{\beta}}
+ \Gamma_i \big(g^2 g^{\alpha {\bar{\beta}}} \xi_I {\cal{D}}_{\alpha} X^I
\xi_J {\cal{D}}_{\bar{\beta}} {\bar{X}}^J (1+\kappa \Gamma_5)^2-{i \over 4}
g_{\alpha \bar{\beta}} \epsilon^{mn} \hn_m z^{\bar{\beta}} \hn_n z^\alpha
\big)
\cr&&
-{3 \over 2} g {\rm Im} \big(\xi_I {\cal{D}}_\alpha X^I \hn_i z^\alpha\big)
-{3i \over 2} g \Gamma_5 {\rm Re} \big(\xi_I {\cal{D}}_\alpha X^I \hn_i z^\alpha\big)
\cr&&
+{1 \over 2} g \epsilon_i{}^j {\rm Re} \big(\xi_I {\cal{D}}_\alpha X^I \hn_j z^\alpha\big)-{i \over 2}g \Gamma_5 \epsilon_i{}^j {\rm Im} \big(\xi_I {\cal{D}}_\alpha X^I \hn_j z^\alpha\big)
\cr&&
-\big({i \over 4} h_i +{1 \over 4} \epsilon_{ij} h^j \Gamma_5 \big)
{\rm Im}{\cal{N}}_{IJ} \big({\rm Im}((\Phi^I+iQ^I)X^J)+i \Gamma_5 {\rm Re}((\Phi^I+iQ^I)X^J)\big)
\cr&&
+{i \over 4} \hn_i \bigg({\rm Im}{\cal{N}}_{IJ} \big({\rm Im}((\Phi^I+iQ^I)X^J)+i \Gamma_5 {\rm Re}((\Phi^I+iQ^I)X^J)\big) \bigg)
\cr&&
+{1 \over 4} \epsilon_i{}^j \Gamma_5 \hn_j  \bigg({\rm Im}{\cal{N}}_{IJ} \big({\rm Im}((\Phi^I+iQ^I)X^J)+i \Gamma_5 {\rm Re}((\Phi^I+iQ^I)X^J)\big) \bigg)
\cr&&
+{1 \over 4} A_i
\bigg({\rm Im}{\cal{N}}_{IJ} \big(i{\rm Re}((\Phi^I+iQ^I)X^J)+\Gamma_5 {\rm Im}((\Phi^I+iQ^I)X^J)\big) \bigg)
\cr&&
-{i \over 4} \epsilon_i{}^j A_j
 \bigg({\rm Im}{\cal{N}}_{IJ} \big({\rm Im}((\Phi^I+iQ^I)X^J)+i \Gamma_5 {\rm Re}((\Phi^I+iQ^I)X^J)\big) \bigg) \bigg) \eta_-=0~.
\eea
To simplify this expression further, we make use of the algebraic condition
({\ref{alg6}}), which can be rewritten, using ({\ref{alg5b}}), as
\bea
\label{algsimpm1}
&&\bigg(\Gamma^i \hn_i \big({\rm Re}z^\alpha -i \Gamma_5 {\rm Im}z^\alpha \big)
+2g (1+\kappa \Gamma_5) \xi_I\big({\rm Im}({\cal{D}}_{\bar{\beta}} {\bar{X}}^I
g^{\alpha {\bar{\beta}}})
\cr&&~~~~~~~~~~-i \Gamma_5 {\rm Re}({\cal{D}}_{\bar{\beta}} {\bar{X}}^I
g^{\alpha {\bar{\beta}}}) \big) \bigg) \eta_-=0~.
\eea
Acting on the left-hand-side of this expression with ${\rm Im}(\xi_J {\cal{D}}_\alpha X^J)
-i \Gamma_5 {\rm Re} (\xi_J {\cal{D}}_\alpha X^J)$ gives the condition
\bea
\bigg( \Gamma^j \hn_j \big({\rm Re}z^\alpha -i \Gamma_5 {\rm Im}z^\alpha\big)
\big({\rm Im}(\xi_J {\cal{D}}_\alpha X^J)
+i \Gamma_5 {\rm Re} (\xi_J {\cal{D}}_\alpha X^J)\big)
\nonumber \\
-2g (1+\kappa \Gamma_5) \xi_I {\cal{D}}_\alpha X^I \xi_J {\cal{D}}_{\bar{\beta}}
{\bar{X}}^J g^{\alpha {\bar{\beta}}} \bigg) \eta_-=0~.
\eea
which is used to eliminate the $g^2$ term from ({\ref{udaux3}}).
Also, a further useful identity is obtained by acting on the
left-hand-side of ({\ref{algsimpm1}}) with ${\rm Im} g_{\alpha {\bar{\lambda}}}
-i \Gamma_5 {\rm Re} g_{\alpha {\bar{\lambda}}}$, to obtain
\bea
&&\bigg(g(1+\kappa \Gamma_5) \big({\rm Im}(\xi_J {\cal{D}}_\alpha X^J)
+i \Gamma_5 {\rm Re} (\xi_J {\cal{D}}_\alpha X^J)\big)
\cr&&~~~~
+{i \over 2} \Gamma_5 \big({\rm Im} g_{\alpha {\bar{\beta}}}
+i \Gamma_5 {\rm Re} g_{\alpha {\bar{\beta}}}\beta)
\Gamma^i \hn_i ({\rm Re} z^{\bar{\beta}} +i \Gamma_5 {\rm Im} z^{\bar{\beta}}) \bigg)
\eta_- =0~.
\eea
Using this expression, ({\ref{udaux3}}) can be rewritten as
\bea
\label{udaux4}
\bigg(S_i + \Gamma_5 T_i \bigg) \eta_-=0~,
\eea
where
\bea
\label{seqn}
S_i &=& -{i \over 4}h_i {\rm Im} {\cal{N}}_{IJ} {\rm Im} ((\Phi^I+iQ^I)X^J)
-{i \over 4} \epsilon_i{}^j h_j {\rm Im} {\cal{N}}_{IJ} {\rm Re} ((\Phi^I+iQ^I)X^J)
\nonumber \\
&+&{i \over 2} \kappa g {\rm Re}(\xi_I {\cal{D}}_\alpha X^I \hn_i z^\alpha)
-{i \over 2} \kappa g \epsilon_i{}^j  {\rm Im}(\xi_I {\cal{D}}_\alpha X^I \hn_j z^\alpha)
\nonumber \\
&+&{i \over 4} \hn_i \bigg({\rm Im} {\cal{N}}_{IJ} {\rm Im} ((\Phi^I+iQ^I)X^J)\bigg)
+{i \over 4} \epsilon_i{}^j \hn_j \bigg( {\rm Im} {\cal{N}}_{IJ} {\rm Re} ((\Phi^I+iQ^I)X^J) \bigg)
\nonumber \\
&+&{i \over 4} A_i {\rm Im} {\cal{N}}_{IJ} {\rm Re} ((\Phi^I+iQ^I)X^J)
-{i \over 4} \epsilon_i{}^j A_j {\rm Im} {\cal{N}}_{IJ} {\rm Im} ((\Phi^I+iQ^I)X^J)~,
\eea
and
\bea
\label{teqn}
T_i &=& {1 \over 4}h_i {\rm Im} {\cal{N}}_{IJ} {\rm Re} ((\Phi^I+iQ^I)X^J)
-{1 \over 4} \epsilon_i{}^j h_j {\rm Im} {\cal{N}}_{IJ} {\rm Im} ((\Phi^I+iQ^I)X^J)
\nonumber \\
&+&{1 \over 2} \kappa g {\rm Im}(\xi_I {\cal{D}}_\alpha X^I \hn_i z^\alpha)
+{1 \over 2} \kappa g \epsilon_i{}^j  {\rm Re}(\xi_I {\cal{D}}_\alpha X^I \hn_j z^\alpha)
\nonumber \\
&-&{1 \over 4} \hn_i \bigg({\rm Im} {\cal{N}}_{IJ} {\rm Re} ((\Phi^I+iQ^I)X^J)\bigg)
+{1 \over 4} \epsilon_i{}^j \hn_j \bigg({\rm Im} {\cal{N}}_{IJ} {\rm Im} ((\Phi^I+iQ^I)X^J)\bigg)
\nonumber \\
&+&{1 \over 4} A_i {\rm Im} {\cal{N}}_{IJ} {\rm Im} ((\Phi^I+iQ^I)X^J)
+{1 \over 4} \epsilon_i{}^j A_j {\rm Im} {\cal{N}}_{IJ} {\rm Re} ((\Phi^I+iQ^I)X^J)~.
\eea
As $S_i$ is imaginary and $T_i$ is real, the condition ({\ref{udaux4}})
is equivalent to
\bea
\label{udaux5}
\bigg(T_i -S_i\bigg) \eta_-=0~,
\eea
where
\bea
\label{stsimp}
T_i-S_i &=& {1 \over 4} (h_i+i \epsilon_i{}^j){\rm Im}{\cal{N}}_{IJ}
(\Phi^I+iQ^I)X^J
-{i \over 2} \kappa g \xi_I {\cal{D}}_\alpha X^I \hn_i z^\alpha
+{1 \over 2} \kappa g \epsilon_i{}^j \xi_I {\cal{D}}_\alpha X^I \hn_j z^\alpha
\nonumber \\
&-& {1 \over 4} \hn_i \bigg( {\rm Im} {\cal{N}}_{IJ}(\Phi^I+iQ^I)X^J \bigg)
-{i \over 4} \epsilon_i{}^j \hn_j \bigg( {\rm Im} {\cal{N}}_{IJ}(\Phi^I+iQ^I)X^J \bigg)
\nonumber \\
&-&{i \over 4} A_i{\rm Im}{\cal{N}}_{IJ} (\Phi^I+iQ^I)X^J
+{1 \over 4} \epsilon_i{}^j A_j{\rm Im}{\cal{N}}_{IJ} (\Phi^I+iQ^I)X^J~.
\eea
However, the conditions we have found on the fields in the previous section imply that $T_i-S_i=0$. In particular, this can be seen
by writing
\bea
{\rm Im}{\cal{N}}_{IJ}(\Phi^I+iQ^I) X^J = 2(A-iB) -2ig \kappa \xi_I X^I~,
\eea
and then by making use of ({\ref{alg4a}}), ({\ref{kappader}}), and ({\ref{alg5b}}).
After some manipulation, one obtains $T_i-S_i=0$.

Hence, it follows that the $u$-dependent part of ({\ref{pt1}})
is implied by ({\ref{pt2}}), ({\ref{alg6}}) and the bosonic conditions.

\subsubsection{The  ({\ref{alg2}}) KSE}

To analyse ({\ref{alg2}}) we begin by contracting ({\ref{udaux2}}) with
$\Gamma^i$ to obtain

\bea
\label{udaux6}
\bigg({1 \over 2} \Delta +{1 \over 8} h_i h^i +{i \over 8} \epsilon^{ij} dh_{ij}
\Gamma_5 -{1 \over 2} g_{\alpha {\bar{\beta}}} \hn_i z^\alpha \hn^i z^{\bar{\beta}}
+{1 \over 2}{\rm Im}{\cal{N}}_{IJ} (\Phi^I \Phi^J+Q^I Q^J)
\nonumber \\
+{i \over 2} g_{\alpha {\bar{\beta}}} \epsilon^{mn} \hn_m z^{\bar{\beta}} \hn_n
z^\alpha -g \xi_I Q^I \Gamma_5
-\Gamma^j \hn_j {\hat{\Theta}}_+
\nonumber \\
+i \Gamma^j A_j {\hat{\Theta}}_+ \Gamma_5 - \Gamma^j {\hat{\Theta}}_+
\Gamma_j {\hat{\Theta}}_+ \bigg) \eta_-=0~.
\eea
The $\Gamma^j \hn_j {\hat{\Theta}}_+$ term is evaluated by making use of
({\ref{alg4a}}) together with ({\ref{kappader}}), and the terms quadratic
in $\Phi^I$ and $Q^I$ are rewritten using ({\ref{alg5b}}). Then
({\ref{udaux6}}) is equivalent to
\bea
\label{udaux7}
&&\bigg({1 \over 2} \Delta +{1 \over 8} h_i h^i +{i \over 8} \epsilon^{ij} dh_{ij}
\Gamma_5 -{1 \over 2} g_{\alpha {\bar{\beta}}} \hn_i z^\alpha \hn^i z^{\bar{\beta}}
+{i \over 2} g_{\alpha {\bar{\beta}}} \epsilon^{mn} \hn_m z^{\bar{\beta}} \hn_n
z^\alpha
\cr &&
-4g^2 \kappa^2 g^{\alpha {\bar{\beta}}} \xi_I {\cal{D}}_\alpha X^I
\xi_J {\cal{D}}_{\bar{\beta}} {\bar{X}}^J -g \xi_I Q^I \Gamma_5
-2g^2 |\xi_I X^I|^2 -{1 \over 2}|{\rm Im}{\cal{N}}_{IJ}(\Phi^I+iQ^I)X^J|^2
+ig \xi_i \Phi^I
\cr &&
-i \Gamma^j \big(-{1 \over2} {\rm Im}{\cal{N}}_{IJ} {\rm Im}((\Phi^I+iQ^I)X^J)
h_j +{1 \over 2} {\rm Im}{\cal{N}}_{IJ} {\rm Re}((\Phi^I+iQ^I)X^J) \epsilon_j{}^k h_k
\big)
\cr &&
-i \Gamma^j\big({\rm Im}{\cal{N}}_{IJ} {\rm Re}X^J \epsilon_j{}^k d_h\Phi_k^I-{\rm Im}{\cal{N}}_{IJ} {\rm Im}X^J d_h\Phi_j^I \big)
\cr &&
-g \Gamma^j \big({\rm Im}(\xi_I {\cal{D}}_\alpha X^I \hn_j z^\alpha)
+i \Gamma_5 {\rm Re}(\xi_I {\cal{D}}_\alpha X^I \hn_j z^\alpha)\big)\big(-1+\kappa \Gamma_5\big) \bigg) \eta_-=0
\eea
This expression can be further simplified in several ways.
Firstly, using ({\ref{alg6}}), the final line can be written as
\bea
\label{uxsimp1}
-g \Gamma^j \big({\rm Im}(\xi_I {\cal{D}}_\alpha X^I \hn_j z^\alpha)
+i \Gamma_5 {\rm Re}(\xi_I {\cal{D}}_\alpha X^I \hn_j z^\alpha)\big) \big(-1+\kappa \Gamma_5\big)  \eta_-
\nonumber \\
\ \ \  =2g^2(1+\kappa \Gamma_5)^2
g^{\alpha {\bar{\beta}}} \xi_I {\cal{D}}_\alpha X^I \xi_J {\cal{D}}_{\bar{\beta}}
{\bar{X}}^J \eta_-~.
\eea
Also, using ({\ref{alg5b}}) we have
\bea
\label{uxsimp2}
-g \xi_I Q^I \Gamma_5 +4 \kappa g^2 g^{\alpha {\bar{\beta}}} \xi_I {\cal{D}}_\alpha X^I \xi_J {\cal{D}}_{\bar{\beta}}
{\bar{X}}^J \Gamma_5
=2g {\rm Im} \bigg( \xi_L {\bar{X}}^L {\rm Im} {\cal{N}}_{IJ}
\big(\Phi^I+iQ^I)X^J \bigg) \Gamma_5.
\eea
Furthermore, it is useful to note the following identity:
\bea
&& \bigg( {1 \over 2} g_{\alpha {\bar{\beta}}} \hn_i z^\alpha \hn^i z^{\bar{\beta}}
-{i \over 2} g_{\alpha {\bar{\beta}}} \epsilon^{mn} \hn_m z^{\bar{\beta}} \hn_n
z^\alpha
-{i \over 2} \Gamma^i \Gamma^j \big({\rm Re} \hn_i z^\alpha +i \Gamma_5 {\rm Im}
\hn_i z^\alpha \big)
\cr &&
~~~~~~~~~ \Gamma_5 \big({\rm Im} g_{\alpha {\bar{\beta}}}
-i \Gamma_5 {\rm Re} g_{\alpha {\bar{\beta}}} \big)
\big({\rm Re} \hn_j z^{\bar{\beta}}+i \Gamma_5 {\rm Im} \hn_j z^{\bar{\beta}} \big)
\bigg) \eta_-=0~,
\eea
and on making repeated use of ({\ref{alg6}}) this expression implies that
\bea
\label{uxsimp3}
&&\bigg({1 \over 2} g_{\alpha {\bar{\beta}}} \hn_i z^\alpha \hn^i z^{\bar{\beta}}
-{i \over 2} g_{\alpha {\bar{\beta}}} \epsilon^{mn} \hn_m z^{\bar{\beta}} \hn_n
z^\alpha
\cr &&~~~~~~~~~~~~~-2g^2(1-\kappa^2)g^{\alpha {\bar{\beta}}} \xi_I {\cal{D}}_\alpha X^I \xi_J {\cal{D}}_{\bar{\beta}}
{\bar{X}}^J \bigg) \eta_-=0~.
\eea
On substituting ({\ref{uxsimp1}}), ({\ref{uxsimp2}}) and ({\ref{uxsimp3}})
info ({\ref{udaux7}}) in order to rewrite the final line of
({\ref{udaux7}}), and then eliminate the $\xi_I Q^I$ term and the terms
quadratic in $\hn z$, we find that ({\ref{udaux7}}) is equivalent to
\bea
\label{udaux8}
&&\bigg({1 \over 2} \Delta +{1 \over 8} h_i h^i +{i \over 8} \epsilon^{ij}
dh_{ij} \Gamma_5
+2g {\rm Im}  \big( \xi_L {\bar{X}}^L {\rm Im} {\cal{N}}_{IJ}
\big(\Phi^I+iQ^I)X^J \big) -2g^2 |\xi_I X^I|^2
\cr && -{1 \over 2}|{\rm Im}{\cal{N}}_{IJ} (\Phi^I+iQ^I)X^J|^2
+ig \xi_I \Phi^I
+ h_j \Gamma^j
\big({i \over 2} {\rm Im}{\cal{N}}_{IJ} {\rm Im} ((\Phi^I+iQ^I)X^J)
\cr &&
-{1 \over 2} \Gamma_5 {\rm Im}{\cal{N}}_{IJ} {\rm Re} ((\Phi^I+iQ^I)X^J)\big)
\cr &&~~~~~~+i {\rm  Im}{\cal{N}}_{IJ}
\Gamma^i {\rm Im} \big( d_h\Phi^I_i -i \epsilon_i{}^j
d_h\Phi^I_j)X^J \big) \bigg) \eta_-=0~.
\eea
After some straightforward rearrangement of terms, we find that
({\ref{udaux8}}) is equivalent to ({\ref{alg2}})

\appendix{Lichnerowicz type theorems for $\phi_\pm$}

In this Appendix, we provide a more detailed description of the proof of the Lichnerowicz type theorems   for $\phi_\pm$ spinors. Note that $\phi_-=\eta_-$ and that the Lichnerowicz type theorem on $\eta_+$
is implied from that on $\phi_+$.

To begin, the covariant derivatives  associated with the gravitino KSE (\ref{ptx}) have been defined
in (\ref{lichd1x}). Next upon using the condition ({\ref{alg5b}}), the algebraic operators
 (\ref{algx}) which define the gaugini KSEs (\ref{lichd2x}) can be
rewritten as
\bea
{\cal A}^\alpha_{(\pm)} &\equiv& \Gamma^i \hn_i {\rm Re} z^\alpha +i \Gamma_5 \Gamma^i \hn_i {\rm Im} z^\alpha
\cr
&+&2(1 \mp \kappa \Gamma_5) \bigg( g \xi_I {\rm Im} \big({\cal{D}}_{\bar{\beta}} {\bar{X}}^I g^{\alpha {\bar{\beta}}}\big)
-ig  \Gamma_5 {\rm Re} \big({\cal{D}}_{\bar{\beta}} {\bar{X}}^I g^{\alpha {\bar{\beta}}}\big) \bigg)~.
\eea
The horizon Dirac operators $\cD^{(\pm)}$ are
\bea
\cD^{(\pm)} \equiv \Gamma^i {\hat{\nabla}}^{(\pm)}_i
\eea
and we assume that $\phi_\pm$ are zero modes, ie they satisfy
\bea
\cD^{(\pm)} \phi_\pm =0 \ .
\eea
We also assume all of the conditions on the fields (\ref{kappabos}), (\ref{kappader}), (\ref{alg5b}), (\ref{alg5c}), (\ref{alg1a}), (\ref{alg1b}), (\ref{alg7a}), (\ref{alg3a}), (\ref{alg4a})  obtained in appendix C. However, we do not assume that the function $\kappa$ is related to the zero mode $\phi_+$. We then have
\bea
\hn_i \hn^i \parallel \phi_\pm \parallel^2
= 2 {\rm Re} \langle \phi_\pm, \hn_i \hn^i \phi_\pm \rangle + 2 \langle \hn^i \phi_\pm,
\hn_i \phi_\pm \rangle
\eea
where after making use of ${\cal{D}}^{(\pm)} \phi_\pm =0$
\bea
&&2 {\rm Re} \langle \phi_\pm, \hn_i \hn^i \phi_\pm \rangle
= {1 \over 2} {\hat{R}} \parallel \phi_\pm \parallel^2
\cr
&&~~~+2 {\rm Re} \langle \phi_\pm , \Gamma^i \hn_i \bigg( \big(-{i \over 2}
A_j \Gamma^j \Gamma_5 -ig \xi_I B^I_j \Gamma^j +2 \hT_\mp \pm {1 \over 4} h_j \Gamma^j \big)
\phi_\pm \bigg) \rangle~.
\eea
It follows that we can write
\bea
\label{biglichn1}
&&\hn_i \hn^i \parallel \phi_\pm \parallel^2
= \bigg({1 \over 2}{\hat{R}} \pm {1 \over 2} \hn^i h_i \bigg) \parallel \phi_\pm \parallel^2
+2 \langle {\hat{\nabla}}^{(\pm)i}  \phi_\pm, {\hat{\nabla}}^{(\pm)}_i \phi_\pm \rangle
\cr
&&~~~~+2 {\rm Re} \langle \phi_\pm , \Gamma^i (-{i \over 2} \Gamma^j A_j \Gamma_5
-ig \xi_I B^I_j \Gamma^j +2 \hT_\mp \pm {1 \over 4} h_j \Gamma^j) \hn_i \phi_\pm \rangle
\cr
&&~~~~-4 {\rm Re} \langle \phi_\pm , \big(-{i \over 2} A^i \Gamma_5 -ig \xi_I B^{Ii} - \hT_\mp^\dagger \Gamma^i \mp {1 \over 4} h^i \big) \hn_i \phi_\pm \rangle
\nonumber \\
&&~~~~+ {\rm Re} \langle \phi_\pm , \big(-{i \over 2} \Gamma^{ij} (dA)_{ij} \Gamma_5
-ig \xi_I dB^I_{ij} \Gamma^{ij} +4 \Gamma^i \hn_i \hT_\mp \big) \phi_\pm \rangle
\cr
&&~~~~-2 \langle \phi_\pm, \big(-{i \over 2} A^i \Gamma_5 -ig \xi_I B^{Ii}-\hT_\mp^\dagger \Gamma^i \mp {1 \over 4} h^i \big)
 \cr
 &&~~~~~~~~~~\big({i \over 2} A_i \Gamma_5 +ig \xi_J B^J_i -\Gamma_i \hT_\mp \mp {1 \over 4} h_i \big) \phi_\pm \rangle~.
\eea
The terms in the second and third lines of the above expression which are linear
in $\hn_i \phi_\pm$ can then be rewritten using ${\cal{D}}^{(\pm)} \phi_\pm =0$ as
\bea
\label{biglichn2}
\pm h^i \hn_i \parallel \phi_\pm \parallel^2
+ {\rm Re} \langle \phi_\pm, & \big(-i \Gamma^j A_j \Gamma_5 +2ig \xi_I B^I_j \Gamma^j
\mp {1 \over 2} h_j \Gamma^j -8g \xi_I ({\rm Im} X^I -i \Gamma_5 {\rm Re}X^I) \big)
\nonumber \\
& \times \big(-{i \over 2} \Gamma^i A_i \Gamma_5 -ig \xi_J B^J_i \Gamma^i \pm
{1 \over 4} h_i \Gamma^i +2 \hT_\mp \big) \phi_\pm \rangle~.
\eea
Furthermore, we also have
\bea
\label{biglichn3}
(dA)_{ij}=-2i g_{\alpha {\bar{\beta}}} \hn_{[i} z^{\bar{\beta}} \hn_{j]}z^\alpha~.
\eea
On substituting these expressions into ({\ref{biglichn1}}), we find for $\phi_+$:
\bea
&&\hn_i \hn^i \parallel \phi_+ \parallel^2 -h^i \hn_i \parallel \phi_+ \parallel^2
= 2 \langle {\hat{\nabla}}^{(+)i}  \phi_+, {\hat{\nabla}}^{(+)}_i \phi_+ \rangle
\cr
&&~~~~~~+ \bigg(4g^2 (1+\kappa^2)g^{\alpha {\bar{\beta}}} \xi_I {\cal{D}}_\alpha X^I \xi_J {\cal{D}}_{\bar{\beta}}{\bar{X}}^J + g_{\alpha {\bar{\beta}}} \hn_i z^\alpha \hn^i z^{\bar{\beta}} \bigg)
\parallel \phi_+ \parallel^2
\cr
&&~~~~~~+ {\rm Re} \langle \phi_+, \bigg(i g_{\alpha {\bar{\beta}}} \epsilon^{ij} \hn_i z^{\bar{\beta}}
\hn_j z^\alpha
-8 \kappa g^2 g^{\alpha {\bar{\beta}}} \xi_I {\cal{D}}_\alpha X^I \xi_J {\cal{D}}_{\bar{\beta}}{\bar{X}}^J \Gamma_5
\cr
 &&~~~~~~-4g \Gamma^i {\rm Im}(\xi_I {\cal{D}}_\alpha X^I \hn_i z^\alpha)
 -4ig \Gamma^i \Gamma_5 {\rm Re}(\xi_I {\cal{D}}_\alpha X^I \hn_i z^\alpha) \bigg) \phi_+ \rangle~,
\eea
and for $\phi_-$ we find
\bea
&&\hn_i \hn^i \parallel \phi_- \parallel^2 + \hn^i \bigg(h_i \parallel \phi_- \parallel^2\bigg)
= 2 \langle {\hat{\nabla}}^{(-)i}  \phi_-, {\hat{\nabla}}^{(-)}_i \phi_- \rangle
\cr
&&~~~~~~+ \bigg(4g^2 (1+\kappa^2)g^{\alpha {\bar{\beta}}} \xi_I {\cal{D}}_\alpha X^I \xi_J {\cal{D}}_{\bar{\beta}}{\bar{X}}^J + g_{\alpha {\bar{\beta}}} \hn_i z^\alpha \hn^i z^{\bar{\beta}} \bigg)
\parallel \phi_- \parallel^2
\cr
&&~~~~~~+ {\rm Re} \langle \phi_-, \bigg(-i g_{\alpha {\bar{\beta}}} \epsilon^{ij} \hn_i z^{\bar{\beta}}
\hn_j z^\alpha +8 \kappa g^2 g^{\alpha {\bar{\beta}}} \xi_I {\cal{D}}_\alpha X^I \xi_J {\cal{D}}_{\bar{\beta}}{\bar{X}}^J \Gamma_5
\cr
&&~~~~~~
-4g \Gamma^i {\rm Im}(\xi_I {\cal{D}}_\alpha X^I \hn_i z^\alpha)
-4ig \Gamma^i \Gamma_5 {\rm Re}(\xi_I {\cal{D}}_\alpha X^I \hn_i z^\alpha) \bigg) \phi_- \rangle~,
\eea
where we have made use of the Einstein equation
\bea
{\hat{R}} = - \hn^i h_i +{1 \over 2} h_i h^i +2 g_{\alpha {\bar{\beta}}}
\hn_i z^\alpha \hn^i z^{\bar{\beta}} +2 V - {\rm Im}{\cal{N}}_{IJ}
\big( \Phi^I \Phi^J + Q^I Q^J \big)~,
\eea
obtained from taking the trace of ({\ref{ein4}}), as well as ({\ref{alg5b}}).

To complete the proof after some computation one can show that
\bea
&&\bigg(4g^2 (1+\kappa^2) g^{\alpha {\bar{\beta}}} \xi_I {\cal{D}}_\alpha X^I \xi_J {\cal{D}}_{\bar{\beta}}{\bar{X}}^J + g_{\alpha {\bar{\beta}}} \hn_i z^\alpha \hn^i z^{\bar{\beta}} \bigg)
\parallel \phi_\pm \parallel^2
\nonumber \\
&+& {\rm Re} \langle \phi_\pm, \bigg(\pm i g_{\alpha {\bar{\beta}}} \epsilon^{ij} \hn_i z^{\bar{\beta}}
\hn_j z^\alpha \mp 8 \kappa g^2 g^{\alpha {\bar{\beta}}} \xi_I {\cal{D}}_\alpha X^I \xi_J {\cal{D}}_{\bar{\beta}}{\bar{X}}^J \Gamma_5
\nonumber \\
&&-4g \Gamma^i {\rm Im}(\xi_I {\cal{D}}_\alpha X^I \hn_i z^\alpha)
-4ig \Gamma^i \Gamma_5 {\rm Re}(\xi_I {\cal{D}}_\alpha X^I \hn_i z^\alpha) \bigg) \phi_\pm \rangle
\nonumber \\
&&= \langle {\cal A}_{(\pm)}^\beta \phi_\pm, \big( {\rm Re}(g_{\alpha \bar{\beta}}) +i \Gamma_5 {\rm Im} (g_{\alpha \bar{\beta}}) \big) {\cal A}_{(\pm)}^\alpha \phi_\pm \rangle~.
\eea
The positive definiteness of this term follows from positive definiteness of
the K\"ahler metric on the scalar manifold and after  further decomposing
${\cal A}_{(\pm)}^\alpha \phi_\pm$ into positive and negative chiralities with
respect to $\Gamma_5$.

\appendix{Properties of the isometry $W$}

In this appendix, we shall consider the case for which the vector field $W$ given in (\ref{extraiso})
does not vanish,
$W \not \equiv 0$, and we shall prove that it
is a symmetry of the full solution.

First $W$ is an isometry of the metric on ${\cal S}$.  This can be seen from either (\ref{concon}) or
verified directly using  ({\ref{pt1}}) and ({\ref{pt2}}) which imply that
\bea
\label{isoexp}
\hn_j W_i ={\rm Re} \bigg(-2i \langle \Gamma_+ \eta_-, \Gamma_5
{\hat{\Theta}}_- \eta_+ \rangle \bigg) \epsilon_{ij}~,~~~\eta_+=\Gamma_+\Theta_-\eta_-,
\eea
and hence
\bea
\hn_{(i} W_{j)} =0~.
\eea

To proceed, consider the algebraic conditions
\bea
{\cal A}_{(\pm)}^\alpha \eta_\pm =0~.
\eea
In particular, on comparing the conditions
\bea
\langle \Gamma_+ \eta_-, {\cal A}_{(+)}^\alpha \eta_+ \rangle =0,
\qquad {\rm and} \qquad \langle {\cal A}_{(-)}^\alpha \eta_-, \Gamma_- \eta_+ \rangle =0~,
\eea
one obtains the condition
\bea
\cL_W {\rm Re} z^\alpha =0~,
\eea
and on comparing the conditions
\bea
\langle \Gamma_+ \eta_-, i \Gamma_5 {\cal A}_{(+)}^\alpha \eta_+ \rangle =0,
\qquad {\rm and} \qquad \langle i \Gamma_5 {\cal A}_{(-)}^\alpha \eta_-, \Gamma_- \eta_+ \rangle =0~,
\eea
one finds that
\bea
\cL_W {\rm Im} z^\alpha =0~,
\eea
and hence
\bea
\cL_W z^\alpha =0 \ .
\eea
The components of $W$ can be rewritten as
\bea
W_i =-{1 \over 2} \parallel \eta_- \parallel^2 h_i
+2g \xi_I {\rm Im}X^I \tau_i +2g \xi_I {\rm Re}X^I \epsilon_{ij}\tau^j~,
\eea
where
\bea
\tau_i = \langle \eta_-, \Gamma_i \eta_- \rangle~,
\eea
and $\tau$ satisfies
\bea
\tau_i \tau^i =(\parallel \eta_- \parallel^2)^2-\langle \eta_-, \Gamma_5 \eta_- \rangle ^2~.
\eea
Then ({\ref{auxd1}}) implies
\bea
\label{parconst1}
\cL_W \parallel \eta_- \parallel^2 = 4g^2 |\xi_I X^I|^2
\bigg( \langle \eta_-, \Gamma_5 \eta_- \rangle^2
-(\parallel \eta_- \parallel^2)^2 \kappa^2 \bigg)~.
\eea
The condition ({\ref{isoexp}}) implies,
on expanding out the expression for
${\rm Im} \langle \Gamma_+ \eta_-, \Gamma_5 {\hat{\Theta}}_-
\eta_+ \rangle$, that
\bea
\label{dWa}
dW_{ij} \epsilon^{ij} &=&
4 \bigg(-g \xi_I {\rm Im} X^I h^i \epsilon_{ij} \tau^j
+g \xi_I {\rm Re} X^I h^i \tau_i
\nonumber \\
&-&4g {\rm Re} \big(\xi_L {\bar{X}}^L
{\rm Im}{\cal{N}}_{IJ} (\Phi^I+iQ^I)X^J \big)
\langle \eta_-, \Gamma_5 \eta_- \rangle \bigg)~.
\eea
However, on taking the exterior derivative of
({\ref{auxd1}}), one finds
\bea
\label{dWb}
dW = W \wedge h - \parallel \eta_- \parallel^2 dh~.
\eea
On comparing the components of $dW$ between ({\ref{dWa}})
and ({\ref{dWb}}), making use of ({\ref{alg1b}}), one finds
that if $\xi_I \Phi^I \neq 0$, then the RHS of ({\ref{parconst1}}) vanishes. So, if $\xi_I \Phi^I \neq 0$
then
\bea
\cL_W \parallel \eta_- \parallel^2=0~.
\eea
Then, taking the Lie derivative of ({\ref{auxd1}}) with respect
to $W$ implies that
\bea
\cL_W h=0~,
\eea
and taking the Lie derivative of the trace of the
Einstein ({\ref{ein4}}) with respect to $W$ gives
\bea
\cL_W \bigg( {\rm Im} {\cal{N}}_{IJ} (\Phi^I \Phi^J + Q^I Q^J) \bigg) =0~,
\eea
and taking the Lie derivative of the Einstein equation
({\ref{ein1}}) with respect to $W$ implies that
\bea
\cL_W \Delta =0~.
\eea
The condition ({\ref{kappabos}}) implies also that
\bea
\cL_W \kappa =0~.
\eea
Next, on taking the Lie derivative of
({\ref{kappader}}) with respect to $W$ gives
\bea
-{\rm Im} \bigg({1 \over \xi_I X^I} \cL_W(A-iB) \bigg)
h_i + {\rm Re}\bigg({1 \over \xi_I X^I} \cL_W(A-iB) \bigg)
\epsilon_i{}^j h_j =0~.
\eea
We remark that it is not consistent to have $h \equiv 0$,
because if $h \equiv 0$ then ({\ref{kappabos}}) implies
that $\kappa^2=1$, and then the condition
({\ref{alg1b}}) is inconsistent with our assumption that
$\xi_I \Phi^I \neq 0$. Hence, we must have
\bea
\cL_W A = \cL_W B =0~,
\eea
which further implies that
\bea
\cL_W \Phi^I = \cL_W Q^I =0~,
\eea
as a consequence of ({\ref{alg5b}}).
Hence, if $\xi_I \Phi^I \neq 0$, then $W$ is a symmetry of the
full solution.

Next, we consider the case for which $\xi_I \Phi^I \equiv 0$.
On taking the Lie derivative of ({\ref{alg5c}}) with respect to $W$, it follows that as $\cL_W z^\alpha=0$, one must have either
$\xi_I \cD_\alpha X^I =0$, or $\cL_W h=0$.
Suppose then that $\xi_I \Phi^I \equiv 0$, but
$\xi_I \cD_\alpha X^I \neq 0$. Then
\bea
\cL_W h=0~.
\eea
As before, the trace of ({\ref{ein4}}), ({\ref{ein1}}) and ({\ref{kappabos}})
imply that
\bea
\cL_W \bigg( {\rm Im} {\cal{N}}_{IJ} (\Phi^I \Phi^J + Q^I Q^J) \bigg) =0, \qquad \cL_W \Delta=0, \qquad \cL_W \kappa =0~,
\eea
and taking the Lie derivative of ({\ref{kappader}}) with respect to $W$ gives
\bea
\cL_W A = \cL_W B =0, \qquad {\rm or} \qquad  h=0~.
\eea
Suppose that $\cL_W A = \cL_W B=0$. Then
({\ref{alg5b}}) implies that
\bea
\cL_W \Phi^I = \cL_W Q^I =0~,
\eea
and hence $W$ is a symmetry of the full solution.

Alternatively, if $h \equiv 0$, then the Einstein equation
({\ref{ein2}}) implies that $
\Delta= {\mathrm{const}}$, $\Phi^I={\mathrm{const}}$
and ({\ref{alg5c}}) implies that $z^\alpha= {\mathrm{const}}$.
The gauge field equation ({\ref{geq2}}) then implies that
$Q^I = \mathrm{const}$ as well. So if $h \equiv 0$, it follows
again that $W$ must be an symmetry of the full solution.
Hence, if $\xi_I \Phi^I =0$ but $\xi_I \cD_\alpha X^I \neq 0$,
then $W$ is a symmetry of the full solution.

It remains to consider the case for which
$\xi_I \Phi^I =0$ and $\xi_I \cD_\alpha X^I =0$.
For such solutions $dh=0$, and $z^\alpha = {\mathrm{const}}$.
To proceed in this case, consider the gravinito integrability conditions ({\ref{intc1}}) and ({\ref{udaux6}}), which
imply
\bea
\label{csc1}
\bigg(2g(\Gamma_5-\kappa) \xi_I Q^I
-2 \Gamma^i \hn_i {\hat{\Theta}}_- \bigg) \eta_+=0~,
\eea
and
\bea
\label{csc2}
\bigg(2g(-\Gamma_5-\kappa) \xi_I A^I +2 \Gamma^i \hn_i {\hat{\Theta}}_- \bigg) \eta_- =0~.
\eea
On taking the inner product of ({\ref{csc1}}) with $\Gamma_+ \eta_-$, and comparing this with the (complex conjugate of)
the inner product of ({\ref{csc2}}) with $\Gamma_- \eta_+$,
we obtain
\bea
\cL_W \bigg( {\rm Im}{\cal{N}}_{IJ} {\rm Im} \big((\Phi^I+iQ^I)X^J \big) \bigg) =0~,
\eea
and on taking the inner product of $i\Gamma_5$({\ref{csc1}}) with $\Gamma_+ \eta_-$, and comparing with the
(complex conjugate of) the inner product of $i \Gamma_5$({\ref{csc2}}) with $\Gamma_- \eta_+$, we find
\bea
\cL_W \bigg( {\rm Im}{\cal{N}}_{IJ} {\rm Re} \big((\Phi^I+iQ^I)X^J \big) \bigg) =0~.
\eea
So, we have
\bea
\cL_W  \bigg( {\rm Im}{\cal{N}}_{IJ} \big((\Phi^I+iQ^I)X^J \big) \bigg) =0 \ .
\eea
The condition ({\ref{alg5b}}) then implies that
\bea
\cL_W \Phi^I = \cL_W Q^I =0~.
\eea
On taking the Lie derivative with respect to $W$ of
the gauge equation ({\ref{geq2}}), we find
\bea
{\rm Im} {\cal{N}}_{IJ} \Phi^J (\cL_W h)_j =
{\rm Im} {\cal{N}}_{IJ} Q^J \epsilon_j{}^k (\cL_W h)_k~,
\eea
which implies that either $\Phi^I=Q^I=0$, or $\cL_W h=0$.
If $\cL_W h=0$, then on taking the Lie derivative of
({\ref{ein1}}) with respect to $W$ gives
\bea
\cL_W \Delta=0~,
\eea
and hence $W$ is a symmetry of the full solution.

It remains to consider the case for which $\Phi^I=0$, $Q^I=0$ and $\xi_I \cD_\alpha X^I=0$. In this case, the Einstein equation ({\ref{ein1}}) can be rewritten as
\bea
\hn^i h_i +8g^2(1+\kappa^2) |\xi_I X^I|^2 =0~.
\eea
On integrating this expression over ${\cal{S}}$, we
see that it admits no solution. Hence, the case for which
$\Phi^I=0$, $Q^I=0$ and $\xi_I \cD_\alpha X^I=0$
is excluded{\footnote{We remark that this excludes the solution
$AdS_4$ with constant $z^\alpha$, and $F^I=0$.}}.

Hence, in all of the above cases, we have shown that the Lie derivative
of all near-horizon data (i.e. the metric on ${\cal{S}}$, $h$,
$z^\alpha$, $\Phi^I$, $Q^I$, and $\Delta$) with respect to $W$ vanishes. We remark that these conditions, together with
({\ref{kappabos}}) imply that in all cases $\cL_W \kappa=0$
as well. Furthermore, one also has $\cL_W \parallel \eta_- \parallel^2=0$ in all cases as well. To see this, take the
Lie derivative of ({\ref{auxd1}}) with respect to $W$ to obtain
\bea
d \big( \cL_W \parallel \eta_- \parallel^2 \big)
= - \big( \cL_W \parallel \eta_- \parallel^2 \big) h~.
\eea
As $\cL_W \parallel \eta_- \parallel^2$ must vanish at some point in ${\cal{S}}$, this condition implies that
$\cL_W \parallel \eta_- \parallel^2=0$ everywhere on ${\cal{S}}$.

\appendix{$1/2$ BPS Near-Horizon Geometries}

It is instructive to describe the  half-supersymmetric near-horizon geometries constructed in
\cite{kzor} in terms of  Gaussian null co-ordinates,
and  extract all the near-horizon data
associated with the solutions. This will incorporate these solutions into our classification scheme
and so there will be a unified description of all near horizon geometries of ${\cal N}=2$ gauged
supergravity coupled to any number of multiplets.

In the spacetime coordinates $(t, z, x, v)$  the metric of the solutions  given in  \cite{kzor}
is
\bea
\label{nmet1}
ds^2 &=& -z^2 e^v \bigg(dt + 4(e^{-2 v}-L)z^{-1} dx \bigg)^2
+4e^{-v} z^{-2} dz^2
\nonumber \\
&+&16e^{-v}(e^{-2 v}-L)dx^2 +{4e^{-2v} \over
Y^2(e^{-v}-Le^v)} d v^2~,
\eea
where $L>0$ is constant, and
\bea
Y^2 = 64 g^2 e^{-v} |\xi_I X^I|^2-1~.
\eea
The scalars depend only on $v$, and satisfy
\bea
\label{scvar}
{dz^\alpha \over d v} = {i \over 2  \xi_I {\bar{X}}^I Y}
(1-iY) g^{\alpha {\bar \beta}} {\cal{D}}_{\bar{\beta}} \bigg( \xi_J {\bar{X}}^J \bigg)~.
\eea
Hence the scalars are constant if and only if
\bea
{\cal{D}}_\alpha \big(\xi_I X^I\big) =0
\eea
Note in particular that ({\ref{scvar}}) implies that
\bea
{d \over dv} \big(|\xi_I X^I|^2 \big)
=  g^{\alpha {\bar{\beta}}}
{\cal{D}}_\alpha (\xi_I X^I) {\cal{D}}_{\bar{\beta}} (\xi_J {\bar{X}}^J)~.
\eea
The gauge field strengths are given by
\bea
\label{gfst}
F^I &=& 8ig \bigg( {\xi_J {\bar{X}}^J \over 1-iY} X^I
- {\xi_J X^J \over 1+iY} {\bar{X}}^I \bigg) dt \wedge dz
\nonumber \\
&+&{4 \over Y} \bigg({2\xi_J {\bar{X}}^J \over 1-iY} X^I
+ {2\xi_J X^J \over 1+iY} {\bar{X}}^I + {\rm Im}{\cal{N}}^{-1IJ} \xi_J \bigg)
(z dt -4L dx) \wedge dv~.
\eea

In order to rewrite the metric ({\ref{nmet1}}) in Gaussian null co-ordinates, we set
\bea
w=e^v, \quad t=u+{4 \over wr},\quad x={1 \over 2  \sqrt{L}}
(\psi+\log (wr)), \quad z=-{{\sqrt{L}} \over 2} wr~.
\eea
Then in the co-ordinates $(u,r,\psi,w)$ the metric is
\bea
ds^2 &=&-{1 \over 4} L w^3 r^2 du^2 +2 du dr +2r du \big((1-Lw^2)d \psi
+w^{-1}dw \big)
\nonumber \\
&+&4(w^{-1}-Lw) d\psi^2 + {4 w^{-4} \over Y^2(w^{-1}-Lw)} dw^2~.
\eea
It follows that the near-horizon data are given by
\bea
\Delta = {L \over 4}w^3, \quad h=(1-Lw^2)d \psi +w^{-1} dw~,
\eea
and
\bea
ds^2_{\cal{S}}= 4(w^{-1}-Lw) d\psi^2 + {4 w^{-4} \over Y^2(w^{-1}-Lw)} dw^2~.
\eea
We choose the volume form on ${\cal{S}}$ to be
\bea
{\rm dvol}_{\cal{S}} = -4 w^{-2}Y^{-1} d\psi \wedge dw~,
\eea
and with this convention, it is straightforward to prove that the scalars in
({\ref{scvar}}) satisfy ({\ref{alg5c}}).

It is also straightforward to compute $\Phi^I$ and $Q^I$ from ({\ref{gfst}});
one finds
\bea
\Phi^I+iQ^I = 4i \sqrt{L} gw \bigg({1-iY \over 1+iY} \bigg) \xi_J X^J {\bar{X}}^I
+2i \sqrt{L} gw \bigg(\xi_J {\rm Im} {\cal{N}}^{-1IJ}+2 \xi_J X^J {\bar{X}}^I\bigg)~.
\eea
In particular, this expression implies that
\bea
{\rm Im}{\cal{N}}_{IJ} X^I (\Phi^J+i Q^J)
= -2i \sqrt{L} gw \xi_J X^J \bigg({1-iY \over 1+iY} \bigg)~,
\eea
and hence
\bea
\Phi^I +i Q^I = -2 {\rm Im}{\cal{N}}_{JN} X^J (\Phi^N+i Q^N) {\bar{X}}^I
+2i \sqrt{L} gw \bigg(\xi_J {\rm Im} {\cal{N}}^{-1IJ}+2 \xi_J X^J {\bar{X}}^I\bigg)~,
\eea
which is consistent with ({\ref{alg5b}}) on setting
\bea
{\langle \phi_+, \Gamma_5 \phi_+ \rangle \over \parallel \phi_+ \parallel^2}
=-\sqrt{L} w~.
\eea

For convenience, we shall also list here a number of useful identities
associated with this class of solutions:
\bea
\label{identx1}
\xi_I \Phi^I = 8 \sqrt{L} gw {|\xi_I X^I|^2 Y \over 1+Y^2}~,
\eea
\bea
\label{identx2}
{dz^\alpha \over dw} = {i \over 2 w \xi_I {\bar{X}}^I Y} (1-iY)
\xi_J {\cal{D}}_{\bar{\beta}} {\bar{X}}^J g^{\alpha \bar{\beta}}~,
\eea
and
\bea
\label{identx3}
A+iB= {2 \sqrt{L} iwg \over 1-iY} \xi_I {\bar{X}}^I~,
\eea
where $A$ and $B$ are defined in ({\ref{extrasc}}).  Furthermore, one can establish
\bea
\label{identx4}
{dY \over dw} = 32 g^2 w^{-2} Y^{-1} \bigg(-{1 \over 2} {\rm Im}{\cal{N}}^{-1 IJ} \xi_I \xi_J -2 |\xi_I X^I|^2 \bigg)~,
\eea
and
\bea
\label{identx5}
{d \Phi^I \over dw}= -4 \sqrt{L} g Y^{-1} \bigg( {1 \over 2} {\rm Im}{\cal{N}}^{-1 IJ} \xi_J + \bigg({1+iY \over 1-iY}\bigg) \xi_J {\bar{X}}^J X^I + \bigg({1-iY \over 1+iY}\bigg) \xi_J X^J {\bar{X}}^I \bigg)\ .
\eea
These formulae provide a useful check on our computations.

\newsection{Geometry of the Near-Horizon Solutions}

The description of the local geometry of horizons depends on whether the
vector field $W$ associated with (\ref{extraiso}) vanishes or not. As it has been presented in detail in appendix E, $W$
is a symmetry of the full solution. In what follows it is useful to consider the identity
\bea
\label{auxd1}
d \parallel \eta_- \parallel^2 = - \parallel \eta_- \parallel^2 h -W~.
\eea
This is one of the identities presented in (\ref{concon}).  It can also been proven directly using  ({\ref{pt2}}).
We shall first consider the special case when $W \equiv 0$.

\subsection{Solutions with $W \equiv 0$}

All these solutions are warped products  $\mathrm{AdS}_2\times_w {\cal S}$.
In this case,  ({\ref{auxd1}}) implies that
\bea
\label{vaniso1}
d \parallel \eta_- \parallel^2 = - \parallel \eta_- \parallel^2 h~,
\eea
and as $\parallel \eta_- \parallel^2$ is nowhere vanishing, one concludes that $dh=0$.
We remark that these solutions are distinct from the class of
half-supersymmetric BPS near-horizon solutions in \cite{kzor},
because for those solutions $dh \neq 0$.

Next ({\ref{ein2}}) can be rewritten as
\bea
\hn^i \hn_i \big( \Delta \parallel \eta_- \parallel^2\big)
+{1 \over \parallel \eta_- \parallel^2} \hn^i \big(\parallel \eta_- \parallel^2) \hn_i \big( \Delta \parallel \eta_- \parallel^2\big)
=
\nonumber \\-2 \parallel \eta_- \parallel^2 {\rm Im} \delta^{ij} {\cal{N}}_{IJ}
d_h\Phi^I_i d_h\Phi^J_j~.
\eea
As ${\rm Im} {\cal{N}}_{IJ}$ is negative definite, an application of the
maximum principle gives the conditions
\bea
\Delta \parallel \eta_- \parallel^2 = \mathrm{const}~,
\eea
and
\bea
d \Phi^I - \Phi^I h =0~.
\eea
Also, ({\ref{ein3}}) implies that
\bea
d \Delta - \Delta h=0~.
\eea
This condition implies that either $\Delta=0$ everywhere, or together with (\ref{alg1a})
$\Delta>0$ everywhere.
Also, ({\ref{alg3}}) implies that
\bea
\xi_I \Phi^I \Theta_+ \phi_+ =0~.
\eea
It follows using (\ref{concon}) that either $\Delta=0$ or $\xi_I \Phi^I=0$.

There are no solutions with $\Delta=0$ and   $\xi_I \Phi^I \neq 0$. To see this observe that
 ({\ref{ein1}}) can be rewritten as
\bea
\label{lapl2}
\hn^i \hn_i \parallel \eta_- \parallel^2
=-2 \parallel \eta_- \parallel^2 \bigg({1 \over 2} {\rm{Im}}{\cal{N}}_{IJ}
(\Phi^I \Phi^J + Q^I Q^J)+V \bigg) \ .
\eea
As the right-hand-side of this expression is non-negative, an application of the
maximum principle implies that $\parallel \eta_- \parallel^2=const$
and that $\Phi^I=0$. However, this
is in contradiction to the assumption that $\xi_I \Phi^I \neq 0$.

Furthermore there are no solutions with $\Delta=\xi_I \Phi^I=0$.
 If $\Delta=0$ then ({\ref{lapl2}}) again holds, which implies
\bea
\Phi^I=Q^I=0, \qquad V=0~,
\eea
and $\parallel \eta_- \parallel^2=const.$
The latter condition implies that $h=0$ as a consequence of
({\ref{vaniso1}}).
In addition, $\Delta=0$
implies that $\Theta_+ \eta_+=0$ as a consequence of ({\ref{alg1a}}) and ({\ref{thetnorm}}).
This, together with the previous conditions, implies
\bea
\xi_I ({\rm Im}X^I +i\Gamma_5 {\rm Re}X^I) \eta_+=0~,
\eea
and hence
\bea
\xi_I X^I=0 \ .
\eea
However, the conditions $\xi_I X^I=0$ and $V=0$ then lead to a contradiction. So we must have $\Delta>0$ everywhere and $\xi_I \Phi^I=0$.

The condition $\xi_I \Phi^I=0$ implies that
\bea
{\rm Re} \bigg({ A-iB \over \xi_I X^I } \bigg) =0 \ .
\eea
Also, as $\Delta >0$ everywhere, $A+iB \neq 0$. Then
({\ref{alg7a}}) and ({\ref{alg5c}}) imply that
\bea
\label{simpsc1}
\hn_i z^\alpha = {1 \over 2 \xi_J {\bar{X}}^J} \xi_I {\cal{D}}_{\bar{\beta}} {\bar{X}}^I
g^{\alpha {\bar{\beta}}} h_i~.
\eea
It will be convenient to define
\bea
\tau= \star_{\cal{S}} h~.
\eea
Then ({\ref{simpsc1}}) implies that
\bea
\cL_\tau z^\alpha=0~.
\eea
In turn, using ({\ref{alg4a}}) and ({\ref{alg1a}}), respectively, one has  that
\bea
\cL_\tau A = \cL_\tau B =0~,
\eea
and
\bea
\cL_\tau \Delta =0~.
\eea
Also, as $dh=0$ and $i_\tau h=0$, we also have
\bea
\cL_\tau h=0~,
\eea
and ({\ref{kappader}}) implies
\bea
\cL_\tau \kappa =0~,
\eea
as well. These conditions, together with ({\ref{extrasc}}) imply
\bea
\cL_\tau \bigg( {\rm Im}{\cal{N}}_{IJ} (\Phi^I +iQ^I) X^J \bigg) =0~,
\eea
and it therefore follows from ({\ref{alg5b}}) that
\bea
\cL_\tau \Phi^I = \cL_\tau Q^I=0~.
\eea
In addition, $\cL_\tau \kappa=0$ and $\cL_\tau X^I=0$ imply, together with
({\ref{kappabos}}) that
\bea
\cL_\tau h^2=0 \ .
\eea
It then follows from ({\ref{ein1}}) that
\bea
\cL_\tau \big( \hn^i h_i \big) =0 \ .
\eea
We shall consider two subcases, corresponding to $h \equiv 0$ and $h \not \equiv 0$.

\subsubsection{Solutions with $W \equiv 0$ and $h \equiv 0$}

For solutions with $W \equiv 0$ and $h=0$, the previously obtained conditions on
the bosonic fields imply that $z^\alpha$, $\kappa$, $A$, $B$, $\Delta$, $\Phi^I$
and $Q^I$ are all constant, with $\Delta>0$. The spacetime geometry is a product
 $AdS_2 \times {\cal{S}}$ described in section 6.1.1.

\subsubsection{Solutions with $W \equiv 0$ and $h \not \equiv 0$.}

For solutions with $W \equiv 0$ and $h \not \equiv 0$, it is convenient to introduce
local co-ordinates $\psi$ and $x$ on ${\cal{S}}$ so that
\bea
\tau = {\partial \over \partial \psi},  \qquad h= dx~.
\eea
A local basis for ${\cal{S}}$ is then given by
\bea
\bbe^1= {1 \over \sqrt{h^2}} dx \qquad \bbe^2 = \sqrt{h^2} \big(d \psi + q(x,\psi) dx\big)~,
\eea
where $h^2=h^2(x)$.
The condition $\cL_\tau (\hn^i h_i)=0$ then implies that
\bea
{\partial^2 q \over \partial \psi^2}=0~,
\eea
and so we have
\bea
q=q_0(x)+\psi q_1(x)~.
\eea
A co-ordinate transformation of the form
\bea
\psi = f_1(x) \psi'+f_2(x)~,
\eea
for appropriately chosen functions $f_1, f_2$ can  be used to further simplify
the basis for ${\cal{S}}$:
\bea
\bbe^1 = {1 \over \sqrt{h^2}} dx, \qquad \bbe^2=\sqrt{h^2}P d \psi'~,
\eea
with $\tau= h^2 P d \psi'$, where $P=P(x)$. We shall now drop the prime on $\psi'$.
The scalars $z^\alpha$, together with $\kappa$, $\Delta$, $h^2$, $P$, $\Phi^I$ and $Q^I$
are independent of the co-ordinate $\psi$, as are all components of the metric.

After some calculation, the Einstein equations ({\ref{ein1}}) and ({\ref{ein4}}) imply that
\bea
\label{h2der1}
\bigg(16g^2 \kappa^2 |\xi_I X^I|^2
-8g \kappa {\rm Im} \bigg(\xi_L {\bar{X}}^L {\rm Im}{\cal{N}}_{IJ}(\Phi^I+iQ^I)X^J \bigg)
\nonumber \\
-{1 \over |\xi_L X^L|^2} g^{\alpha \bar{\beta}} \xi_I \cD_\alpha X^I
\xi_J \cD_{\bar{\beta}} {\bar{X}}^J h^2 \bigg) h_i + \hn_i h^2=0~,
\eea
and ({\ref{simpsc1}}) implies that
\bea
\label{simpsc2}
\hn_i |\xi_I X^I|^2 = g^{\alpha {\bar{\beta}}} \xi_I \cD_\alpha X^I
\xi_J \cD_{\bar{\beta}} {\bar{X}}^J h_i~.
\eea
Furthermore, note that
\bea
\label{Peqn}
\hn^i h_i = {d h^2 \over dx} + {h^2 \over P} {dP \over dx}~.
\eea
On making use of ({\ref{ein1}}) and ({\ref{h2der1}}),
together with ({\ref{alg5b}}) and ({\ref{alg1a}}), we
find that
\bea
\hn^i h_i -{d h^2 \over dx } = -{1 \over 2} h^2 \bigg(1+{1 \over |\xi_L X^L|^2} \xi_I \cD_{\alpha} X^I \xi_J \cD_{\bar{\beta}} {\bar{X}}^J
g^{\alpha \bar{\beta}} \bigg)~.
\eea
It follows that ({\ref{Peqn}}) implies
\bea
P^{-1} {dP \over dx} = -{1 \over 2} \bigg(1+{d \over dx} \log |\xi_I X^I|^2 \bigg)~,
\eea
and so
\bea
P = {L e^{-{x \over 2}} \over |\xi_I X^I|}~,
\eea
for constant $L$.

Next note that ({\ref{alg4a}}) implies
that
\bea
{d \over dx} \log \bigg( \big({A-iB \over \xi_I X^I} \big)^2 |\xi_J X^J|^2 \bigg) =1~,
\eea
so that
\bea
{A-iB \over \xi_I X^I} = {i \nu \over |\xi_J X^J|} e^{x \over 2}~,
\eea
for $\nu \in \bR$ constant, and ({\ref{alg1a}}) then implies
\bea
\Delta =4 \nu^2 e^x~.
\eea
As we require that $\Delta \neq 0$, we must take $\nu \neq 0$. The scalar $\kappa$
then satisfies
\bea
\label{kappader2}
{d \kappa \over dx} = \kappa -{\nu \over 2g |\xi_I X^I|} e^{x \over 2}~,
\eea
as a consequence of ({\ref{kappader}}), and $h^2$ is then given by ({\ref{kappabos}}) as
\bea
h^2 = 16g^2 |\xi_I X^I|^2 (1-\kappa^2) \ .
\eea

The near-horizon data for this class of solutions have been collected in (\ref{solwzero}).  The dependence of the fields
in terms of $x$ is determined  by the equations (\ref{wzeropq}) and (\ref{wzerozk}).

\subsection{Solutions with $W \not \equiv 0$}

As we have already mentioned $W$ leaves all the fields invariant. In addition, the Lie derivatives of $\kappa$, and $\parallel \eta_- \parallel^2$ with respect to $W$ also vanish.
We present the proof of these in Appendix E.

\subsubsection{Solutions with $W \not \equiv 0$, and $\kappa=\mathrm{const}$ with $|\kappa|\not=1$ }

First we consider the special case for which $\kappa=\mathrm{const}$.
Then ({\ref{auxd1}}) implies that if $h \equiv 0$,
then $W \equiv 0$. So it follows that $h \not \equiv 0$,
and hence ({\ref{kappader}}) implies
\bea
A-iB =2ig \kappa \xi_I X^I~.
\eea

Then ({\ref{alg5b}}) gives that
\bea
\label{simpcharges1}
\Phi^I+iQ^I =-2ig \kappa \bigg( \xi_J {\rm Im}{\cal{N}}^{-1IJ}
+4 \xi_J X^J {\bar{X}}^I \bigg)~,
\eea
and ({\ref{alg1a}}) implies that
\bea
\Delta = 16 g^2 \kappa^2 |\xi_I X^I|^2~.
\eea
In particular, ({\ref{simpcharges1}}) implies that $\xi_I \Phi^I=0$, and hence ({\ref{alg1b}}) implies that
\bea
dh=0~.
\eea
The Einstein equation ({\ref{ein1}}) implies that
\bea
\label{hderspec}
\hn^i h_i = 2(1-\kappa^2) \bigg(4g^2 g^{\alpha {\bar{\beta}}}
\xi_I \cD_\alpha X^I \xi_J \cD_{\bar{\beta}} {\bar{X}}^J
-4g^2 |\xi_I X^I|^2 \bigg)~,
\eea
and ({\ref{alg5c}}) implies that
\bea
g_{\alpha {\bar{\beta}}} \hn_i z^{\alpha}
\hn^i z^{\bar{\beta}} = 4g^2 (1-\kappa^2) g^{\alpha {\bar{\beta}}}
\xi_I \cD_\alpha X^I \xi_J \cD_{\bar{\beta}} {\bar{X}}^J~.
\eea
So, on taking the trace of the Einstein equation
({\ref{ein4}}) we find
\bea
{\hat{R}}=8g^2(1+\kappa^2)
\bigg( g^{\alpha {\bar{\beta}}}
\xi_I \cD_\alpha X^I \xi_J \cD_{\bar{\beta}} {\bar{X}}^J
- |\xi_I X^I|^2 \bigg)~,
\label{topt}
\eea
and hence
\bea
{\hat{R}}={(1+\kappa^2) \over (1-\kappa^2)} \hn^i h_i~.
\eea
Thus ${\cal S}$ is topologically $T^2$.

There are two different cases to consider, corresponding as
to whether $\parallel \eta_- \parallel^2$ is constant,
or not constant.

If $\parallel \eta_- \parallel^2$ is constant, then
({\ref{auxd1}}) implies that
\bea
\parallel \eta_- \parallel^2 h +W=0~.
\eea
As $dh=0$ this implies that $dW=0$. Hence, it follows that
both $h$ and $W$ are covariantly constant on ${\cal{S}}$.
Therefore ${\cal{S}}=T^2$, and ${\hat{R}}=0$ implies that
\bea
g^{\alpha {\bar{\beta}}}
\xi_I \cD_\alpha X^I \xi_J \cD_{\bar{\beta}} {\bar{X}}^J
= |\xi_I X^I|^2, \quad {\rm and} \quad  {\rm Im}{\cal{N}}^{-1 IJ} \xi_I \xi_J
= -4 |\xi_I X^I|^2~.
\eea
As $h^2$ is constant, it follows from ({\ref{kappabos}})
that $|\xi_I X^I|^2$ is constant, and also $\Delta$ is constant.
Furthermore, ({\ref{alg5c}}) implies that
\bea
\hn_i z^\alpha = {i \over 2 \xi_J {\bar{X}}^J} \xi_I \cD_{\bar{\beta}} {\bar{X}}^I g^{\alpha {\bar{\beta}}} \epsilon_i{}^j h_j~.
\eea
It is straightforward to obtain local co-ordinates for the metric;
as $h$ is covariantly constant, we can introduce local co-ordinates
$x$, $y$ on ${\cal{S}}$ such that
\bea
h=dx, \qquad \star_{\cal{S}} h =dy~,
\eea
so that the $z^\alpha$, $\Phi^I$ and $Q^I$ depend only on $y$. The metric and equations that determine the dependence of
the remaining fields on $x$ are summarized in section 6.2.1.

Next, consider the case for which $\parallel \eta_- \parallel^2$
is not constant.
As $\cL_W h=0$ and $dh=0$ it follows that
$i_W h=const.$ Furthermore, from ({\ref{auxd1}}), together with
$\cL_W \parallel \eta_- \parallel^2 =0$, it follows that
\bea
\parallel \eta_- \parallel^2 i_W h + W^2=0~,
\eea
and hence $i_W h<0$. We shall set $i_W h=-\mu^2$, and we shall
furthermore introduce local co-ordinates $x$ and $\psi$ on
${\cal{S}}$ such that
\bea
W = {\partial \over \partial \psi}, \qquad x= \parallel \eta_- \parallel^2~,
\eea
with
\bea
h=-\mu^2 d \psi \ .
\eea
Then ({\ref{auxd1}}) implies that
\bea
\label{norm1}
\hn_i x \hn^i x =h^2 x^2-\mu^2 x~,
\eea
and moreover
\bea
\label{norm2}
W^2=\mu^2 x \ .
\eea
As $i_W dx=0$, it follows that
\bea
dx=\beta \star_{\cal{S}} W~,
\eea
for some function $\beta$, and on taking the norm of both
sides of this expression, using ({\ref{norm1}}) and ({\ref{norm2}}) one finds  that
\bea
dx= \mu^{-1} \sqrt{h^2 x - \mu^2} \star_{\cal{S}} W~.
\eea
On substituting this expression back into ({\ref{auxd1}}) it follows that
\bea
\label{hdecomp1}
h=-x^{-1} W - \mu^{-1} x^{-1} \sqrt{h^2 x-\mu^2} \star_{\cal{S}} W~.
\eea
Next,substituting this expression into ({\ref{alg5c}}) and using the fact that
$d z^\alpha$ must be proportional to $\star_{\cal{S}} W$, we get that
\bea
\label{zcondauxb}
\hn_i z^{\alpha}= -{i \over 2x} \bigg(1-{i \over \mu} \sqrt{h^2x-\mu^2}
\bigg) {1 \over \xi_J {\bar{X}}^J} \xi_I \cD_{\bar{\beta}} {\bar{X}}^I g^{\alpha \bar{\beta}}
\epsilon_{ij} W^j~,
\eea
or equivalently
\bea
{dz^\alpha \over dx} = -{i \over 2x} \bigg({\mu \over \sqrt{h^2 x-\mu^2}}-i \bigg)
{1 \over \xi_J {\bar{X}}^J} \xi_I \cD_{\bar{\beta}} {\bar{X}}^I g^{\alpha \bar{\beta}}~.
\eea

Next we shall consider the conditions ({\ref{alg7a}}), ({\ref{alg3a}}) and ({\ref{alg4a}}).
In evaluating these expressions, we make use of ({\ref{zcondauxb}}), together with
\bea
\label{phider2}
\hn_i \Phi^I &=& 2 g \kappa x^{-1} \bigg( -\big(1+i\mu^{-1} \sqrt{h^2 x-\mu^2}\big)
{1 \over \xi_J X^J} \xi_L \cD_{\bar{\alpha}} {\bar{X}}^L \xi_N \cD_\beta X^N
g^{{\bar{\alpha}} \beta} X^I
\nonumber \\
&+& \big(1-i\mu^{-1} \sqrt{h^2 x-\mu^2}\big) \cD_\alpha X^I \xi_N \cD_{\bar{\beta}} {\bar{X}}^N g^{\alpha
\bar{\beta}}
\nonumber \\
&-&\big(1-i\mu^{-1} \sqrt{h^2 x-\mu^2}\big) {1 \over \xi_J {\bar{X}}^J} \xi_L \cD_\alpha X^L
\xi_N \cD_{\bar{\beta}} {\bar{X}}^N g^{\alpha \bar{\beta}} {\bar{X}}^I
\nonumber \\
&+& \big(1+i\mu^{-1} \sqrt{h^2 x-\mu^2}\big) \cD_{\bar{\alpha}} {\bar{X}}^I \xi_N \cD_\beta X^N
g^{\bar{\alpha} \beta} \bigg) \epsilon_{ij} W^j~,
\eea
and
\bea
h_i -i \epsilon_{ij} h^j = -x^{-1} \big(1+i\mu^{-1} \sqrt{h^2 x-\mu^2}\big) \bigg(W_i -i \epsilon_{ij} W^j \bigg)~,
\eea
and
\bea
\label{auxpein2}
\hn_i |\xi_I X^I|^2 = -x^{-1} \mu^{-1} \sqrt{h^2 x-\mu^2} \xi_I \cD_\alpha X^I
\xi_J \cD_{\bar{\beta}} {\bar{X}}^J g^{\alpha {\bar{\beta}}} \epsilon_{ij} W^j~.
\eea
Then on decomposing ({\ref{alg4a}}) into directions parallel and orthogonal to $W$, we find the condition
\bea
\label{auxcondv2}
\kappa \bigg( {\rm Im}{\cal{N}}^{-1 IJ} \xi_I \xi_J +4 |\xi_I X^I|^2 \bigg) =0~.
\eea
This condition is sufficient to ensure that ({\ref{alg7a}}), ({\ref{alg3a}}) and ({\ref{alg4a}})
are satisfied.

Suppose that $\kappa \neq 0$. Then the condition ({\ref{auxcondv2}}), together with
({\ref{hderspec}}) implies that
\bea
\hn^i h_i=0~,
\eea
and it follows on taking the divergence of ({\ref{auxd1}}) that
\bea
\hn^i \hn_i \parallel \eta_- \parallel^2 + h^i \hn_i \parallel \eta_- \parallel^2 =0~.
\eea
An application of the maximum principle then implies that $\parallel \eta_- \parallel^2 = \mathrm{const}$,
but this is in contradiction to our assumption that $\parallel \eta_- \parallel^2$ is not constant.
So, for this class of solutions, we must have $\kappa=0$, which in turn implies that
\bea
\Delta=0, \qquad \Phi^I=Q^I=0, \qquad h^2=16g^2 |\xi_I X^I|^2~.
\eea

It remains to choose a local basis for ${\cal{S}}$; we take
\bea
\bbe^1 = \mu^{-1} x^{-{1 \over 2}} W = \mu^{-1} x^{-{1 \over 2}} (\mu^2 x d \psi -dx)~,
\eea
and
\bea
\bbe^2 = \mu^{-1} x^{-{1 \over 2}} \star_{\cal{S}} W = {x^{-{1 \over 2}} \over \sqrt{h^2 x-\mu^2}}
dx~,
\eea
so that
\bea
ds^2_{\cal{S}}={1 \over x} \bigg(\mu^{-2} (\mu^2 x d \psi -dx)^2+{1 \over h^2 x-\mu^2} dx^2 \bigg)~.
\eea
This metric can be simplified further by changing co-ordinates as
\bea
x = \mu^2 x', \qquad \psi = \mu^{-2} \psi'~,
\eea
to obtain (on dropping primes)
\bea
ds^2_{\cal{S}} = {1 \over x} \bigg( \big(x d \psi-dx\big)^2 +{1 \over 16 g^2 |\xi_I X^I|^2 x -1}
dx^2 \bigg)~,
\eea
with
\bea
h = -d \psi~.
\eea
The results have been summarized in section 6.2.1.
The spacetime metric and the equations that determine the dependence of the scalars on $x$ are given in
(\ref{wconstm}) and (\ref{wconstz}), respectively.

\subsubsection{Solutions with $W \not \equiv 0$ and $\kappa \neq const.$}

To proceed with the analysis, we first make use of
({\ref{kappader}}) in order to write $h$ in terms of
$d \kappa$ and $\star_{\cal{S}} d \kappa$. We find
\bea
h = {1 \over \chi} \bigg( \big(\kappa-{\rm Im}({A-iB \over 2g \xi_I X^I}) \big) d \kappa - {\rm Re} ({A-iB \over 2g \xi_I X^I}) \star_{\cal{S}} d \kappa \bigg)~,
\eea
where
\bea
\chi = \big(\kappa-{\rm Im}({A-iB \over 2g \xi_I X^I}) \big)^2+
\big({\rm Re} ({A-iB \over 2g \xi_I X^I})\big)^2~.
\eea
As $d z^\alpha$ must be proportional to $d \kappa$,
({\ref{alg5c}}) implies that
\bea
\label{zzder}
\hn_i z^\alpha ={1 \over 2 \chi \xi_J {\bar{X}}^J}
\bigg(\kappa+{i(A-iB) \over 2g \xi_I X^I}\bigg)
 \xi_L \cD_{\bar{\beta}} {\bar{X}}^L g^{\alpha \bar{\beta}}
\hn_i \kappa~.
\eea
Next, we consider ({\ref{alg4a}}), and decompose the resulting
expression into terms parallel and orthogonal to $\hn_i \kappa$,
by noting that
\bea
h+i \star_{\cal{S}} h
={1 \over \chi} \bigg(\kappa+{i(A-iB) \over 2g \xi_I X^I} \bigg)
\bigg(d \kappa +i \star_{\cal{S}} d \kappa \bigg)~.
\eea
On eliminating the terms involving $\hn \Phi^I$ from the two
expressions obtained in this fashion, we find
\bea
\label{compder}
\hn_i (A+iB)
&=&{1 \over \chi} \bigg(\kappa -{i (A+iB) \over 2g
\xi_L {\bar{X}}^L}\bigg) \bigg({1 \over 2}(A+iB)
-{\xi_J \Phi^J \over 4 \xi_I X^I} \bigg) \hn_i \kappa
\nonumber \\
&+&i(A+iB) A_i~.
\eea
In fact, the remaining parts of ({\ref{alg7a}}),
({\ref{alg3a}}) and ({\ref{alg4a}}) also hold automatically.
This makes use of ({\ref{alg1a}}) and ({\ref{zzder}}).
Furthermore, using ({\ref{alg5b}}) together with ({\ref{compder}}) and ({\ref{zzder}}), we find that
\bea
\Phi^I =-2(A-iB) {\bar{X}}^I -2(A+iB)X^I~.
\eea
 One then finds
\bea
\label{phder}
\hn_i \Phi^I &=&-{1 \over \chi}
{\rm Re} \bigg(4 (A+iB) (\kappa-{i(A+iB) \over 2g \xi_N {\bar{X}}^N}) X^I
- {\kappa (A+iB) \over \xi_L {\bar{X}}^L} {\rm Im} {\cal{N}}^{-1 IJ} \xi_J \bigg) \hn_i \kappa~.
\nonumber \\
\eea
Using these expressions, the remaining content of ({\ref{alg7a}}),
({\ref{alg3a}}) and ({\ref{alg4a}}) holds automatically.

To proceed, we return to the condition ({\ref{compder}}). Motivated by the expression
for $A+iB$ in ({\ref{identx3}}) for the example in Appendix F, we set
\bea
A+iB =\kappa \xi_I {\bar{X}}^I \cG~.
\eea
Then ({\ref{compder}}) can be rewritten as
\bea
\label{compderv2}
{d \cG \over d \kappa} = {\kappa^{-1} \over 1+{i \over 2g} \cbG}
\bigg({1 \over 2} \cbG (1-{i \over g} \cG) -{1 \over 2|\xi_L X^L|^2}
\cG \xi_I \cD_\alpha X^I \xi_J \cD_{\bar{\beta}} {\bar{X}}^J g^{\alpha \bar{\beta}} \bigg)~.
\eea
On taking the complex conjugate of ({\ref{compderv2}}), one obtains the following
condition
\bea
\label{argode}
{d \over d \kappa} \log \bigg({ {i \over 2g}+{1 \over \cbG} \over -{i \over 2g}+{1 \over \cG}}\bigg)
={\kappa^{-1} \over 2 |\cG|^2 |1-{i \over 2g}\cG|^2} \big(\cG+\cbG\big)\big(\cbG-\cG-{i \over g} |\cG|^2\big)~.
\eea
To proceed further, we shall  set, see appendix F,
\bea
\cG = -{2ig \over 1-iY}~,
\eea
for $Y$ a complex function, where $Y \not \equiv 0$,
and $Y \not \equiv -i$~.
Then ({\ref{argode}}) is equivalent to
\bea
{d \over d \kappa} \bigg({ {\bar{Y}} \over Y} \bigg)
= {1 \over 2} \kappa^{-1} \bigg(1- \bigg({ {\bar{Y}} \over Y} \bigg)^2\bigg)~,
\eea
which has the general solution
\bea
\label{argsol}
{\bar{Y} \over Y} = {\kappa+ic \over \kappa-ic}~,
\eea
for constant $c \in \bR$.
Using this expression, we can eliminate $\cbG$ in favour of $\cG$ in ({\ref{compderv2}}) to
find
\bea
\label{compderv3}
{d \cG \over d \kappa}
={1 \over 2\kappa (\kappa+ic)} \bigg({\kappa \cG +ig(\kappa+ic) \over {1 \over 2} \cG +ig} \bigg)
\bigg({ig(\kappa-ic) \cG \over \kappa \cG+ig (\kappa+ic)} (1-{i \over g}\cG)
\nonumber \\
-{1 \over |\xi_L X^L|^2} \cG \xi_I \cD_\alpha X^I \xi_J \cD_{\bar{\beta}} {\bar{X}}^J g^{\alpha \bar{\beta}} \bigg)~,
\eea
and moreover, on using ({\ref{zzder}}) we also have
\bea
\label{compderv4}
{d \over d \kappa} |\xi_I X^I|^2
={1 \over 2 \kappa (\kappa+ic)} \bigg( {\kappa \cG + 2ig(\kappa+ic)
\over {1 \over 2} \cG +ig} \bigg)  \xi_I \cD_\alpha X^I \xi_J \cD_{\bar{\beta}} {\bar{X}}^J g^{\alpha \bar{\beta}}~.
\eea

We shall consider the cases for which $\kappa \cG +2ig(\kappa+ic)$ vanishes identically, and is non-zero, separately.

Suppose first that $\kappa \cG +2ig(\kappa+ic) \not \equiv 0$.
Then the conditions ({\ref{compderv3}}) and ({\ref{compderv4}})
can be combined to give
\bea
\label{compderv5}
{d \over d \kappa} \log \bigg({1 \over |\xi_I X^I|^2}
(1-iY)(i\kappa (1+iY)+c(1-iY)) \bigg) = {c \over \kappa(\kappa+ic)} (Y^{-1}+i)~.
\eea
Furthermore, we recall that $W= \beta \star_{\cal{S}} d \kappa$
for some function $\beta=\beta(\kappa)$. On substituting this
into the condition ({\ref{auxd1}}), one obtains
\bea
\label{compderv6}
\beta = {\parallel \eta_- \parallel^2 \over 2 g \kappa}
{{\rm Re} \cG \over (1+{i \over 2g}\cbG)(1-{i \over 2g}\cG)}~,
\eea
and
\bea
\label{compderv7}
{d \parallel \eta_- \parallel^2 \over d \kappa}
=-{\parallel \eta_- \parallel^2 \over 2 \kappa}
\bigg({1 \over 1-{i \over 2g} \cG} +{1 \over 1+{i \over 2g} \cbG}
\bigg)~.
\eea
Then ({\ref{compderv7}}) can be rewritten in terms of $Y$
as
\bea
\label{compderv8}
{d \over d \kappa} \log \parallel \eta_- \parallel^2 =
{c \kappa^{-1} Y^{-1} \over \kappa+ic}-\kappa^{-1}~.
\eea
Next  on combining ({\ref{compderv8}}) and
({\ref{compderv5}}), we find that the resulting condition
can be integrated up to give
\bea
\label{compderv9}
{(\kappa+ic)(1-iY)(i\kappa(1+iY)+c(1-iY)) \over \kappa^2
\parallel \eta_- \parallel^2 |\xi_I X^I|^2} =ip~,
\eea
for $p \in \bR$ constant, $p \neq 0$. To see that $p \neq 0$,
we rewrite ({\ref{compderv9}}) using ({\ref{argsol}}) as
\bea
{(1+c^2 \kappa^{-2}) |1-iY|^2 \over \parallel \eta_- \parallel^2
|\xi_I X^I|^2} = p \ .
\eea
To obtain local expressions for all the near-horizon data,
we take local co-ordinates $\kappa, \psi$ with $W={\partial \over \partial \psi}$ and take, without loss of generality
\bea
\label{compderv10}
W=S d \psi~,
\eea
for $S=S(\kappa)${\footnote{This can always be done by making use
of a co-ordinate transformation of the form $\psi=\psi'+H(\kappa)$}. Then
\bea
\label{compderv11}
S=W^2= \beta^2 \hn_i \kappa \hn^i \kappa
={4 (\parallel \eta_- \parallel^2)^2 (1-\kappa^2)|\xi_I X^I|^2 (
{\rm Re} \cG)^2 \over (1+{i \over 2g} \cbG)(1-{i \over 2g}\cG)}~,
\eea
where we have used ({\ref{compderv6}}) together with
({\ref{kappabos}}) and ({\ref{kappader}}).
This implies that
\bea
\star_{\cal{S}} d \kappa = 8g \kappa \parallel \eta_- \parallel^2
(1-\kappa^2)|\xi_I X^I|^2 ({\rm Re} \cG) d \psi~.
\eea
In addition, $\Delta$ is given by ({\ref{alg1a}}) as
\bea
\label{nhd1}
\Delta = {16g^2 \kappa^2 |\xi_I X^I|^2 \over |1-iY|^2}~,
\eea
and ({\ref{zzder}}) implies that
\bea
\label{nhd2}
{d z^\alpha \over d \kappa} = {1 \over 2 \kappa \xi_J {\bar{X}}^J}(1+iY^{-1}) \xi_I \cD_{\bar{\beta}} {\bar{X}}^I g^{\alpha \bar{\beta}}~.
\eea
It is convenient to set $\psi={p \over 16g^2}\phi$,
then the metric on ${\cal{S}}$ can be written
as
\bea
\label{nhd3}
ds^2_{\cal{S}} = \Delta^{-1} \bigg({1 \over |Y|^2 (1-\kappa^2)}
d \kappa^2 +(\kappa^2+c^2)(1-\kappa^2)d \phi^2 \bigg)~.
\eea
The expression for $h$ is obtained by using
({\ref{auxd1}}), together with ({\ref{compderv10}})
and ({\ref{compderv11}}) and ({\ref{compderv8}}),
to find
\bea
\label{nhd4}
h=\kappa^{-1} \bigg(1-{c \over (\kappa+ic)Y}\bigg)d \kappa
-(1-\kappa^2)d \phi~.
\eea
Furthermore, ({\ref{alg5b}}) implies
\bea
\label{nhd5}
\Phi^I+iQ^I = -{8ig \kappa \over 1+i {\bar{Y}}}
\xi_J X^J {\bar{X}}^I
-2ig \kappa {\rm Im}{\cal{N}}^{-1 IJ} \xi_J~.
\eea
The spacetime metric and the equations that determine the  near horizon fields are summarized in section 6.2.2.
The special case for which $\kappa \cG +2ig(\kappa+ic)=0$ is summarized in section 6.2.2

\appendix{Gauge Field Equations}

Here, we list the non-trivial content of the gauge field
equations ({\ref{geq2}}). In a number of cases, these hold automatically.
In the remaining cases, only one non-trivial component of ({\ref{geq2}})
needs to be checked  as the others can be shown to hold automatically.

The cases to be considered are

\begin{itemize}

\item[(1)] The class of solution in section (9.1.2).
The gauge field equation is
\bea
{d \over dx} \bigg( {\rm Im}{\cal{N}}_{IJ} Q^J \bigg)
+{d \over dx} \bigg( {\rm Re}{\cal{N}}_{IJ}  \bigg) \Phi^J
- {\rm Im}{\cal{N}}_{IJ} Q^J =0 \ .
\eea

\item[(2)] The first class of solutions in section (9.2.1) - for which
$\parallel \eta_- \parallel^2  = \mathrm{const}$, i.e. up to equation ({\ref{stopb}}).
For this case, the gauge field equation content is:

\bea
{d \over dy} \bigg( {\rm Im}{\cal{N}}_{IJ} Q^J \bigg)
+{d \over dy} \bigg( {\rm Re}{\cal{N}}_{IJ}  \bigg) \Phi^J
- {\rm Im}{\cal{N}}_{IJ} \Phi^J =0~.
\eea

\item[(3)] The solution of section (9.2.2). For this case, the gauge field equation content is:

\bea
&&{d \over d \kappa } \bigg( {\rm Im}{\cal{N}}_{IJ} Q^J \bigg)
+{d \over d \kappa} \bigg( {\rm Re}{\cal{N}}_{IJ}  \bigg) \Phi^J
- {\rm Im}{\cal{N}}_{IJ} \bigg( {1 \over (\kappa+ic)Y} \Phi^J
\cr &&~~~~~~~~~~~~~~~~~~~
- {c -(\kappa+ic)Y \over \kappa(\kappa+ic)Y} Q^J \bigg) =0~.
\eea

\end{itemize}

To evaluate these equations it is useful to first note that
({\ref{alg5b}}) implies that
\bea
Q^J = {\hat{Q}}^J -2g \kappa {\rm Im} {\cal{N}}^{-1 JL} \xi_L
\eea
where
\bea
\Phi^J-i {\hat{Q}}^J = W X^J
\eea
for some complex function $W$ whose precise form
depends on the case under consideration.
For all of the gauge field equations, we must evaluate a
term of the type
\bea
\label{specialgauge}
d({\rm Im}{\cal{N}}_{IJ} Q^J) +\Phi^J d \bigg({\rm Re}{\cal{N}}_{IJ} \bigg)
=-2g \xi_I d \kappa + {\rm Im}{\cal{N}}_{IJ} d {\hat{Q}}^J
+ {\rm Re} \bigg(WX^J d {\cal{N}}_{IJ} \bigg)~.
\eea

The final term in the
above expression can be rewritten using the conditions of special geometry.
In particular we have
\bea
X^J d {\cal{N}}_{IJ} =-2i {\rm Im}{\cal{N}}_{IJ} \cD_\alpha X^J dz^\alpha~,
\eea
where we have made use of the special K\"ahler geometry  identities (\ref{gcouple}) in
 appendix B.
On using these identities one obtains
\bea
&&d({\rm Im}{\cal{N}}_{IJ} Q^J) +\Phi^J d \bigg({\rm Re}{\cal{N}}_{IJ} \bigg)
=-2g \xi_I d \kappa + {\rm Im}{\cal{N}}_{IJ} d {\hat{Q}}^J
\cr &&~~~~~~~~~~~~~
+2 {\rm Im}{\cal{N}}_{IJ} {\rm Im} \bigg( W \cD_\alpha X^J dz^\alpha \bigg)~.
\eea
All of the terms in this expression can then be directly calculated
using the conditions we have found on the solutions.  In particular,
the dependence of $\kappa$ is known, the $d {\hat{Q}}^I$ term
can be calculated directly, as can $W$ and $dz^\alpha$.

\end{document}